\documentclass[reprint,twocolumn,prx]{revtex4-2}
\usepackage{amsmath,amssymb,braket,fancybox}
\usepackage{amsfonts}
\usepackage{graphicx}
\usepackage{comment,color}
\usepackage{ascmac}
\usepackage{ascmac}
\usepackage{amsthm}

\theoremstyle{definition}

\newtheorem*{theorem*}{Theorem}

\newtheorem*{definition*}{Definition}

\newcommand{\av}[1]{\overline{#1}}
\newcommand{\mc}[1]{\mathcal{#1}}
\newcommand{\mr}[1]{\mathrm{#1}}
\newcommand{\mbb}[1]{\mathbb{#1}}
\newcommand{\mbf}[1]{\mathbf{#1}}

\newcommand{\lrs}[1]{\left( #1 \right)}
\newcommand{\lrm}[1]{{\left\{ #1 \right\}}}
\newcommand{\lrl}[1]{\left[ #1 \right]}
\newcommand{\lrv}[1]{\left| #1 \right|}
\newcommand{\braketL}[1]{\left\langle #1 \right\rangle}

\newcommand{\fracd}[2]{\frac{\mathrm{d} #1 }{\mathrm{d} #2 }}

\newcommand{\fracpd}[2]{\frac{\partial #1 }{\partial #2 }}

\newcommand{\aln}[1]{
\begin{align}
#1
\end{align}
}

\newcommand{\ra}{\rightarrow}

\newcommand{\Tr}{\mr{Tr}}

\begin{document}
\title{
Speed Limits for Macroscopic Transitions
}
\date{\today}
\author{Ryusuke Hamazaki}
\affiliation{
Nonequilibrium Quantum Statistical Mechanics RIKEN Hakubi Research Team, RIKEN Cluster for Pioneering Research (CPR), RIKEN iTHEMS, Wako, Saitama 351-0198, Japan
}

\begin{abstract}
Speed of state transitions in macroscopic systems is a crucial concept for foundations of nonequilibrium statistical mechanics as well as various applications in quantum technology represented by optimal quantum control.
While extensive studies have made efforts to obtain rigorous constraints on dynamical processes since Mandelstam and Tamm, speed limits that provide tight bounds for macroscopic transitions have remained elusive.
Here, by employing the local conservation law of probability, the fundamental principle in physics, we develop a general framework for deriving qualitatively tighter speed limits for macroscopic systems than many conventional ones.
We show for the first time that the speed of the expectation value of an observable defined on an arbitrary graph, which can describe general many-body systems, is bounded by the ``gradient" of the observable, in contrast with conventional speed limits depending on the entire range of the observable.
This framework enables us to derive novel quantum speed limits for macroscopic unitary dynamics.
Unlike previous bounds, the speed limit decreases when the expectation value of the transition Hamiltonian increases; this intuitively describes a new tradeoff relation between time and  quantum phase difference.
Our bound is dependent on instantaneous quantum states and thus can achieve the equality condition, which is conceptually distinct from the Lieb-Robinson bound.
We also find that, beyond expectation values of macroscopic observables, the speed of macroscopic quantum coherence can be bounded from above by our general approach.
The newly obtained bounds are verified in transport phenomena in particle systems and nonequilibrium dynamics in many-body spin systems.
We also demonstrate that our strategy can be applied for finding new speed limits for macroscopic transitions in stochastic systems, including quantum ones, where the bounds are expressed by the entropy production rate.
Our work elucidates novel speed limits on the basis of local conservation law, providing fundamental limits to various types of nonequilibrium quantum macroscopic phenomena.

\end{abstract}

\maketitle

\section{Introduction}

\begin{figure*}
    \centering
    \includegraphics[width=\linewidth]{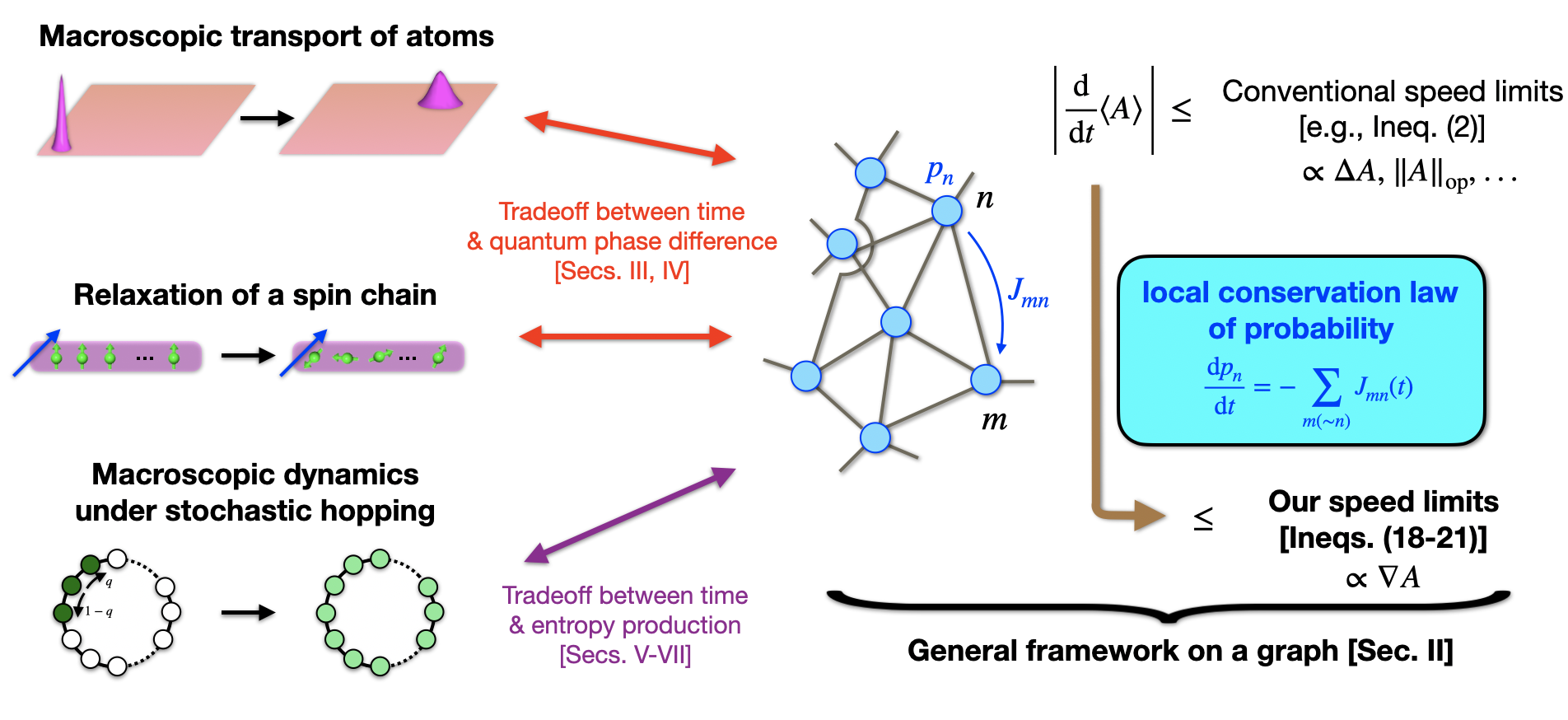}
    \caption{Schematic illustration of our achievements.
    We establish a general framework for deriving qualitatively tighter speed limits of a quantity $A$ than many conventional ones, which depend on the entire range of $A$, such as $\Delta A$ or $\|A\|_\mr{op}$. Our strategy is to map general dynamics of our interest to dynamics on a graph, where we use the local conservation of probability. In contrast with conventional bounds, our speed limits involve the gradient $\nabla A$ of $A$ on the graph, which can significantly tighten the bound when $\nabla A\ll \Delta A\text{ or }\|A\|_\mr{op}$.
    When applied to macroscopic quantum systems (such as macroscopic transport of atoms or relaxation of a locally perturbed spin chain), our theory indicates a novel tradeoff relation between time and quantum phase difference.
    When applied to macroscopic stochastic dynamics, including quantum one, our theory indicates a  tradeoff relation between time and quantities such as entropy production.
    }
    \label{figbegin}
\end{figure*}

Understanding how fast a state changes in time is a fundamental problem in nonequilibrium physics.
In 1945, Mandelstam and Tamm showed in their seminal work~\cite{mandelstam1945energy} that, in an isolated quantum system, time for an initial state to relax to a state orthogonal to it is rigorously lower bounded as
\aln{\label{MT}
T\geq T_\mr{MT}:=\frac{\pi\hbar}{2\Delta H},
}
where $\Delta H$ is the energy fluctuation of the system.
The appearance of the energy fluctuation is deeply related to the quantum uncertainty relation between energy and time.
Indeed, the derivation of inequality~\eqref{MT} can be carried out from a relation for a more general observable $\hat{A}$: the expectation value of its speed $\mr{d}{\hat{A}}/\mr{d}t=i[\hat{H},\hat{A}]/\hbar$ is evaluated as a consequence of the uncertainty relation,
\aln{\label{ur}
\lrv{\braket{\dot{\hat{A}}}}\leq \mc{B}_\mr{ur}:=\frac{2}{\hbar}\Delta A\cdot \Delta H,
}
where $\Delta A$ is the quantum fluctuation of $\hat{A}$.
Such bounds on speed of quantum transitions are nowadays called quantum speed limits and are generalized in many ways with various applications as one of the central issues on quantum dynamics~\cite{deffner2017quantum}.
The measure breakthrough of the quantum speed limit includes the  Margolus-Levitin bound~\cite{margolus1998maximum,PhysRevA.67.052109,PhysRevLett.103.160502}, which is based on the energy expectation value rather than the fluctuation~\cite{fleming1973unitarity,bhattacharyya1983quantum,PhysRevLett.65.1697,PhysRevLett.70.3365}, and
generalization to dissipative quantum systems and mixed states~\cite{PhysRevLett.110.050402,PhysRevLett.111.010402,PhysRevLett.110.050403,PhysRevA.89.012307,zhang2014quantum,PhysRevA.94.052125,mondal2016quantum,PhysRevA.93.052331,PhysRevLett.120.060409,PhysRevA.103.062204}, to name a few.
The quantum speed limits are also related to information theory (for example, inequality~\eqref{ur} is a special case for the  quantum Cram\'er-Rao inequality~\cite{liu2019quantum}) and geometry of quantum states~\cite{PhysRevA.82.022107,PhysRevX.6.021031,garcia2021unifying}, indicating its fundamental mathematical structure.
Furthermore, quantum speed limits turn out to impose essential constraints on quantum technologies, such as optimal quantum control~\cite{PhysRevA.74.030303,PhysRevLett.103.240501,PhysRevA.82.022318,PhysRevA.85.052327,PhysRevX.9.011034,PhysRevX.11.011035}, shortcuts to adiabaticity~\cite{PhysRevLett.118.100601,PhysRevLett.118.100602,RevModPhys.91.045001}, and quantum metrology~\cite{giovannetti2001quantum,giovannetti2004quantum,toth2014quantum}.
Recently, speed limits have been found to exist even for classical systems~\cite{PhysRevLett.120.070402,PhysRevLett.120.070401}, and various bounds for classical transitions are obtained in light of information theory~\cite{PhysRevLett.121.030605,PhysRevX.10.021056,nicholson2020time} and irreversible thermodynamics~\cite{PhysRevLett.121.070601,PhysRevLett.125.120604}.

\begin{figure}
    \centering
    \includegraphics[width=\linewidth]{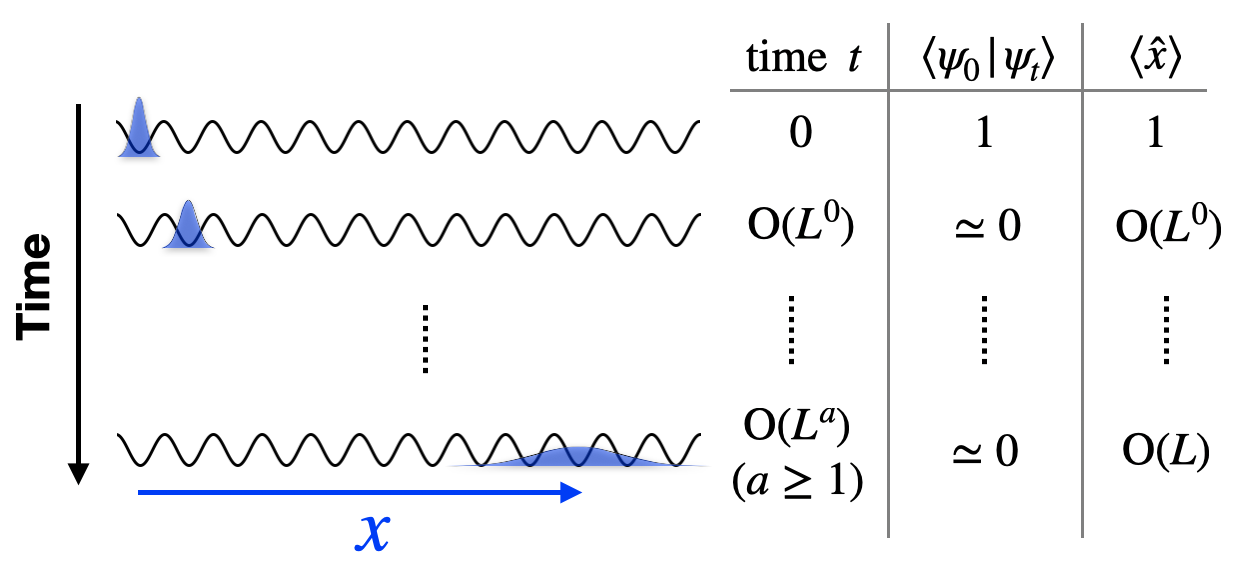}
    \caption{A simple example of a process with a macroscopic transition. For a single particle at the left end ($\braket{\hat{x}}=1$) to relax to the extended region ($\braket{\hat{x}}=\mr{O}(L)$) on a one-dimensional lattice with size $L$, it takes at least time $t=\mr{O}(L^a)\:(a\geq 1)$.
    On the other hand, the bound in inequality~\eqref{MT} is loose and cannot capture this timescale, since the time-evolved state can be nearly orthogonal to the initial state even for short times with $t=\mr{O}(L^0)$.
    }
    \label{figintro}
\end{figure}

Despite their success for a  wide range of situations, many  established speed limits 
fail to capture physically relevant bounds for certain processes, i.e., processes with macroscopic transitions (Fig.~\ref{figbegin}).
To see this, let us consider a single quantum particle that is initially located at the left end on a one-dimensional lattice with system size $L\gg 1$ (see Fig.~\ref{figintro}).
If we consider the average position $\hat{x}$ of a particle counted from the left, the time for the initial particle  ($\braket{\hat{x}(0)}=1$)
to be transferred to a distant position with $\braket{\hat{x}(T)}=\mr{O}(L)$ is expected to be $T\sim \mr{O}(L^a)\:\:(a\geq 1)$, assuming the short-ranged hopping~\footnote{Here, $a=1$ for a ballistic and $a=2$ for a diffusive process.}.
From a different viewpoint, the instantaneous speed $\fracd{\braket{{\hat{x}}}}{t}$ is always below the $\mr{O}(L^0,t^0)$ quantity.
On the other hand, many conventional bounds cannot describe these scales.
For example, direct application of inequalities~\eqref{MT} and \eqref{ur} leads to $T_\mr{MT}= \mr{O}(L^0)$ and $\mc{B}_\mr{ur}=\mr{O}(t^b)\:(b>0)$~\footnote{In a typical situation, $b=1$ for a ballistic and $b=1/2$ for a diffusive process. For long times, $\mc{B}_\mr{ur}$ saturates to a value $\mr{O}(L)$.} in this case, which are quite loose for large $L$ and $t$.
The situation becomes even worse when we consider many-body systems, where $T_\mr{MT}$ becomes smaller for larger system size~\cite{PhysRevX.9.011034,PhysRevA.103.062204}.
These problems are relevant for the issue of foundation of nonequilibrium statistical mechanics, i.e., quantum unitary dynamics after quench~\cite{RevModPhys.83.863,yukalov2011equilibration,eisert2015quantum,d2016quantum}, which attracts recent intensive attention by the experimental development of artificial quantum systems~\cite{bloch2012quantum,gross2017quantum}.
In fact, while the timescale of dynamics is a fundamental quantity, many conventionally known timescales cannot be used for processes involving macroscopic transitions, such as  particles' transport from an inhomogeneous initial state, since they do not grow even with increasing  system size~\cite{PhysRevLett.111.140401,goldstein2015extremely,reimann2016typical,de2018equilibration}.
Another important field of study concerning this problem is the optimal quantum control represented by the quantum state transfer~\cite{PhysRevLett.91.207901,kay2010perfect}.
While previous studies try to estimate the speed of the process analyzing simple settings~\cite{PhysRevA.74.030303,PhysRevLett.103.240501,PhysRevA.82.022318,PhysRevA.85.052327} and proposing conjectures~\cite{PhysRevX.9.011034,PhysRevX.11.011035},
rigorous and general relations on speed limits for macroscopic transitions are seldom known.

From a formal point of view, there are two reasons why many conventional bounds do not work for the above setting.
One is that they rely on statistical measures that do not take into account the macroscopic geometric structure of the setup.
For example, the Mandelstam-Tamm and  other similar bounds rely on quantum fidelity $|\braket{\psi(t)|\psi(0)}|$ (for pure states) or the Bures angle to distinguish initial and final states.
However, such measures are not suitable for characterizing the spatial distance for the particle transport $x$ in Fig.~\ref{figintro};
indeed, the quantum fidelity rapidly decays even for short times $t\simeq \mr{O}(L^0)$ provided that the wavepacket overlap between the two states become small.
The other reason is that, since the spectral range of $\hat{x}$ diverges with $L$, the bound in inequality~\eqref{ur} and the one based on, e.g., the trace distance $\|\mr{d}{\hat{\rho}}/\mr{d}t\|_1$~\cite{deffner2017geometric,funo2019speed} do not lead to the finite speed limit for such a divergent observable in general (notice $|\mr{d}\braket{{\hat{x}}}/\mr{d}t|\leq \|\hat{x}\|_\infty\cdot \|\mr{d}{\hat{\rho}}/\mr{d}t\|_1$).
We note in passing that these discussions are related to the optimal transport problem~\cite{villani2009optimal}.
In this problem, one introduces the so-called Wasserstein distance, which is a distance between two probability distributions that 
takes account of the underlying geometric structure for the random variables (see Appendix~\ref{wasapp}).
While beautiful relations between Wasserstein distances and certain thermodynamic speed limits in stochastic systems are recently known~\cite{dechant2019thermodynamic,PhysRevLett.126.010601,nakazato2021geometrical,dechant2021geometric,dechant2021minimum}, the distances are often practically complicated and hard to calculate in general.
Moreover, the extension of the distance to quantum realm is still controversial despite various efforts~\cite{carlen2014analog,datta2020relating,deffner2017geometric,chow2019discrete,PhysRevLett.126.010601,de2021quantum,PhysRevA.101.042107}.

We note that the Lieb-Robinson bound~\cite{lieb1972finite}, which  describes the general bound of information propagation in quantum many-body systems,
is often not satisfactory for the precise evaluation of the speed.
Indeed, the Lieb-Robinson bound only treats the maximal velocity independent of the quantum state.
Therefore, the bound is typically not tight and cannot attain the equality condition, as opposed to state-dependent speed limits such as the Mandelstam-Tamm bound.
In addition, unlike the Lieb-Robinson bound, the state-dependent bound often indicates a notable tradeoff relation, e.g., the tradeoff relation between time and energy fluctuation as in Eq.~\eqref{MT}.

\subsection{Summary of the results}

\subsubsection{General framework for deriving speed limits on macroscopic transitions}
In this work, we develop a new, general, and rigorous \textit{framework} to obtain state-dependent speed limits applicable to processes with macroscopic transitions on the basis of the local conservation law of probability, the  fundamental principles of physics (see Fig.~\ref{figbegin}).
We first demonstrate for the first time that
the speed of the expectation value of an observable $A$ (either classical or quantum) defined on vertices of a graph, which describes arbitrary systems including many-body ones, is bounded like
\aln{\label{smpc}
\lrv{\fracd{\braket{{A}}}{t}}\leq &\text{\:(a term involving $\nabla A$)} \nonumber\\
&\times
\text{(a term involving local probability current)},
}
which takes place of, e.g., inequality~\eqref{ur}.
Here, the gradient $\nabla A$ mathematically corresponds to the derivative-like operation on a discrete graph (see Appendix~\ref{lapapp}).
The appearance of the gradient can dramatically tighten the speed limit for the case
\aln{
\nabla A\ll \Delta A\text{ or }\|A\|_\mr{op}.
}
For the example in Fig.~\ref{figintro}, while $\Delta x$ in inequality~\eqref{ur} becomes large before saturating to $\mr{O}(L)$, the term involving $\nabla x$ in inequality~\eqref{smpc} is always $\mr{O}(L^0)$.
We also show that the inequality \eqref{smpc} also provides a reasonable timescale for the macroscopic transition.
Furthermore, going beyond the expectation value, we show that our method can be used for obtaining the speed limits for, e.g., the variance of the observable and entropy of the state.
We also discuss how our results on general graphs to connected to continuous systems. We also point out the new relation of our speed limits and the optimal transport problem, which indicates the connection
between nonequilibrium statistical mechanics and underlying mathematical structures.

Our theory only relies on the local probability conservation and thus are applied to any physically normal systems, as demonstrated for quantum unitary dynamics, nonlinear dynamics, classical stochastic dynamics, and quantum stochastic dynamics in this manuscript.
Furthermore, our results are even applicable to discrete many-body systems, which is concisely formulated with the language of the graph theory.
Note that related speed limits based on the local probability conservation were  obtained for, e.g., the continuous classical Fokker-Planck equation~\cite{PhysRevE.97.062101,PhysRevX.10.021056} before; however, our work is fundamentally distinct from previous work by demonstrating that the local probability conservation generally provides useful and insightful speed limits even for quantum systems, discrete systems, and many-body systems, pointing out that it holds for any physical processes.

\subsubsection{Speed limits in quantum unitary dynamics as a new tradeoff relation}

As a primary application of our general framework, we derive a novel type of speed limits applicable to unitary quantum dynamics.
Unlike conventional quantum speed limits, we find the importance of the expectation value $E_\mr{trans}$ of the transition Hamiltonian of the system (see Eq.~\eqref{Etrans}) for our speed limit.
In particular, instead of inequality~\eqref{ur}, we show the following inequality
\aln{\label{qname}
\lrv{\fracd{\braket{\hat{A}}}{t}}\leq &\text{\:(a term involving $\nabla{A}$)}\times\sqrt{R^2-E_\mr{trans}^2}
}
for an observable $\hat{A}$ given in Eq.~\eqref{aform}.
Here, $R$ is bounded by the probability-distribution-dependent factor (see Eq.~\eqref{glRpos}), which is accessible in state-of-the-art experiments (such as the quantum gas microscope in ultracold atomic gases~\cite{bakr2009quantum}).
It is further bounded by the state-dependent constant  that is easily known from the Hamiltonian structure, which is also equivalent to the maximum degree of the weighted graph (see Eq.~\eqref{inftycon}).
In addition, beyond conventional expectation values, we prove similar speed limits for dynamics of macroscopic coherence, a key quantity for quantum information, which previous literature seldom discussed.
We also discuss the relation with the continuous-space nonlinear Schr\"{o}dinger equation and elucidate that the inequality \eqref{qname} for a general discrete system reduces to a previously unknown speed limit (see inequality~\eqref{contene}) based on the kinetic energy in the continuous system.
We verify our speed limits for the transport in single-particle and  interacting many-particle systems and the nonequilibrium process of 
an interacting many-body spin system.

Interestingly, inequality~\eqref{qname} suggests that the increase of $E_\mr{trans}$ can decrease the speed provided that $R$ is the same, which we discuss is due to the suppression of the phase difference of the quantum state.
Thus, our inequality~\eqref{qname} (or its continuous version in \eqref{contene}) intuitively represents the novel tradeoff relation between \textit{time and quantum phase difference}, in stark contrast with the tradeoff relation between time and energy fluctuation in~\eqref{MT}. 

Let us briefly discuss the conceptual distinctions between previous works.
Our bound is in general state-dependent through $E_\mr{trans}$ and can be tighter than the bound indicated by the Lieb-Robinson velocity~\cite{lieb1972finite}, which is independent of the quantum state and provides only maximal velocity.
Quite importantly, in contrast with the Lieb-Robinson bound, 
our bound can satisfy the equality condition in some situations, as discussed in Sec.~\ref{SecIV}.
In addition, our bound is different from 
Refs.~\cite{PhysRevX.9.011034,PhysRevX.11.011035}, which proposed quantum-geometry-based conjectures.
Our theory instead derives fundamental and rigorous laws that govern speed limits of general macroscopic dynamics using the distinct principle of local conservation of probability.

\subsubsection{Speed limits in stochastic dynamics by entropy production}

To demonstrate the broad applicability of our general approach, we also prove speed limits for macroscopic systems involving Markovian dissipation.
We first derive a speed limit based on the irreversible entropy production; while the entropy production is recently found to play an important role for state transitions~\cite{PhysRevLett.105.170402,PhysRevLett.107.140404,PhysRevLett.117.190601,PhysRevLett.114.158101,PhysRevLett.116.120601,horowitz2020thermodynamic,dechant2018current,PhysRevE.97.062101,PhysRevLett.121.070601,funo2019speed}, we show that a related speed limit of macroscopic observables on a general graph is obtained from our framework.
The bound can be qualitatively better for macroscopic systems than that proposed in Ref.~\cite{PhysRevLett.121.070601}.
In addition, we obtain a modified speed limit, which is valuable even without the detailed balance condition, using the Hatano-Sasa entropy production~\cite{PhysRevLett.86.3463}.
We verify our bounds for the dynamics of the simple exclusion processes.
We also show that a similar speed limit is obtained even for macroscopic quantum open systems described by the Gorini-Kossakowski-Sudarshan-Lindblad equation~\cite{gorini1976completely,lindblad1976generators}.

\subsection{Organization of the paper}

The rest of the paper is organized as follows (also see Fig.~\ref{figbegin}).
In Sec.~\ref{SecII}, we present a general framework of deriving the speed limit using the local conservation law of probability on a general graph. We also discuss the limit for transition times and the relation with continuous systems.
In Sec.~\ref{SecIII}, on the basis of the general framework, we derive  speed limits that are useful for unitary quantum dynamics with macroscopic transitions, noting the importance of the transition Hamiltonian.
In Sec.~\ref{SecIV}, our quantum speed limits  are confirmed for particle systems and a many-body spin chain, which are relevant for state-of-the-art experiments of artificial quantum systems.
In Sec.~\ref{SecV}, we apply our general framework to classical Markovian systems and obtain speed limits for macroscopic transitions based on the entropy production rate. In particular, we derive a useful speed limit even without the detailed balance condition based on the Hatano-Sasa entropy production rate.
In Sec.~\ref{SecVI}, our speed limits for classical stochastic systems are confirmed with many-particle systems obeying the simple exclusion process.
In Sec.~\ref{SecVII}, we show the corresponding speed limit for dissipative quantum systems described by the Gorini-Kossakowski-Sudarshan-Lindblad equation.
In Sec.~\ref{SecVIII}, we briefly discuss several miscellaneous topics that our approach can investigate.
In Sec.~\ref{SecIX}, we conclude our results by suggesting some directions for future studies.

\section{Macroscopic speed limit constrained by currents\label{SecII}}
We begin with a general formulation of our speed limit based on the local conservation law of probability for a given graph structure, which represents a general system including a many-body one.
We show that the instantaneous speed of the expectation value of an observable is bounded by the gradient of an observable on the graph and the sum of the magnitude of the local probability current.
We demonstrate that this bound can dramatically improve the speed limit for certain observables compared with the bound based on, e.g., the uncertainty relation in inequality~\eqref{ur}.

\begin{figure*}
    \centering
  \includegraphics[width=\linewidth]{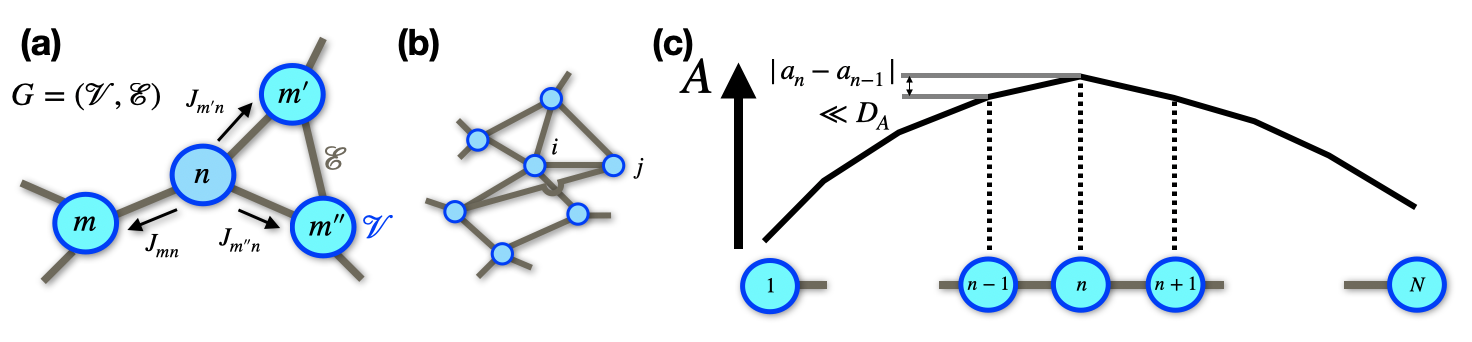}
    \caption{
  (a)  Schematic illustration of a graph composed of the vertices $\mc{V}$ and the edges $\mc{E}$. Each of the vertices satisfies the continuity equation as shown in Eq.~\eqref{conteq}.
  (b) An example of a graph, where each vertex indicates the microscopic states $i,j,\cdots$ such as the basis of the Hilbert space.
  (c) Another example of a graph, which is one dimension and has edges $\mathcal{E}=\{(1,2),\cdots,(n,n+1),\cdots,(N-1,N)\}$.
  We are interested in a situation where the observable $A$ is smooth on a graph, i.e., $|a_n-a_m|$ with $(n,m)\in\mc{E}$ are much smaller than the maximum range of $A$, denoted by $D_A$.
    }
    \label{fig_graph}
\end{figure*}


\subsection{Instantaneous speed limit on a general graph}
We consider a graph $G$ consisting of a set of vertices $\mc{V}$ and  edges $\mc{E}$ (see Figure~\ref{fig_graph}(a)), which is obtained from the system of our interest (see Fig.~\ref{figbegin}).
The vertices are labeled as, e.g., $n\in \mc{V}$ and $m\in \mc{V}$, and the edges are denoted by, e.g., $(n, m)\in \mc{E}$ (we assume that $(n,n)\notin \mc{E}$).
We consider a time-dependent probability distribution on the graph, $p(t)=\lrm{p_n(t)}_n$, whose time evolution is assumed to obey the continuity equation,
\aln{\label{conteq}
\fracd{p_n(t)}{t}=-\sum_{m(\sim n)}J_{mn}(t),
}
where $J_{mn}(t)$ satisfies $J_{mn}(t)=-J_{nm}(t)$ and $m\: (\sim n)$ means that we take a sum of $m$ connected to $n$ by the edge, i.e., $m:(n,m)\in\mc{E}$.
Physically, $n\in\mc{V}$ labels some (possibly coarse-grained) subspace obtained from the decomposition of the total state space, e.g., the Hilbert space in quantum systems.
For example, we can decompose the total space into non-coarse-grained states characterizing the fundamental microscopic dynamics (Fig.~\ref{fig_graph}(b)).
Instead, the space can be decomposed such that the corresponding graph becomes one dimension with $\mathcal{E}=\{(1,2),\cdots,(n,n+1),\cdots,(N-1,N)\}$ (Fig.~\ref{fig_graph}(c)).

We define a time-independent observable $A$ whose expectation value with respect to $p(t)$ can be written as 
\aln{
\braket{A}(t)=\sum_na_np_n(t).
}
We also define another observable-dependent graph generated from $\mc{E}$ and $A$,
\aln{
\mc{E}_A=\lrm{(n,m)|((n,m)\in \mc{E}) \cap (a_n\neq a_m)}.
}

The speed  for $\braket{A}$, i.e.,  $\braket{\dot{A}}:=\fracd{\braket{A}}{t}$, is given as
\aln{
\braket{\dot{A}}&= -{\sum_{n\sim m}a_nJ_{mn}}\nonumber\\
&= -\frac{1}{2}{\sum_{n\sim m}(a_n-a_m)J_{mn}}\nonumber\\
&= -\frac{1}{2}{\sum_{n\sim_A m}(a_n-a_m)J_{mn}}.
}
Here, $n\sim m$ [$n\sim_A m$] means the sum of $n$ and $m$ satisfying $(n,m)\in\mc{E}\:[\mc{E}_A]$.
To derive this equation, we use the continuity equation and the fact that $J_{mn}$ is anti-symmetric.
Now, using H\"{o}lder's inequality, we have
a set of speed limits, such as
\aln{\label{cs}
|\braket{\dot{A}}|
\leq \frac{1}{2}\sqrt{\sum_{n\sim m}(a_n-a_m)^2r_{nm}\cdot
\sum_{n\sim_A m}\frac{J_{nm}^2}{r_{nm}}}
}
and
\aln{\label{ho}
|\braket{\dot{A}}|
\leq \frac{1}{2}\max_{n\sim m}|a_n-a_m|\cdot
\sum_{n\sim_A m}|J_{nm}|
}
for some symmetric real numbers $r_{nm}$, which are assumed to be positive if and only if $n\sim_A m$.

The inequalities \eqref{cs} and \eqref{ho} are the first main results of our paper.
They have a simple meaning: the transition rate of  $A$ is upper bounded using the probability current $|J_{mn}|$ and the difference between $a_n$ and $a_m$,
which is regarded as the gradient of the observable $A$ on the graph.
These relations  are particularly useful when
$|a_n-a_m|$ with $n\sim m$ is (typically) much smaller than $D_A:=\max_{n,m\in\mc{V}}|a_n-a_m|$.
To see this, let us consider the following inequality  instead of inequality~\eqref{ho}:
\aln{\label{ho2}
|\braket{\dot{A}}|&=\lrv{\sum_{n\in\mc{V}}(a_n-\alpha)\dot{p}_n}\nonumber\\
&\leq \max_{n\in\mc{V}}|a_n-\alpha|\cdot
\sum_{n\in\mc{V}}|\dot{p}_n|
}
for some $\alpha\in\mbb{R}$.
The inequality becomes optimal when  $\alpha=(\max_{n\in\mc{V}}a_n+\min_{n\in\mc{V}} a_n)/2$, for which we obtain
\aln{\label{dame}
|\braket{\dot{A}}|\leq \frac{D_A}{2}\cdot
\sum_{n\in\mc{V}}|\dot{p}_n|
}
Then, if
$\max_{n\sim m}|a_n-a_m|$ is  much smaller than $D_A$, 
inequality~\eqref{ho} becomes much tighter than inequality~\eqref{dame}.
As an example for this situation, 
 let us consider a one-dimensional graph $1, \cdots, N\in \mc{V}$ and $(n,n+1)\in \mc{E}\:(1\leq n\leq N-1)$  (see Fig~\ref{fig_graph}(c)).
If  $A$ describes the position on the graph counted from the left, i.e., $a_n=n$,  we have
\aln{
1=\max_{n\sim m}|a_n-a_m|\ll D_A=N-1,
}
meaning that 
\eqref{ho} is much tighter than \eqref{dame} when the order of $\sum_{n\sim_A m}|J_{nm}|$  and $\sum_{n\in\mc{V}}|\dot{p}_n|$ is the same.

As indicated above, the importance of inequalities \eqref{cs} and \eqref{ho} is that the dependence of observables appears as their gradient, i.e., $a_n-a_m$ with $n$ and $m$ being connected by the graph.
Since $\max_{n\sim m}|a_n-a_m|$ measures the maximum variation of $A$ concerning the change of the neighboring vertices, it is regarded as the discrete version of the Lipschitz constant. 
To describe it, we introduce the following notation 
\aln{
\|\nabla A\|_\infty:=\max_{n\sim m}|a_n-a_m|.
}
Similarly, we can write the first factor on the right-hand side in inequality~\eqref{cs} with the graph Laplacian known in the graph theory~\cite{merris1994laplacian}, which is analogous to the Laplacian for continuous functions (see Appendix~\ref{lapapp} for details).
To desribe it, we  define
\aln{
[A^\mathsf{T}\nabla_r^2 A]:=\frac{1}{2}\sum_{n\sim m}(a_n-a_m)^2r_{nm},
}
where $\nabla_r^2$ is the graph Laplacian matrix for a graph weighted by $r_{nm}$, whose elements are given by
\aln{
(\nabla_r^2)_{nm}=-r_{nm}+\delta_{nm}\sum_{m'(\sim_A n)}r_{nm'}.
}
Note that $A$ is regarded as a vector whose elements are $a_n$.
With these notations, inequalities~\eqref{cs} and ~\eqref{ho} are simply written as
\aln{\label{css}
|\braket{\dot{A}}|\leq \sqrt{\frac{[A^\mathsf{T}\nabla_r^2 A]}{2}\cdot\sum_{n\sim_A m}\frac{J_{nm}^2}{r_{nm}}}
}
and
\aln{\label{hos}
|\braket{\dot{A}}|\leq \frac{1}{2}\|\nabla A\|_\infty\cdot\sum_{n\sim_A m}|J_{nm}|.
}

\subsection{Speed limit for other quantities}
We can obtain Similar speed limits for other quantities not written as the expectation value of an observable.
For example, the general scalar function of the probability distribution written as $F(p)$ has the following speed limit:
\aln{\label{entF}
|\dot{F}|\leq \sqrt{\frac{[(\partial F)^\mathsf{T}\nabla_r^2 (\partial F)]}{2}\cdot\sum_{n\sim_{\partial F} m}\frac{J_{nm}^2}{r_{nm}}}
}
and
\aln{\label{entF2}
|\dot{F}|\leq \frac{1}{2}\|\nabla (\partial F)\|_\infty\cdot\sum_{n\sim_{\partial F} m}|J_{nm}|,
}
where 
$
\partial F(p)=\lrs{\fracpd{F}{p_1},\cdots, \fracpd{F}{p_{|\mc{V}|}}}^\mathsf{T}.
$
Taking $F=\sum_na_np_n$ leads to the inequalities \eqref{css} and \eqref{hos}.

The general result is useful when we consider the speed limit of entropy-like quantity,
\aln{\label{shannonentropy}
\mathcal{S}(p,X)=-\sum_n p_n\ln\frac{p_n}{X_n},
}
where $X=\{X_n\}$ is independent of $p$ and $t$.
When we take $X=p^\mr{ref}$, where $p^\mr{ref}$ is some reference probability distribution, $\mathcal{S}(p,X)$ reduces to the Kullback-Leibler divergence $-D(p||p^\mr{ref})$, which quantifies the difference between the two probability distributions~\cite{cover1999elements}.
Instead, if we take $X_n$ as the dimension of the subspace $n$, $\mathcal{S}(p,X)$ reduces to the observational entropy $S_\mr{obs}^\mc{V}(\hat{\rho})$, which is the promising candidate for entropy in isolated quantum systems~\cite{PRXQuantum.2.030202}.
In this case, we find the speed limit, e.g.,
\aln{\label{shannonentropybound}
|\dot{\mathcal{S}}|\leq\frac{1}{2}\sqrt{\sum_{n\sim m}r_{nm}\lrs{\ln\frac{p_nX_m}{p_mX_n}}^2} 
\sqrt{\sum_{n\sim m}\frac{|J_{nm}|^2}{r_{nm}}}.
}

Moreover, as detailed in the following sections, the speed of  more nontrivial quantities, such as macroscopic quantum coherence and variance of observables, is bounded by a similar quantity. 
For example, let us consider a variance of ${A}$,
\aln{\label{varmoto}
\mbb{V}[A]=\Delta A^2=\braket{A^2}-\braket{A}^2=\frac{1}{2}\sum_{nm}(a_n-a_m)^2p_np_m.
}
Using inequality~\eqref{entF}, we have
\aln{\label{vars}
&|\dot{\mbb{V}}[A]|\nonumber\\
&\leq\frac{1}{2}\sqrt{\sum_{n\sim_A m} (a_n-a_m)^2(a_n+a_m-2\braket{A})^2r_{nm}}\sqrt{\sum_{n\sim_A m} \frac{|J_{nm}|^2}{r_{nm}}}\nonumber\\
&\leq\frac{\|\nabla A\|_\infty}{2}\sqrt{\sum_{n\sim_A m} (a_n+a_m-2\braket{A})^2r_{nm}}\sqrt{\sum_{n\sim_A m} \frac{|J_{nm}|^2}{r_{nm}}},
}
meaning that the speed of the variance is also bounded using the gradient of the observable.

\subsection{Limit for transition time}
Next, we discuss that our approach also enables us to bound the  timescales for transition processes.
Let us consider a situation for which the expectation value of an observable $A$ changes from $A_\mr{ini}$ at $t=0$ to 
$A_\mr{fin}$ at $t=T$.
Since
$|A_\mr{fin}-A_\mr{ini}|\leq\int_0^Tdt\lrv{\braket{\dot{A}(t)}}$,
we readily have a lower bound for the transition time
\aln{\label{weakb}
T\geq \frac{|A_\mr{fin}-A_\mr{ini}|}{\av{|\braket{\dot{A}}|}},
}
which can be evaluated with inequalities \eqref{css} and \eqref{hos},
where
$
\av{\cdots}=\frac{1}{T}\int_0^Tdt\cdots
$
denotes the temporal average.
We also obtain a better lower limit by noticing
$
A_\mr{fin}-A_\mr{ini}=-\frac{T}{2}\sum_{(n,m)\in\mc{E}_A}(a_n-a_m)\av{J_{mn}},
$
from which we have
\aln{\label{csta}
T\geq \frac{|A_\mr{fin}-A_\mr{ini}|}{\sqrt{\frac{[A^\mathsf{T}\nabla_r^2 A]}{2}\cdot\sum_{n\sim_A m}\frac{\av{J_{nm}}^2}{r_{nm}}}}
}
and
\aln{\label{hota}
T\geq \frac{2|A_\mr{fin}-A_\mr{ini}|}{\|\nabla A\|_\infty\cdot\sum_{n\sim_A m}|\av{J_{nm}}|}.
}
These inequalities are useful when 
$|A_\mr{fin}-A_\mr{ini}|$ is much larger than $[A\nabla_r^2 A]$ or $\|\nabla A\|_\infty$.
For example, let us again consider a one-dimensional graph $1, \cdots, N\in \mc{V}$ and $(n,n+1)\in \mc{E}\:(1\leq n\leq N-1)$.
For $a_n=n$, by choosing $A_\mr{ini}=1$ and $A_\mr{fin}=N$, we have (from inequality~\eqref{hota})
$
T\geq \frac{2(N-1)}{\sum_{n\sim_Am}\lrv{\av{J_{mn}}}}.
$
assuming that $\sum_{(n,m)\in\mc{E}_A}\lrv{\av{J_{mn}}}=\mr{O}(T^{-\gamma})\:(\gamma\geq 0)$, we have a proper macroscopic timescale $T\gtrsim \mr{O}(N^{\frac{1}{1-\gamma}})$.
Note that $\gamma=0$ and $\gamma=1/2$ correspond to the ballistic and diffusive timescale, respectively~\footnote{Assuming that the current fluctuates randomly in time with zero mean, we obtain $\gamma=1/2$.}.
We note that this macroscopic timescale cannot be attained by the speed limit similar to
inequality~\eqref{dame}, i.e., 
\aln{
T\geq \frac{2|A_\mr{fin}-A_\mr{ini}|}{D_A\cdot\sum_{n\in\mc{V}}\av{|\dot{p}_n|}}.
}
Indeed, if we consider the above example for this inequality, the right-hand side is $\mr{O}(1)$ when $\sum_{n\in\mc{V}}\av{|\dot{p}_n|}=\mr{O}(1)$, which is quite loose.

\subsection{Continuous case}
We can consider a similar speed limit for  continuous systems as the ones obtained for the discrete graph.
To see this, let us consider the continuous system whose  space coordinate is denoted by $\mbf{x}$.
We consider time evolution for the probability distribution $P(\mbf{x},t)$,
which obeys the continuity equation
\aln{
\fracd{P(\mbf{x},t)}{t}=-\nabla\cdot \mbf{J}(\mbf{x},t).
}
We assume that $\mbf{J}(\mbf{x})$ becomes zero for $|\mbf{x}|\ra 0$.
Using the integration by parts, the instantaneous speed limit for  $\braket{{A}(t)}=\int d\mbf{x}P(\mbf{x},t)A(\mbf{x})$ reads
\aln{\label{contJ}
|\braket{\dot{A}}|\leq \sqrt{\int d\mbf{x}r(\mbf{x})(\nabla A(\mbf{x}))^2}\sqrt{\int d\mbf{x}\frac{|\mbf{J}(\mbf{x})|^2}{r(\mbf{x})}}
}
for $r(\mbf{x})>0$ and
\aln{
|\braket{\dot{A}}|\leq \max_{\mbf{x}}|\nabla A(\mbf{x})|\int d\mbf{x}|\mbf{J}(\mbf{x})|,
}
which are the continuous version of inequalities~\eqref{cs} and \eqref{ho}.
Note that this is generalized to arbitrary function written as $F=F[P(\mbf{x},t)]$,
where $\nabla A(\mbf{x})$ is replaced by $\nabla\lrs{\frac{\delta F}{\delta P}}$.
We also note that, while Ref.~\cite{PhysRevE.97.062101} uses 
inequality \eqref{contJ} for specific $r(\mbf{x})$ in classical Fokker-Planck equation, we consider more general systems, including, e.g., a system obeying the nonlinear Schr\"{o}dinger equation (see Sec.~\ref{SecIII}).

Moreover, when we consider a situation where the expectation value of the observable $A$ changes from $A_\mr{ini}$ at $t=0$ to 
$A_\mr{fin}$ at $t=T$,
we have a relation (cf. inequality~\eqref{hota})
\aln{\label{Was1}
|A_\mr{fin}-A_\mr{ini}|\leq \max_{\mbf{x}} |\nabla A(\mbf{x})| \int d\mbf{x} T|\av{\mbf{J}(\mbf{x})}|.
}
Here, we note that $\max |\nabla A(\mbf{x})|$ is the Lipschitz constant for a differentiable function $A(\mbf{x})$.

We note that inequality~\eqref{Was1} is closely tied to the order-one Wasserstein distance.
The Wasserstein distance plays a crucial role in characterizing the distance between two probability distributions $P(\mbf{x})$ and $Q(\mbf{x})$
by taking account of the underlying geometric structure, such as the Euclidean distance between $\mbf{x}$ and $\mbf{y}$ (see Appendix~\ref{wasapp} for the definition).

As shown in Appendix~\ref{wasapp}, in one dimension, 
we can show that the order-one Wasserstein distance with respect to the Euclidean distance $|x-y|$ becomes
\aln{
W_1(P(0),P(T))=T\int d{x}\lrv{\av{{J}}(x)},
}
where we have assumed $\av{{J}}(-\infty)=0$.
Thus, combined with inequality~\eqref{Was1}, we have
\aln{
|A_\mr{fin}-A_\mr{ini}|\leq \max |\partial_xA({x})|\cdot W_1(P(0),P(T)).
}
Thus, in this case, our approach to derive the speed limit is directly connected to the optimal transport problem.

\section{Unitary quantum dynamics: theory\label{SecIII}}
In this section, we apply the general formalism obtained in the previous section to unitary quantum systems.
We introduce the decomposition of the Hilbert space for general quantum many-body systems with discrete states and show several speed limits for certain observables.
In particular, by bounding the probability current from above,
we show that the speed limit is bounded using the expectation value $E_\mr{trans}$ of the transition Hamiltonian, which is a standard observable.
Interestingly, the speed limit can be tighter when the magnitude of $E_\mr{trans}$ becomes increased because of the suppression of the quantum phase difference.
This intuitively means the novel tradeoff relation between time and quantum phase difference, instead of the  tradeoff relation between time and energy fluctuation  by Mandelstamm and Tamm.
We also show that, beyond expectation values, we can bound the speed of the change of entropy and macroscopic quantum coherence.
Furthermore, we discuss the case for  continuous systems and elucidate the relation to speed limits for discrete systems.
The detailed derivation of each result is given in Appendices~\ref{Appderexp}-\ref{Appvarfint}.

\subsection{General setting for discrete quantum systems}
Let us consider the von Neumann equation
\aln{
\fracd{\hat{\rho}(t)}{t}=-i[\hat{H}(t),\hat{\rho}(t)],
}
where we set the Planck constant as the unity in the following.
We assume that  the Hilbert space $\mc{H}$ is finite dimensional and decompose it as
$
\mc{H}=\bigoplus_{n=1}^{|\mc{V}|}\mc{H}_n
$
and define $\hat{P}_n$ as the projection operator onto $\mc{H}_n$.
From the von Neumann equation, we find a continuity equation for 
\aln{
p_n(t)=\Tr[\hat{\rho}(t)\hat{P}_n]
}
as
\aln{
\fracd{p_n}{t}=-\sum_{m (\sim n)}J_{mn}^q(t),
}
where
\aln{\label{quantcurrent}
J_{mn}^q(t)=i\Tr[\hat{H}(t)_{nm}\hat{\rho}(t)_{mn}-\hat{H}(t)_{mn}\hat{\rho}(t)_{nm}].
}
Here, $\hat{X}_{nm}=\hat{P}_n\hat{X}\hat{P}_m$ for an operator $\hat{X}$
and $(n,m)\in\mc{E}$ if $n\neq m$ and  $\hat{H}(t)_{nm}\neq 0$ for any $t$.

As discussed in Fig.~\ref{fig_graph}(a), two cases are particularly important.
One is the case where $n$ labels the complete basis $\lrm{\ket{x}}\:(1\leq x\leq \dim[\mc{H}])$ of the Hilbert space. In that case, $\hat{P}_x=\ket{x}\bra{x}$.
The other interesting case is that $n$ labels a position in a one-dimensional graph, i.e.,  $\mc{E}=\lrm{(n,n+1)\:|\:1\leq n\leq N-1}$.


\subsection{Speed limit for expectation values of observables}
As a first target, we show the speed limit of the expectation value of an observable given by
\aln{\label{aform}
\hat{A}=\sum_na_n\hat{P}_n.
}
Following the general derivation in the previous section, we find 
\aln{
|\braket{\dot{\hat{A}}}|\leq\mc{B}_\mr{gL}=
\sqrt{\frac{[A^\mathsf{T}\nabla_r^2 A]}{2}}\sqrt{\sum_{n\sim_A m}\frac{|J_{nm}^q|^2}{r_{nm}}}
}
and
\aln{\label{Binfty}
|\braket{\dot{\hat{A}}}|\leq\mc{B}_\mr{Lip}=
{\frac{\|\nabla A\|_\infty}{2}}{\sum_{n\sim_A m}{|J_{nm}^q|}},
}
where $A=(a_1,\cdots, a_{|\mc{V}|})^\mathsf{T}$ is regarded as a vector obtained from $\hat{A}$ (``gL" and ``Lip" represent the graph Laplacian and Lipshitz constant, respectively).

Although the speed limits above contain the sum of the functions of local probability currents, we find that they are further bounded from above using  more physically relevant quantities.
Indeed, we obtain a hierarchy of the bounds
\aln{\label{hier2}
|\braket{\dot{\hat{A}}}|\leq\mc{B}_\mr{gL}
\leq \mc{B}_{p}
\leq \mc{B}_{H},
}
where each of the bounds is explained in the following.
In particular, below we show the importance of the expectation value of the transition Hamiltonian defined as
\aln{\label{Etrans}
E_\mr{trans}=\braketL{\sum_{n\sim_A m}\hat{H}_{nm}}.
}
This quantity is just an expectation value of the globally defined observable $\sum_{n\sim_A m}\hat{H}_{nm}$, unlike $\sum_{n\sim_A m}|J_{nm}|$ which involves information of local current.
We also note that, when $\mc{E}=\mc{E}^A$, we have $E_\mr{trans}=\Tr\lrl{\hat{\rho}\lrs{\hat{H}-\sum_n\hat{H}_{nn}}}$.

First, $\mc{B}_{p}$ 
is given by
\aln{\label{glRpos}
 \mc{B}_{p}
&=
\sqrt{2\frac{[A^\mathsf{T}\nabla^2_{r^p}A]}{{R_p}}}\sqrt{R_p^2-{E_\mr{trans}^2}},
}
where
\aln{
r^p_{nm}=\|\hat{H}_{nm}\|_\mr{op}\sqrt{p_np_m}
}
for $n\sim m$ (otherwise $r_{nm}^p=0$) and
\aln{
R_p=\Tr[\nabla^2_{r^p}]=
\sum_{n\sim_Am}\|\hat{H}_{nm}\|_\mr{op}\sqrt{p_np_m}
}
are dependent on probability distribution $p(t)$.

The probability distribution $\{p_n\}$ in $R_p$ is in principle measured in experiments of artificial quantum systems, such as the quantum-microscope technique in ultracold atomic systems~\cite{bakr2009quantum}.
Since $E_\mr{trans}$ is just a standard expectation value of the observable, the bound in inequality~\eqref{glRpos} is accessible in state-of-the-art experiments.

\begin{figure}
    \centering
    \includegraphics[width=\linewidth]{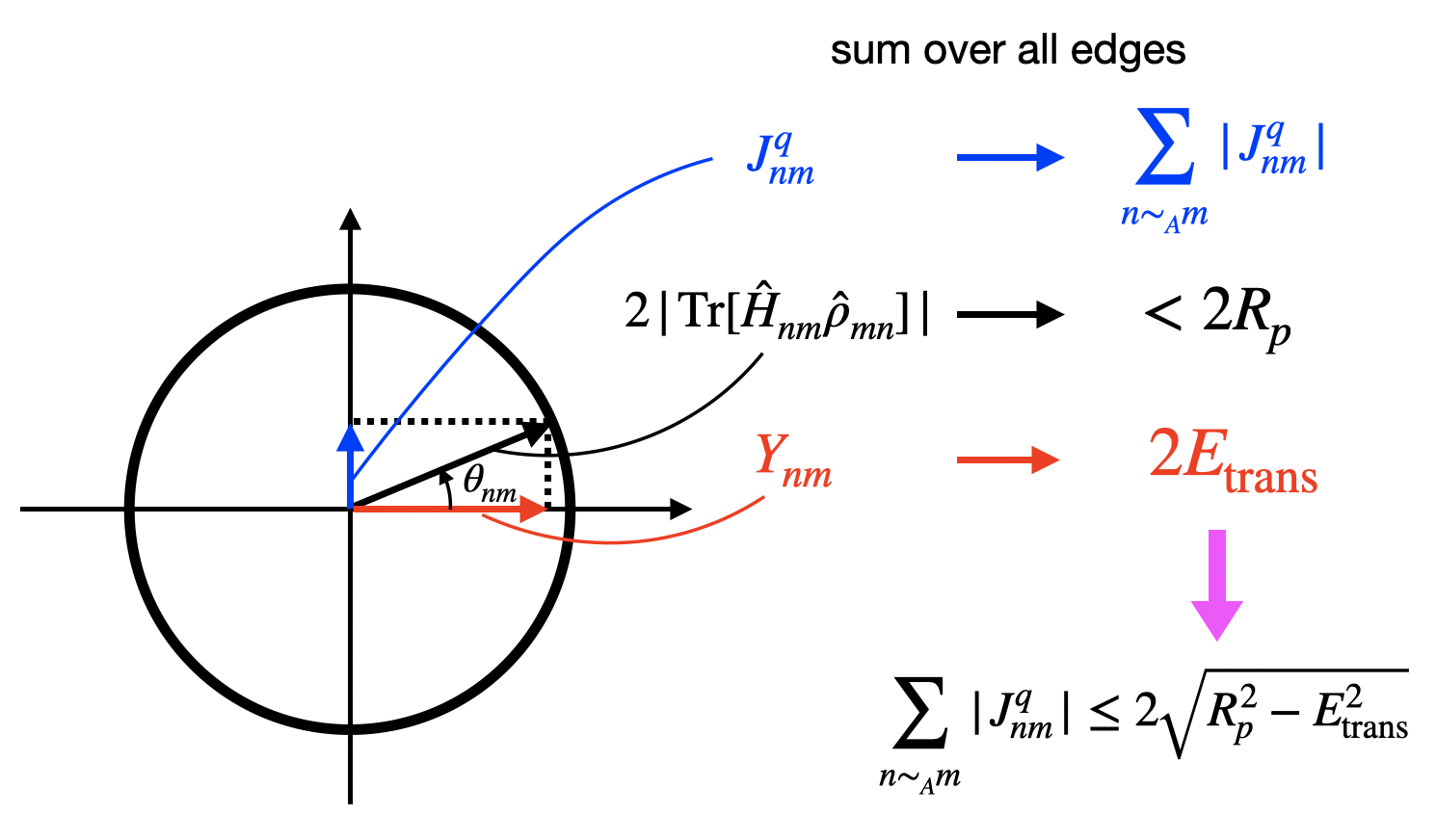}
    \caption{Schematic illustration of the tradeoff relation between the current and the expectation value of the transition energy. The local current $|J_{nm}^q|$ and the expectation value of the local transition Hamiltonian $\propto Y_{nm}$ cannot be simultaneously large when $2|\Tr[\hat{H}_{nm}\hat{\rho}_{mn}]|$ is fixed, since $|J_{nm}^q|=2|\Tr[\hat{H}_{nm}\hat{\rho}_{mn}]|\sin\theta_{nm}$ and $Y_{nm}=2|\Tr[\hat{H}_{nm}\hat{\rho}_{mn}]|\cos\theta_{nm}$, where $\theta_{nm}$ plays the role of difference of quantum phases. Summation over all of the edges $n\sim_Am$ leads to the tradeoff relation between $\sum_{n\sim_Am}|J_{nm}^q|$, $E_\mr{trans}$, and $R_p$, as detailed in Appendix~\ref{Appderexp}.}
    \label{tradeoff}
\end{figure}

This bound contains nontrivial information.
First, the speed limit becomes tighter when the absolute value of the expectation value of the transition Hamiltonian is larger, provided that $R_p$ is almost the same.
It may sound counter-intuitive that large energy concerning the transition Hamiltonian suppresses the speed for transition.
Intuitively, this is attributed to the fact that, while the difference of quantum phases generates the current, such  phase difference should be suppressed for large $E_\mr{trans}$.
To be more precise, the local current $J_{nm}^q$ and  the 
local energy $Y_{nm}$ concerning the transition Hamiltonian are respectively given by  $2|\Tr[\hat{\rho}_{nm}\hat{H}_{mn}]|\sin\theta_{nm}$ and  $2|\Tr[\hat{\rho}_{nm}\hat{H}_{mn}]|\cos\theta_{nm}$, where $\theta_{nm}$ plays the role of difference of quantum phases~\footnote{When we consider a pure state and take the entire basis $\{\ket{i}\}$ of the Hilbert space to describe each subspace, $\theta_{ij}$ indeed becomes the standard phase difference, $\theta_j-\theta_i$. Here, we take $\braket{i|\psi}=|\braket{i|\psi}|e^{i\theta_i}$}.
Thus, if $\sum_{n\sim_Am}|2\Tr[\hat{\rho}_{nm}\hat{H}_{mn}]|$ (which can be bounded by $2R_p$) is given, the sum of the local current $\sum_{n\sim_Am}|J_{nm}^q|$ and energy of the transition Hamiltonian $2E_\mr{trans}=\sum_{n\sim_Am}Y_{nm}$ cannot be large simultaneously (see Fig.~\ref{tradeoff} and Appendix~\ref{Appderexp}).
Note that the correspondence between 
$\sqrt{R_p^2-E_\mr{trans}^2}$ and the quantum phase difference becomes more evident by considering the continuous-space limit (see subsection \ref{cont}).

The above discussion indicates the novel tradeoff relation between time and quantum phase difference:
since small magnitude of the phase difference (i.e., the large expectation value of transition Hamiltonian) suppresses the speed,
we cannot simultaneously decrease the
time required for transitions and the magnitude of quantum phase difference.
This new tradeoff relation takes the place of the conventional tradeoff relation between time and energy fluctuation  in inequalities \eqref{MT} and \eqref{ur}. 
In addition, as discussed in Sec.~\ref{SecVIII},
the term  $\sqrt{R_p^2-E_\mr{trans}^2}$ in our bounds cannot be replaced with $\Delta H$.

Second, since $R_p=\sum_{n\sim_Am}\|\hat{H}_{nm}\|_\mr{op}p_n-\sum_{n\sim_Am}\|\hat{H}_{nm}\|_\mr{op}(\sqrt{p_n}-\sqrt{p_m})^2/2$, the bound can be tight if the gradient of the probability distribution (i.e., the difference between $p_n$ and $p_m$) becomes larger.
This is distinct from typical classical systems, where the speed can increase when the gradient is larger.
The distinction again comes from the fact that the probability current for unitary time evolution can arise owing to the gradient of phases instead of probability distributions.

Next, the bound  
$\mc{B}_{H}$
is simply given by
\aln{\label{inftycon}
\mc{B}_{H}=\|\nabla A\|_\infty\sqrt{C_H^2-E_\mr{trans}^2},
}
where
\aln{
C_H:=\max_{n\in\mc{V}}\sum_{m(\sim_An)}\|\hat{H}_{nm}\|_\mr{op}
}
is the maximum degree of the graph with the edge $\mc{E}_A$ and  weight $\|\hat{H}_{nm}\|_\mr{op}$.
Equation~\eqref{inftycon} does not explicitly contain the probability distribution (it appears only through $E_\mr{trans}$).
Thus, given a Hamiltonian and an observable $\hat{A}$,
the bound is readily obtained if   the expectation value of the transition Hamiltonian can be  measured.

We stress that the bounds 
Eqs.~\eqref{glRpos} and~\eqref{inftycon} universally hold for any isolated quantum systems for an observable given as Eq.~\eqref{aform}.
As highlighted in Sec.~\ref{SecIV}, these bounds indeed turn out to be useful for macroscopic transitions.

We discuss the conceptual distinction between the Lieb-Robinson bound~\cite{lieb1972finite,nachtergaele2006lieb} and our results.
The Lieb-Robinson bound is the speed limit for the information propagation described by the operator norm of the commutator of $\hat{A}(t)$ and $\hat{B}$, where $\hat{A}$ and $\hat{B}$ are spatially distant.
While the Lieb-Robinson bound provides a bound for general operators that may not be restricted to Eq.~\eqref{aform} and leads to various dynamical constraints~\cite{PhysRevLett.97.050401,lashkari2013towards,PhysRevLett.115.256803,kuwahara2016floquet,abanin2017rigorous,PhysRevB.96.060301,PhysRevLett.124.210606,PhysRevA.101.052122}, it only gives a state-independent maximal velocity and cannot achieve the equality condition in general.
Our speed limit is state-dependent through, e.g., $E_\mr{trans}$ and thus can give a tighter bound than the Lieb-Robinson bound for an operator in the form of \eqref{aform}.
Furthermore, our bounds can achieve the equality condition for some situations, as discussed in Sec.~\ref{SecIV}.

We note that we can obtain a completely different type of  inequalities by our approach, i.e., speed limits of acceleration, provided that $\hat{H}$ is independent of time.
As shown in Appendix~\ref{accapp}, we find bounds in the form $|\braket{\Ddot{\hat{A}}}|\leq \mc{Q}$, from which we have
$|\braket{\dot{\hat{A}}}|\leq \mc{B}_\mr{acc}:=\int_0^td\tau\mc{Q}$.
Such bounds $\mc{B}_\mr{acc}$ become zero for short-time limit $t\ra 0$, which may sometimes be better than $\mc{B}_\mr{gL}$ and
$\mc{B}_\mr{Lip}$ since they can be nonzero for $t\ra 0$.

\subsection{Speed limit for macroscopic coherence}
As represented by inequality~\eqref{entF}, the bound is generalized to quantities that are not written as the standard expectation value of an observable.
Here, we show that a reasonable speed limit is obtained even for macroscopic quantum coherence.
While we can consider different measures for coherence~\cite{PhysRevLett.113.140401,PhysRevLett.95.090401,PhysRevA.93.022122,RevModPhys.89.041003,RevModPhys.90.025004},  we introduce the following measure to describe  macroscopic quantum coherence:
\aln{\label{cohdef}
\mc{C}=\sum_{n,m\in\mc{V}}c^{nm}||\hat{\rho}(t)_{nm}||_2^2
}
for given $c^{nm}\geq 0$,
where $\|\hat{X}\|_2=\sqrt{\Tr[\hat{X}^\dag\hat{X}]}$.
While this measure reduces to (the square of) $l_2$-norm coherence for $c^{nm}=-\delta_{nm}+1$, we 
here assume that $c^{nm}$ is a more general symmetric and positive function of $n$ and $m$.
For example, let us consider a single particle on a one-dimensional lattice and take $c^{nm}=|n-m|$\:($1\leq n,m\leq N$), which is regarded as a distance between $n$ and $m$~\footnote{To be precise, $\mc{C}$ obtained from this specific choice of $c_{nm}=|n-m|$ may not satisfy some criteria~\cite{PhysRevLett.113.140401} for  quantum coherence. Nevertheless, the measure obtained by this choice can be useful for discussing the speed of certain transitions between states with different macroscopic coherence. We thus  use this choice in this work as well as the  coherence measure based on the variance.}.
Then, a state $\ket{\psi}=\frac{\ket{l_1}+\ket{l_2}}{\sqrt{2}}$ has $\mc{C}=|l_1-l_2|/2$, whose macroscopicity is controlled by $|l_1-l_2|$.
Another state $|\psi\rangle= \sum_{k=1}^N\frac{|k\rangle}{\sqrt{N}}$ has $\mathcal{C}=\frac{N^2-1}{3N}$, indicating the existence of the macroscopic coherence.

Instead, when we take $c_{nm}=(a_n-a_m)^2$ and assume pure states, $\mc{C}$ essentially reduces to  the variance of the macroscopic observable, i.e.,
$\mc{C}=2\mbb{V}[\hat{A}]=2(\braket{\hat{A}^2}-\braket{\hat{A}}^2)$.
This provides a lower bound of the macroscopic coherence indicator proposed in Ref.~\cite{PhysRevLett.95.090401} and satisfies some plausible criteria for macroscopic quantum coherence~\cite{PhysRevA.93.022122}.

For simplicity, we consider pure states
$\hat{\rho}=\ket{\psi}\bra{\psi}$ to derive the speed limit of macroscopic coherence (we discuss the generalization to mixed states in Appendix~\ref{cohapp}).
In that case, $\mc{C}=\sum_{n,m}c^{nm}p_np_m$ and we can use the framework in inequalities~\eqref{entF} and \eqref{entF2}.
For example, using~\eqref{entF} and~\eqref{entF2}, we have
\aln{\label{avecohb}
|\dot{\mc{C}}|&\leq 2\sqrt{\frac{\sum_{n\sim_\mc{C}l}(\tilde{c}^n-\tilde{c}^l)^2r_{nl}^p}{R_p}}\sqrt{R_p^2-E_\mr{trans}^2}
}
and
\aln{\label{maxcohb}
|\dot{\mc{C}}|&\leq 2\max_{n\sim_\mc{C}l}|\tilde{c}^n-\tilde{c}^l|\sqrt{C_H^2-E_\mr{trans}^2},
}
where $\tilde{c}^n:=\sum_mc^{nm}p_m$ and 
$\sim_A$ in the previous subsection is replaced with $\sim_\mc{C}$, defined from the graph
\aln{
\mc{E}_\mc{C}
=\lrm{(n,l)\:|\:((n,l)\in \mc{E}) \cap (\tilde{c}^n\neq \tilde{c}^l)}.
}
Interestingly, for pure quantum states,
the obtained bounds of macroscopic coherence contain the same factor with $E_\mr{trans}$ as in the case for the expectation values of observables.

When $c^{nm}$ satisfies
the triangle inequality, e.g., $c^{nm}=|n-m|$, we can replace $\tilde{c}^n-\tilde{c}^l$ above with $c^{nl}$. For example, we have (from, e.g., inequality~\eqref{maxcohb})
\aln{\label{cohbound}
|\dot{\mc{C}}|&\leq\mc{\tilde{B}}_p:= 2\max_{n\sim_\mc{C} l}|c^{nl}|\sqrt{R_p^2-E_{\mr{trans}}^2}\nonumber\\ 
&\leq\mc{\tilde{B}}_{H} := 2\max_{n\sim_\mc{C} l}|c^{nl}|\sqrt{C_H^2-E_{\mr{trans}}^2}.
}
The bound in inequality~\eqref{cohbound} becomes particularly useful when $\max_{n\sim_\mc{C} l}|c^{nl}|\ll \max_{n, l\in\mc{V}}|c^{nl}|$.
Indeed, for the example of a particle in the one-dimensional system, 
$\mc{C}=|l_1-l_2|/2$ itself can be large when  $|l_1-l_2|=\mr{O}(N)$,
whereas $\max_{n\sim_\mc{C} l}|c^{nl}|=1$ and remains $\mr{O}(N^0)$.

Instead, we can take $c_{nm}=(a_n-a_m)^2$ and obtain the speed limit for the variance, e.g.,
\aln{\label{varbound}
&\lrv{\dot{\mbb{V}}[\hat{A}]} \leq \mc{B}_\mr{var}\\
&:= \sqrt{\frac{\sum_{n\sim_\mc{C}l}(a_n-a_l)^2(a_n+a_l-2\braket{\hat{A}})^2r_{nl}^p}{R_p}}\sqrt{R_p^2-E_\mr{trans}^2}\nonumber\\
&\leq \|\nabla A\|_\infty\sqrt{\frac{\sum_{n\sim_\mc{C}l}(a_n+a_l-2\braket{\hat{A}})^2r_{nl}^p}{R_p}}\sqrt{R_p^2-E_\mr{trans}^2}\nonumber,
}
where we have used inequality~\eqref{vars}.

We also note that, applying a similar type of the inequality,
we obtain the absolute bound for the square root of the variance (i.e., the standard deviation $\Delta A=\sqrt{\mbb{V}[\hat{A}]}$) at finite times (see Appendix~\ref{Appvarfint}), 
{
\aln{\label{varfint}
\lrv{\Delta A(T)-\Delta A(0)}\leq \|\nabla A\|_\infty \av{\sqrt{C_H^2-E_\mr{trans}^2}}T.
}
}
This means that the graph structure generated by the Hamiltonian can control even the growth of the standard deviation of the observable.

\subsection{Speed limit for entropy-like quantities}
Applying inequality~\eqref{shannonentropybound} to unitary quantum dynamics, we can also derive the bound for the entropy-like quantities
$
\mathcal{S}(p,X),
$
which is defined in Eq.~\eqref{shannonentropy}.
We have, e.g.,
\aln{\label{sfis}
&|\dot{\mathcal{S}}|\nonumber\\
\leq &\sqrt{\frac{\max_{n\sim m}\|\hat{H}_{nm}\|_\mr{op}}{{R_p}}\sum_{n\sim m}p_{n}\lrs{\ln\frac{p_nX_m}{p_mX_n}}^2} 
\sqrt{R_p^2-{E_\mr{trans}^2}},
}
where we set $r=r^p$ and use  $\sqrt{p_np_m}\leq (p_n+p_m)/2$. Here, $R_p$ and $E_\mr{trans}$ are defined from the graph $\mc{E}$ instead of $\mc{E}_A$.
For $X=p^\mr{ref}$, the first term in the right-hand side reduces to the square root of the (discrete version of) Fisher divergence~\cite{lyu2012interpretation},
$\tilde{\mc{F}}(p||p^\mr{ref})=\sum_{n\sim m}p_n(\ln(p_n/p_m)-\ln(p_n^\mr{ref}/p_m^\mr{ref}))^2/2$ (see inequality~\eqref{fisdiv} for the continuous case).

When we consider the Shannon entropy, $S(p)=\mathcal{S}(p,\vec{1})=-\sum_np_n\ln p_n$, 
we have a different type of nontrivial bounds.
As shown in Appendix~\ref{entapp}, we find
\aln{
|\dot{S}|\leq \sqrt{8\lrs{\frac{C_H}{R_p}-1}}\sqrt{{R_p^2-{E_\mr{trans}^2}}}.
}
This is further evaluated by the bound determined from only the graph structure and the expectation value of the transition Hamiltonian:
\aln{
|\dot{S}|\leq  \sqrt{2}\lrs{C_H-\frac{E_\mr{trans}^2}{C_H}}.
}
It is an interesting future problem to compare our result with other speed limits for entropies~\cite{PhysRevResearch.2.013161}, such as
the Bremermann-Bekenstein bound~\cite{bremermann1967quantum, PhysRevLett.46.623} for information transfer.

\subsection{Bound for continuous case}\label{cont}
Before ending this section, let us  discuss a speed limit for continuous systems and its relation to the bound for discrete systems.
For simplicity, we here focus on the following  nonlinear Schr\"{o}dinger equation,
\aln{
i \hbar \frac{\partial \psi}{\partial t}=\left(-\frac{\hbar^{2}}{2 m} \nabla^{2}+V_\mr{e x t}+g|\psi|^{2}\right) \psi,
}
where we explicitly write the Planck constant in this subsection.
This equation can describe, e.g., the Gross-Pitaevskii equation~\cite{RevModPhys.71.463} for Bose-Einstein condensates by imposing the normalization condition $\int d\mbf{x}|\psi(\mbf{x})|^2=M$, where $M$ is the number of Bosons.
In the following, we instead assume $\int d\mbf{x}|\psi(\mbf{x})|^2=1$ without loss of generality, which makes it easier directly to compare with the discussion in the previous subsections.

In this model, the quantum current is given by
\aln{
\mbf{j}=\frac{i \hbar}{2 m}\left(\psi \nabla \psi^{*}-\psi^{*} \nabla \psi\right):=|\psi(\mbf{x},t)|^2\mbf{v}=\rho\mbf{v},
}
where $\rho=|\psi|^2$.
This satisfies the continuity equation
$
\fracpd{|\psi|^2}{t}+\nabla\cdot\mbf{j}=0.
$
Then, assuming that $\mbf{j}(\mbf{x})\ra 0$ for $|\mbf{x}|\ra \infty$, the expectation value of an on-site observable $A(\mbf{x})$ has the following speed limit:
\aln{\label{contene}
&\lrv{\fracd{}{t}\int d\mbf{x} A(\mbf{x})|\psi(\mbf{x},t)|^2}\nonumber\\
&=\lrv{\int d\mbf{x} \nabla A\cdot \mbf{j}}
\leq \sqrt{\int d\mbf{x} \rho|\nabla A|^2}\sqrt{2E_\mr{kin}}.
}
Here, we have introduced the kinetic energy
\aln{
E_\mr{kin}=\int d\mbf{x}\frac{\rho|\mbf{v}|^2}{2}=\int d\mbf{x}\frac{\rho|\nabla\theta|^2\hbar^2}{2m^2}
}
and   the quantum phase $\theta$ of the state (or the condensate), which satisfies $\nabla\theta=m\mbf{v}/\hbar$.
Inequality \eqref{contene} indicates that the speed should be small when the magnitude of the quantum phase difference is small, which represents the tradeoff relation between transition time and the quantum phase difference.
We also note that
$
E_\mr{kin}=\int d\mbf{x}\frac{\hbar^2}{2m}|\nabla \psi|^2-\int d\mbf{x}\frac{\hbar^2}{8m}\rho(\nabla\ln\rho)^2,
$
where the second term is called the quantum pressure~\cite{RevModPhys.71.463} in atomic physics, which is also proportional to (one of the definitions of) the classical Fisher information~\cite{villani2009optimal} known in statistics.
As far as we know, the quantum speed limit for the nonlinear Schr\"{o}dinger equation in the form of inequality~\eqref{contene} has never been  mentioned before.

We can also find the speed limits of the variance and the  entropy-like quantities as in the inequalities~\eqref{varfint} and~\eqref{sfis}.
For example, the variance of $A$ satisfies
\aln{
\lrv{\fracd{\mbb{V}[A]}{t}}\leq 2\|\nabla A\|_\infty \sqrt{\mbb{V}[A]}\sqrt{2E_\mr{kin}},
}
which leads to
\aln{
|\Delta A(T)-\Delta A(0)|\leq \|\nabla A\|_\infty \av{\sqrt{2E_\mr{kin}}}T.
}
In addition, the Kullback-Leibler divergence between $\rho(\mbf{x},t)$ and $\sigma(\mbf{x})$ has a speed limit
\aln{\label{fisdiv}
\lrv{\fracd{}{t}\int d\mbf{x}\rho(\mbf{x},t)\ln\frac{\rho(\mbf{x},t)}{\sigma(\mbf{x})}}\leq\sqrt{\mc{F}(\rho||\sigma)}\sqrt{2E_\mr{kin}},
}
where
$\mc{F}(\rho||\sigma):=\int d\mbf{x}\rho(\mbf{x})\lrv{\nabla\ln\rho(\mbf{x})-\nabla\ln\sigma(\mbf{x})}^2$ is the Fisher divergence~\cite{lyu2012interpretation}.

Notably, for a simple case,  this speed limit corresponds to the continuous version of  $\mc{B}_p$ for a discrete nonlinear Schr\"{o}dinger equation.
To see this, let us focus on the following equation of motion for a one-dimensional system
\aln{
i\fracd{\psi_l}{t}=-K(\psi_{l-1}+\psi_{l+1})+W\psi_l+\tilde{g}|\psi_l|^2\psi_l.
}
In this case, we have
$
E_\mr{trans}=\sum_lK(\psi_l^*\psi_{l-1}+\psi_{l-1}^*\psi_l)
$
and $ \|\hat{H}_{nm}\|_\mr{op}=K$.
With inserting $\psi_l=\sqrt{\rho_l}e^{i\theta_l}$ and taking the continuous limit, we  have (neglecting the boundary term)
\aln{
{E_\mr{trans}}&=K\sum_l\lrm{\rho_l+\rho_{l-1}-(\sqrt{\rho_l}-\sqrt{\rho_{l-1}})^2}\cos(\theta_l-\theta_{l-1})\nonumber\\
&\ra 2K-Ka^2\int dx\rho(\partial_x\theta)^2-Ka^2\int dx \frac{(\partial_x\rho)^2}{4\rho},
}
\aln{\label{Rdiscon}
R_p\ra 2K-Ka^2\int dx \frac{(\partial_x\rho)^2}{\rho},
}
\aln{
R_p-\frac{E_\mr{trans}^2}{R_p}
\ra 2Ka^2\int dx \rho(\partial_x\theta)^2,
}
and
\aln{
2[A^\mathsf{T}\nabla^2_{r^p}{A}]\ra 2Ka^2\int dx (\partial_x A)^2\rho.
}
Here, $a$ is the lattice constant for the discrete system, and we take the leading order of $a$.
Consequently, the transition-energy bound in Eq.~\eqref{glRpos} becomes
\aln{
\lrv{\fracd{}{t}\int d{x} A({x})|\psi({x},t)|^2}
\leq 2Ka^2\sqrt{\int d{x} \rho|\partial_x A|^2}\sqrt{2E_\mr{kin}},
}
which is equivalent to inequality~\eqref{contene} by setting
$
K=\frac{\hbar}{2ma^2}.
$

The comparison with the continuous and discrete systems also clarifies the close relationship between $R_p$ and the Fisher information.
In fact, if $\|\hat{H}_{nm}\|_\mr{op}$ is constant for all $n$ and $m$, $R_p$ involves the negative sign of the discrete version of Fisher information (see Eq.~\eqref{Rdiscon} for the continuous conterpart). 
As a more rigorous statement, in this simple case, we can show that 
\aln{
\frac{R_p}{\|\hat{H}_{nm}\|_\mr{op}}= \sum_{n\sim_Am}p_n-\frac{1}{2}\sum_{n\sim_Am}(\sqrt{p_n}-\sqrt{p_m})^2,
}
where the second term becomes proportional to the Fisher information when we take the continuous limit.

\section{Unitary quantum dynamics: example\label{SecIV}}

\begin{figure*}
    \centering
    \includegraphics[width=\linewidth]{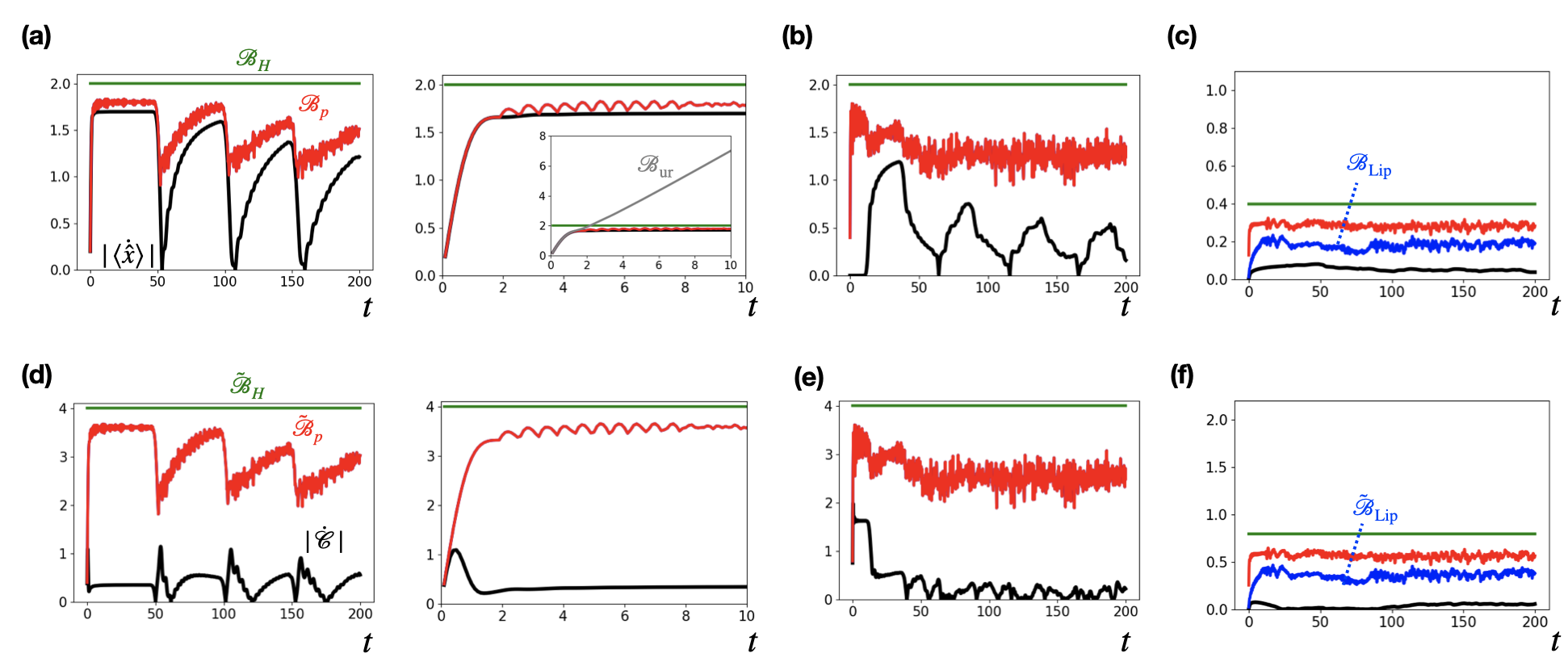}
    \caption{(a-c) Speed of the position, ${\braket{\dot{\hat{x}}}}$ (black), and the speed limits for the initial states (a) $\ket{\psi_1}$ (the right panel shows short-time regime), (b) $\ket{\psi_2}$, and (c) $\ket{\psi_3}$ in a single particle system. The bounds are given by $\mc{B}_\mr{Lip}$ in Eq.~\eqref{Binfty} (blue), $\mc{B}_p$ in Eq.~\eqref{glRpos} (red), and $\mc{B}_H$ in Eq.~\eqref{inftycon} (green). In the inset of the right panel of (a), we also show the  bound  $\mc{B}_\mr{ur}$ based on the uncertainty relation in Eq.~\eqref{ur} (gray).
    Our bounds are not divergent with $L$ and $t$ even for long times and become much tighter than $\mc{B}_\mr{ur}$.
    Furthermore, the bounds become typically tighter in (c) than those for (a) and (b), since $E_\mr{trans}$ is larger. We note that the data for $\mc{B}_\mr{Lip}$ and $\mc{B}_p$ overlap for (a) and (b).
    (d-f) Speed of macroscopic coherence measure  $|{{\dot{\mc{C}}}}|$ defined in Eq.~\eqref{scoh} (black), and the speed limits for the initial states (d) $\ket{\psi_1}$ (the right panel is for short times), (e) $\ket{\psi_2}$, and (f) $\ket{\psi_3}$. The bounds are given by $\mc{\tilde{B}}_\mr{Lip}$  in Eq.~\eqref{cohsumJ} (blue),  $\mc{\tilde{B}}_p$ (red) and $\mc{\tilde{B}}_H$  (green) in Eq.~\eqref{cohbound}.
    We again find that each speed limit provides a physically reasonable bound, which  does not diverge with $L$ and $t$.
    We use $K=1,W_l=0,L=100$, and $M=1$.
    }
    \label{fig100_1}
\end{figure*}

\begin{figure}
    \centering
    \includegraphics[width=\linewidth]{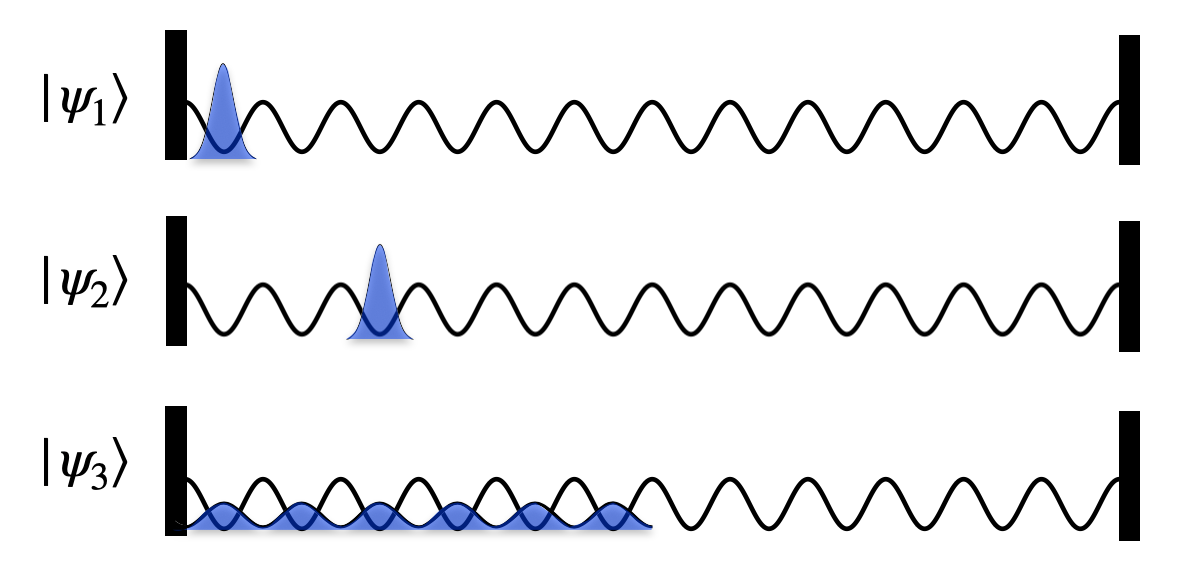}
    \caption{Three initial states for a single-paticle system: $\ket{\psi_1}$ in Eq.~\eqref{ini1}, $\ket{\psi_2}$ in Eq.~\eqref{ini2}, and $\ket{\psi_3}$ in Eq.~\eqref{ini3}
with $L=12$. The state $\ket{\psi_3}$ has a larger expectation value of the transition Hamiltonian than $\ket{\psi_1}$ and $\ket{\psi_2}$.}
    \label{figinitial}
\end{figure}

\subsubsection{Single-particle system}
As a first example for our quantum speed limits, let us  consider a single-particle system on a one-dimensional discrete lattice with length $L$, which obeys the following Schr\"{o}dinger equation:
\aln{
i\fracd{\psi_l}{t}=-K(\psi_{l-1}+\psi_{l+1})+W_l\psi_l,
}
where $K$ is the hopping energy and $W_l$ is a potential.
Assuming the open boundary condition, we can write the corresponding Hamiltonian as
\aln{
\hat{H}=-\sum_{l=1}^{L-1} K(\hat{a}^\dag_{l+1}\hat{a}_l+\mr{h.c.})+\sum_{l=1}^{L}W_l\hat{n}_l,
}
where $\hat{a}_l$ is the annihilation operator of a particle at site $l$ and $\hat{n}_l=\hat{a}_l^\dag\hat{a}_l$ is the number operator.
We define the subspace $n$ as the label for the physical sites measured from the left, for which we have $N=L$.
In this case, we have
$
J_{l,l-1}^q=-J_{l-1,l}^q=-iK(\psi_l^*\psi_{l-1}-\psi_{l-1}^*\psi_l)
$
and $E_\mr{trans}=\sum_{l=1}^{L-1}K(\psi_l^*\psi_{l+1}+\psi_{l+1}^*\psi_l)=E-\sum_lp_lW_l$.
We also have 
$R_p=2K\sum_l\sqrt{p_lp_{l+1}}$ and $C_H=2K$.
As an observable, let us consider the average position
\aln{
\hat{x}=\sum_ll\hat{n}_l=\sum_ll\hat{P}_l
}
here. In this case, we have
$
||\nabla {x}||_\infty=1
$
and
$
[x^\mathsf{T}\nabla_{r^p}^2x]=R_p/2.
$

Figure~\ref{fig100_1}(a-c) shows time evolutions of ${\braket{\dot{\hat{x}}}}$ and our speed limits for several initial states (see Fig.~\ref{figinitial}), i.e.,
\aln{\label{ini1}
\ket{\psi_1}=\hat{a}_1^\dag\ket{\mr{vac}},
}
\aln{\label{ini2}
\ket{\psi_2}=\hat{a}_{L/4}^\dag\ket{\mr{vac}},
}
and
\aln{\label{ini3}
\ket{\psi_3}=\frac{1}{\sqrt{L/2}}\sum_{i=1}^{L/2}\hat{a}_i^\dag\ket{\mr{vac}},
}
where $\ket{\mr{vac}}$ denotes the vacuum state. Here, $|E_\mr{trans}|$ for $\ket{\psi_3}$ is larger  than those for $\ket{\psi_1}$ and $\ket{\psi_2}$.
As shown in the figure, we find that each of the bounds $\mc{B}_\mr{Lip},\mc{B}_p,$ and $\mc{B}_H$ in Eqs.~\eqref{Binfty}, \eqref{glRpos}, and  \eqref{inftycon} provides a physically meaningful bound, which is proportional to the gradient of the quantity and is $\mr{O}(1)$ with respect to $L$ and $t$.
We note that $\mc{B}_\mr{Lip}=\mc{B}_p$ for the initial states $\ket{\psi_1}$ and $\ket{\psi_2}$ when we set $W_l=0$.
Remarkably, these two bounds also become equal to the speed of $\braket{\hat{x}}$ for short times ($t\lesssim 2$) from  $\ket{\psi_1}$ (see Fig.~\ref{fig100_1}(a) right), meaning that the inequalities for $\mc{B}_\mr{Lip}$ and $\mc{B}_p$ satisfy the equality condition  for appropriate situations (see Appendix~\ref{tightapp} for details).

Several remarks are in order.
First, the obtained bounds are much tighter than  the conventional bound $\mc{B}_\mr{ur}$ in~\eqref{ur}, where $\Delta x$  diverges as $\propto t$ for $\ket{\psi_1}$ (inset of Fig.~\ref{fig100_1}(a) right), $\propto t$ for $\ket{\psi_2}$, and $\propto L$ for $\ket{\psi_3}$  (data not shown)~\footnote{
This divergence for the speed is different from the divergence of the transition time discussed in~\cite{PhysRevA.94.052125,PhysRevA.103.022210}, which is caused by the vanishing averaged speed near the stationary state.
}.
Second, as demonstrated in the result for $\ket{\psi_1}$, the bounds $\mc{B}_\mr{Lip} $ and $ \mc{B}_p$ can even capture the non-monotonic behavior of $|\braket{\dot{\hat{x}}}|$ (see, e.g., $t\sim 50$).
Third, as shown in Fig.~\ref{fig100_1}(c), the speed of the observable  for $\ket{\psi_3}$ becomes typically much smaller than those for 
$\ket{\psi_1}$ and $\ket{\psi_2}$ because of large $E_\mr{trans}$, which is correctly captured by our speed limits.

Next, we discuss the speed limit for macroscopic coherence.
For this purpose, we first consider the following measure
\aln{\label{scoh}
\mc{C}=\sum_{l,l'}|l-l'||\rho_{l,l'}|^2
=\sum_{l,l'}|l-l'|p_lp_{l'}.
}
Figure~\ref{fig100_1}(d-f) shows time evolutions of $|\dot{\mc{C}}|$ and our speed limits for the different initial states.
Here, $\mc{\tilde{B}}_p$ and $\mc{\tilde{B}}_H$ are defined in inequalities \eqref{cohbound}, and we have also defined another bound
\aln{\label{cohsumJ}
\tilde{\mc{B}}_\mr{Lip}=2\max_{n\sim_\mc{C}m}|c^{nm}|\cdot\sum_{n\sim_\mc{C}m}|J_{nm}^q|.
}
As demonstrated in the figure, we again find that each bound provides a  physically meaningful bound, which is proportional to the quantities characterizing the gradient and does not diverge with $L$ and $t$.
Note that the speed limits for $\mc{C}$ are just  twice as large as those for $\braket{\hat{x}}$ in the present case, since $\|\nabla x\|_\infty=\max_{n\sim_\mc{C}m}|c^{nm}|=1$.

\begin{figure}
    \centering
    \includegraphics[width=\linewidth]{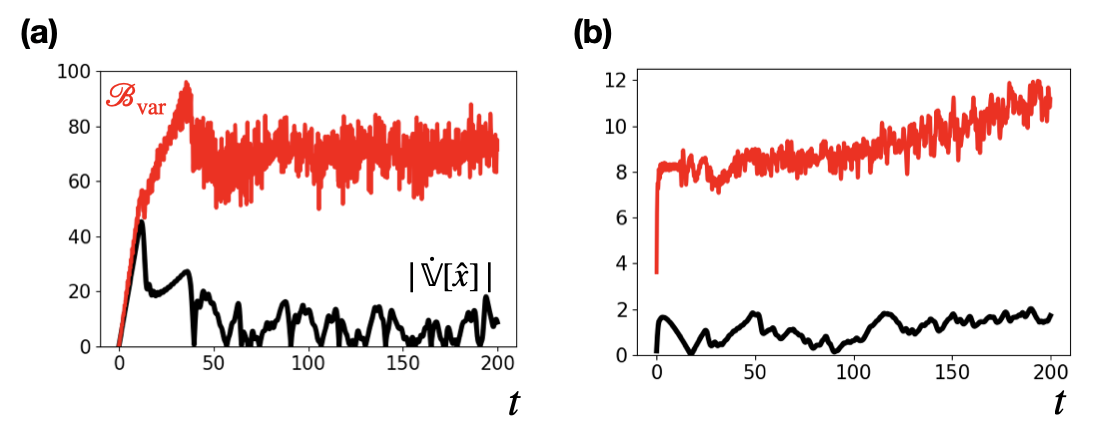}
    \caption{The speed  of the macroscopic coherence measured by the variance of $\hat{x}$, $|\dot{\mbb{V}}[\hat{x}]|$, and the speed limit $\mc{B}_\mr{var}$ in inequality~\eqref{varbound} for the initial states (a) $\ket{\psi_2}$ and (b) $\ket{\psi_3}$ for a single particle system. The speed limit can provide a good bound especially for $t\lesssim L/(4\cdot 2K)$ from $\ket{\psi_2}$ [see (a)].
Moreover, it becomes smaller when the expectation value of the transition Hamiltonian (relative to $R_p$) becomes larger [see (b)].
We use $K=1,W_l=0,L=100$, and $M=1$.}
    \label{figvar}
\end{figure}

As another indicator of macroscopic coherence, we can consider the variance for $\hat{x}$, $\mbb{V}[\hat{x}]$.
In Fig.~\ref{figvar}, we show the speed  of the variance of $\hat{x}$, i.e., $|\dot{\mbb{V}}[\hat{x}]|$ and the speed limit $\mc{B}_\mr{var}$ in inequality~\eqref{varbound} for the initial states $\ket{\psi_2}$ and $\ket{\psi_3}$.
We find that the speed limit (i) can provide an excellent bound even for the variance speed (especially for $t\lesssim L/(4\cdot 2K)$ from $\ket{\psi_2}$)
and (ii) becomes smaller when the expectation value of the transition Hamiltonian (relative to $R_p$) becomes larger, as seen from the result for $\ket{\psi_3}$.

\begin{figure}
    \centering
    \includegraphics[width=\linewidth]{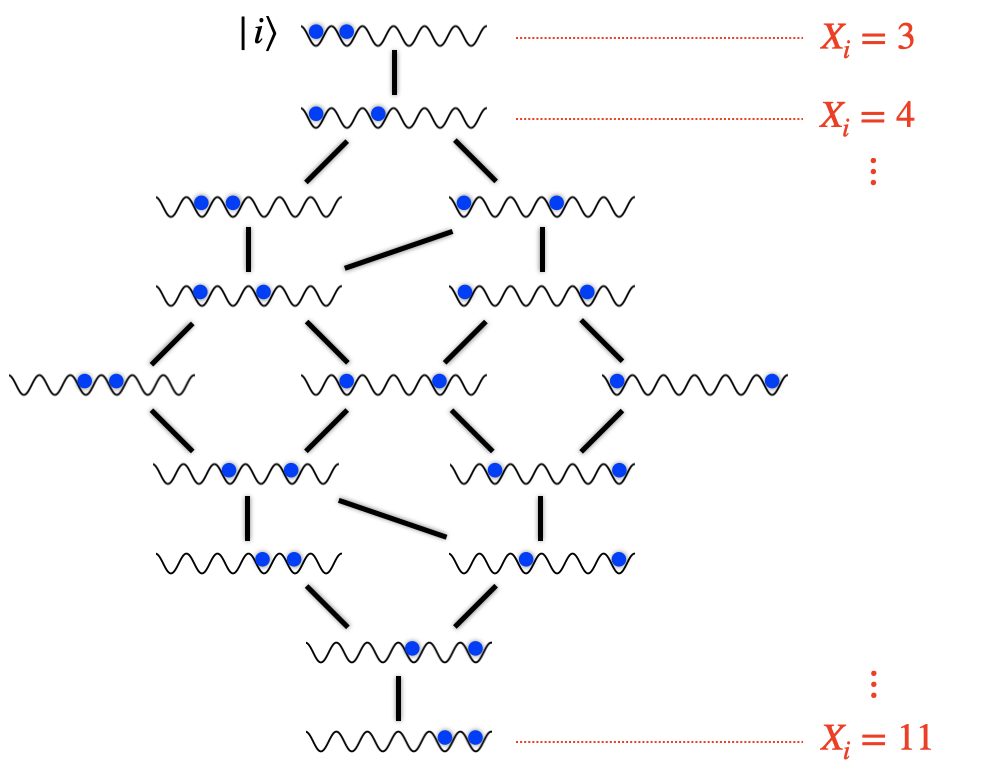}
    \caption{An example of the graph $\mc{E}$ (or equivalently, $\mc{E}_X$) for the system with many hardcore bosons in Eq.~\eqref{mbpart}. We can calculate $X_i$ for each Fock state $\ket{i}$. We have $\|\nabla X\|_\infty=1$ and $C_H=2MK$. The case with $L=6$ and $M=2$ is shown.}
    \label{mbgraph}
\end{figure}

\begin{figure}
    \centering
    \includegraphics[width=\linewidth]{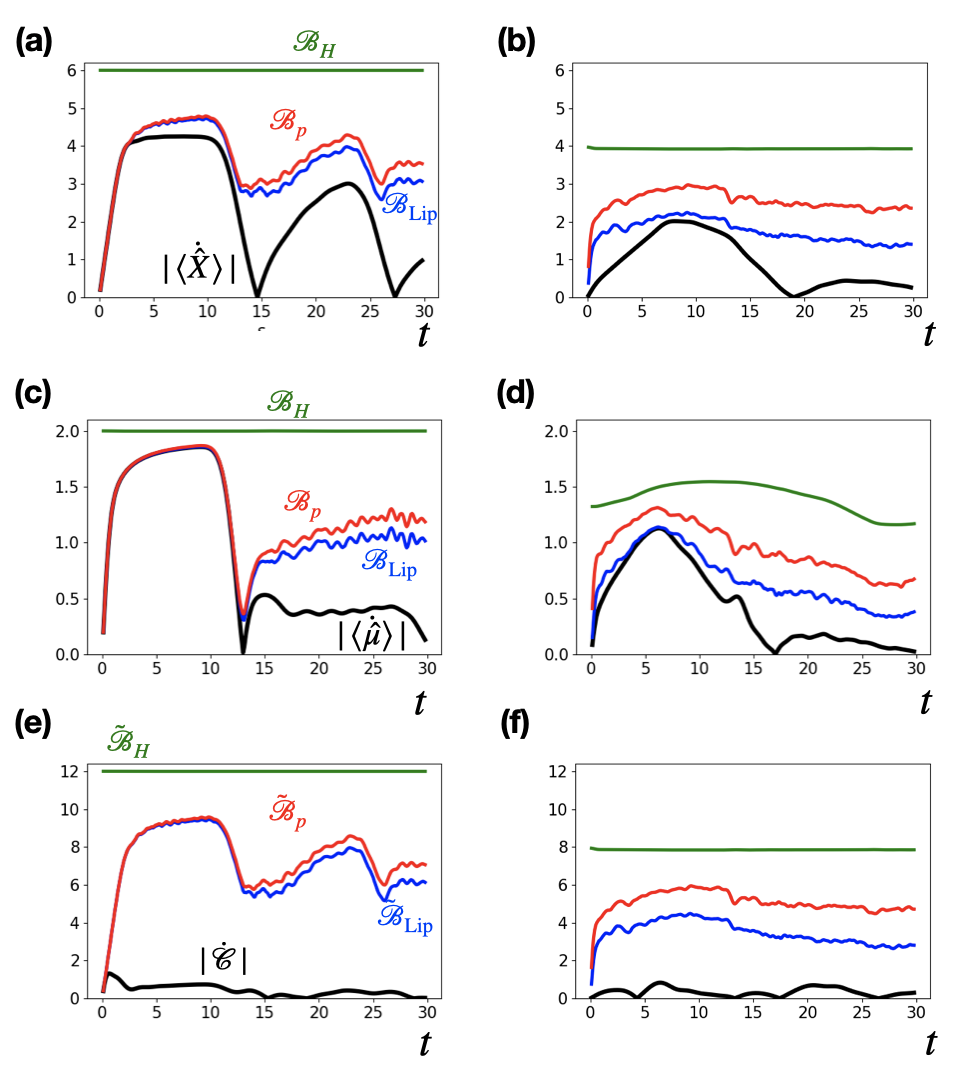}
    \caption{Verification of the speed limits for systems with multiple hardcore bosons from  the initial state (a,c,e) $\ket{\Psi_1}$ or (b,d,f) $\ket{\Psi_3}$. (a,b) Speed of the sum of the positions of the particles, ${\braket{\dot{\hat{X}}}}$ (black), and the speed limits  given by $\mc{B}_\mr{Lip}$ in Eq.~\eqref{Binfty} (blue), $\mc{B}_p$ in Eq.~\eqref{glRpos} (red), and $\mc{B}_H$ in Eq.~\eqref{inftycon} (green). We find that our  speed limits are valid even for many-body systems.
    (c,d) Speed of the  position of the rightmost particle, ${\braket{\dot{\hat{\mu}}}}$ (black) and our speed limits. Remarkably, the bounds 
 $\mc{B}_\mr{Lip}$ and $\mc{B}_p$
 can provide good bounds for $t\lesssim 12$ from $\ket{\Psi_1}$, for which $|\braket{\dot{\hat{\mu}}}|$ exhibits non-monotonic behavior.
 (e,f) Speed of  macroscopic coherence $|{{\dot{\mc{C}}}}|$ (black) and the speed limits  given by $\mc{\tilde{B}}_\mr{Lip}$  in Eq.~\eqref{cohsumJ} (blue),  $\mc{\tilde{B}}_p$ (red) and $\mc{\tilde{B}}_H$  (green) in Eq.~\eqref{cohbound}.
 We use $K=1, V=0.1, W_l=0,L=24$, and $M=3$.
}
    \label{fig24_3}
\end{figure}

\subsubsection{Many-particle system}
Next, we consider a one-dimensional system composed of $M$ hardcore bosons in $L$ lattice sites ($M<L/2$) under the open boundary condition, whose Hamiltonian is described by
\aln{\label{mbpart}
\hat{H}=-\sum_{l=1}^{L-1} K(\hat{a}^\dag_{l+1}\hat{a}_l+\mr{h.c.})+\sum_{l=1}^{L-1}V\hat{n}_l\hat{n}_{l+1}+ \sum_{l=1}^{L}W_l\hat{n}_l,
}
where $\hat{a}_l$ denotes the annihilation operator of a hardcore boson at site $l$ and $\hat{n}_l=\hat{a}_l^\dag\hat{a}_l$.
In the following, we consider that each of $\mc{H}_n$ is the one-dimensional space describing the (many-body) Fock state, $\ket{i}\:(i=1,\cdots,\mr{dim}[\mc{H}])$.
Transport experiments of such multiple-particle systems are realized in cold atomic systems~\cite{schneider2012fermionic,PhysRevLett.110.205301}.

As a quantity of interest, we first consider
the sum of the positions of $M$ particles:
\aln{
\hat{X}=\sum_ll\hat{n}_l=\sum_iX_i\hat{P}_i,
}
where $X_i=\sum_ml_{i,m}$ is the sum of the particle positions $l_{i,1},\cdots,l_{i,M}$ for the Fock basis  $\ket{i}$.
Under this setup, we have $E_\mr{trans}=-\braket{\sum_{l=1}^{L-1} K(\hat{a}^\dag_{l+1}\hat{a}_l+\mr{h.c.})}$, $R_p=2K\sum_{i\sim j}\sqrt{p_ip_j}$, and $C_H=2MK$.
We also have
$\|\nabla X\|_\infty=1$ and $[X^\mathsf{T}\nabla_{r^p}^2X]=R_p/2$.
Figure~\ref{mbgraph} shows the example of the graph $\mc{E}\:(=\mc{E}_X)$ generated by the many-body Hamiltonian for $L=6$ and $M=2$.

Figure~\ref{fig24_3}(a,b) shows time evolutions of ${\braket{\dot{\hat{X}}}}$ and our speed limits for different initial states, i.e.,
\aln{\label{manyini1}
\ket{\Psi_1}=\hat{a}_1^\dag\cdots\hat{a}_M^\dag\ket{\mr{vac}}
}
and
\aln{\label{manyini2}
\ket{\Psi_3}=\sqrt{\frac{M!(L/2-M)!}{(L/2)!}}\sum_{1\leq i_1< \cdots< i_M\leq L/2 }\hat{a}_{i_1}^\dag\cdots\hat{a}_{i_M}^\dag\ket{\mr{vac}},
}
which reduce to $\ket{\psi_1}$ and $\ket{\psi_3}$ for $M=1$, respectively. 
Note that $\ket{\Psi_3}$ has larger $|E_\mr{trans}|$ than $\ket{\Psi_1}$.
From the figure, we find that our  speed limits are valid even for many-body systems.
In particular,  (i) they do not show unbounded increase with  $t$,  (ii)  $\mc{B}_\mr{Lip}$ and $\mc{B}_p$ provide good bounds for short times from 
$\ket{\Psi_1}$, and (iii) the expectation value of the transition Hamiltonian again suppresses the bound (compare (a) and (b)).
For (i), we also note that the bounds are independent of $L$ and approximately proportional to $M$ (data not shown). 
This means that the speed of the average position $\braket{\dot{\hat{X}}}/M$ becomes $\mr{O}(M^0)$, which is natural for locally interacting systems.

Next, we consider  the position of the rightmost particle as another observable of interest:
\aln{
\hat{\mu}=\sum_ll\hat{n}_l\prod_{l'=l+1}^L(1-\hat{n}_{l'})=\sum_i\mu_i\hat{P}_i,
}
where $\mu_i=\max_ml_{i,m}$ is the maximum position of the particle for a given Fock state $\ket{i}$.
Note that we cannot write this observable as a sum of local observables.
In this case, we have $E_\mr{trans}=\braket{\sum_{i\sim_\mu j} H_{ij}}$, $R_p=2K\sum_{i\sim_\mu j}\sqrt{p_ip_j}$, and $C_H=2K$.
We also have
$\|\nabla \mu\|_\infty=1$ and $[\mu^\mathsf{T}\nabla_{r^p}^2\mu]=R_p/2$.

Figure~\ref{fig24_3}(c,d) shows time evolutions of ${\braket{\dot{\hat{\mu}}}}$ and our speed limits for 
$\ket{\Psi_1}$
and
$\ket{\Psi_3}$.
We find that our  speed limits are valid for this non-local observable.
Remarkably, the bounds 
 $\mc{B}_\mr{Lip}$ and $\mc{B}_p$
 can provide good bounds for $\ket{\Psi_1}$ and $t\lesssim 12$, for which $|\braket{\dot{\hat{\mu}}}|$ exhibits non-monotonic behavior.

Finally, we discuss the speed limit for the macroscopic coherence, which we choose as
\aln{
\mc{C}=\sum_{i,j}\lrv{\sum_{m=1}^M l_{i,m}-l_{j,m}}|\rho_{ij}|^2
=\sum_{i,j}\lrv{\sum_{m=1}^M l_{i,m}-l_{j,m}}p_ip_j.
}
This reduces to Eq.~\eqref{scoh} for $m=1$.
Figure~\ref{fig24_3}(e,f) shows time evolutions of $|{{\dot{\mc{C}}}}|$ and our speed limits for 
$\ket{\Psi_1}$
and
$\ket{\Psi_3}$.
 In this case, the bounds 
$\mc{\tilde{B}}_p$ and $\mc{\tilde{B}}_H$ defined in Eq.~\eqref{cohbound} and $\mc{\tilde{B}}_\mr{Lip}$ defined in Eq.~\eqref{cohsumJ} are twice as large as $\mc{{B}}_p, \mc{{B}}_H$, and $\mc{{B}}_\mr{Lip}$ for $\braket{\dot{\hat{x}}}$.
We find that these speed limits indeed bound the speed of coherence.
Except for very short times from $\ket{\Psi_1}$, the bounds do not seem to be tight enough for this case; it remains as future work to investigate whether a tighter universal speed limit for coherence exists.

While we have considered systems with hardcore bosons for simplicity,
we stress that our speed limits can lead to valuable bounds for interacting fermionic and bosonic systems, such as the Fermi- and Bose-Hubbard models.
It is easy to confirm this fact for spinless fermions since the graph's structure describing the Hamiltonian is similar to the case for the hardcore bosons.
For (standard) bosons, the situation becomes more complicated since the local transition strength can diverge as $\max_{i\sim j}|H_{ij}|\sim M$; because of this, studies to find  (state-independent) maximal speeds beyond the conventional Lieb-Robinson bound have recently been active for  bosonic systems~\cite{PhysRevA.84.032309,PRXQuantum.1.010303,PhysRevLett.127.070403,faupin2021maximal}.
In our case, this divergence does not cause much harm, especially when we consider the speed of the average position of atoms, $\hat{X}/M$.
To see this, let $m_{1,i}, \cdots, m_{L,i}$ be the number of particles at sites $1,\cdots, L$ for the state $\ket{i}$.
Then, $M=\sum_{l=1}^Lm_{l,i}$ and $C_H=\max_i\sum_{j(\sim_Xi)}|H_{ij}|\sim K \max_i\sum_l\sqrt{m_{l,i}(m_{l+1,i}+1)}+\sqrt{m_{l,i}(m_{l-1,i}+1)}$.
Thus, $\braket{\dot{\hat{X}}/M}$ is bounded by the factor
$\sqrt{(C_H/M)^2-(E_\mr{kin}/M)^2}$, which does not grow with $M$~\footnote{We have $\sum_l\sqrt{m_{l,i}(m_{l\pm 1,i}+1)}=\sum_{l:m_{l,i}\neq 0}\sqrt{m_{l,i}(m_{l\pm 1,i}+1)}$. 
Using the Cauchy-Schwarz inequality, this is further bounded by
$C_H\lesssim 2K\sqrt{\sum_{l:m_{l,i}\neq 0}{m_{l,i}}\sum_{l:m_{l,i}\neq 0}{(m_{l\pm 1,i}+1)}}\leq 2\sqrt{2}KM$. Thus, $C_M/M$ does not grow with $M$.
}.

\subsubsection{Many-body Spin system}
\begin{figure*}
    \centering
    \includegraphics[width=\linewidth]{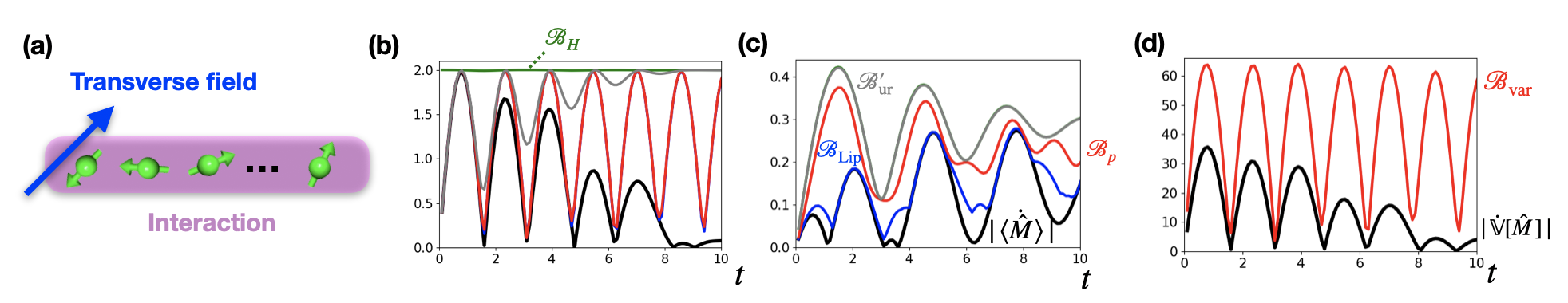}
    \caption{(a) Schematic figure of the many-body spin system in Eq.~\eqref{tfim}.
    The spins consist of the spin-conserving interaction  for the entire spins and the transverse magnetic field only at the first site.
    (b,c) Speed  of the magnetization, $\braket{\dot{\hat{M}}}$ (black), and the speed limits from the initial state (b) $\ket{\Phi_1}$ and (c) $\ket{\Phi_2}$.
    The bounds are given by $\mc{B}_\mr{Lip}$ (blue), $\mc{B}_p$ (red),  $\mc{B}_H$ (green), and $\mc{B}'_\mr{ur}$ (gray).
    The speed limits can qualitatively capture (part of) the oscillations of $\lrv{\braket{\dot{\hat{M}}}}$ and provide nice bounds for certain times, such as $t\simeq p\pi/2\:(p=1,2,\cdots)$ for $\mc{B}_\mr{Lip}$ and $\mc{B}_p$ from $\ket{\Phi_1}$.
    Note that some of the curves are almost overlapped ($\mc{B}_\mr{Lip}\simeq\mc{B}_p$ for $\ket{\Phi_1}$ and  $\mc{B}_H\simeq \mc{B}'_\mr{ur}$ for $\ket{\Phi_2}$).
    (d) Time evolution of the macroscopic coherence measured by the variance $\mbb{V}[\hat{M}]$ (black) and its bound $\mc{B}_\mr{var}$ (red) from the GHZ state $\ket{\Phi_3}$.
We find that the bound works well  especially for $t\simeq p\pi/2\:(p=1,2,\cdots)$.
    We consider the case with $L=10, J_x=J_y=0.1, J_z=0.05,$  and $h=1$ for all of the calculations.
    }
    \label{figspin}
\end{figure*}

Next, we consider an interacting quantum many-body spin system in one dimension with system size $L$ (see Fig.~\ref{figspin}(a)) whose Hamiltonian reads
\aln{\label{tfim}
\hat{H}=h\hat{\sigma}_1^x+\sum_{l=1}^{L-1}J\hat{\sigma}_l^x\hat{\sigma}_{l+1}^x+J\hat{\sigma}_l^y\hat{\sigma}_{l+1}^y+J_z\hat{\sigma}_l^z\hat{\sigma}_{l+1}^z.
}
We note that the interaction terms do not change the total magnetization 
\aln{
\hat{M}=\sum_{l=1}^L\hat{\sigma}_l^z=\sum_{i=1}^{2^L}m_i\hat{P}_i,
}
while the transverse magnetic field $h\hat{\sigma}_1^x$ does.
Here, we consider that each of $\mc{H}_n$ is the one-dimensional space describing the (many-body) computational state, $\ket{i}\:(i=1,\cdots,\mr{dim}[\mc{H}]=2^L)$.

As an observable of interest, we consider $\hat{M}$ itself.
In this case, we have
$E_\mr{trans}=h\braket{\hat{\sigma}_1^x}$,
$R_p=2h\sum_{i\sim_Mj}\sqrt{p_ip_j}$, $C_H=h$, $\|\nabla M\|_\infty=2$, and $[M^\mathsf{T}\nabla_{r^p}^2M]=2R_p$.
As initial states, we take
\aln{
\ket{\Phi_1}=\bigotimes_{l=1}^L\ket{\uparrow}_l
}
and
\aln{
\ket{\Phi_2}=\frac{1}{\sqrt{2}}(\ket{\uparrow}_1+\ket{\downarrow}_1)\otimes\lrs{\bigotimes_{l=2}^L\ket{\uparrow}_l},
}
where $\ket{\uparrow}_l/\ket{\downarrow}_l$ is the eigenstate of $\hat{\sigma}_l^z$ with the eigenvalue $+1/-1$.
Since $\braket{\hat{M}(0)}\simeq L$ for both of the initial states, a macroscopic transition of the magnetization occurs when the system undergoes a unitary time evolution.

Figure~\ref{figspin}(b,c) shows time evolutions of $|{\braket{\dot{\hat{M}}}}|$ and our speed limits for 
$\ket{\Phi_1}$
and
$\ket{\Phi_2}$.
We can see that our speed limits $\mc{B}_\mr{Lip},\mc{B}_p$, and $\mc{B}_H$ are verified even for the many-body spin system.
In particular, these speed limits can qualitatively capture (part of) the oscillations of $\lrv{{\braket{\dot{\hat{M}}}}}$ and provide a good bound for some times (such as $t\simeq p\pi/2\:(p=1,2,\cdots)$ for $\mc{B}_\mr{Lip}$ and $\mc{B}_p$ from $\ket{\Phi_1}$).

We note that, in this specific Hamiltonian and observable, we can also consider a modified version of the conventional bound in Eq.~\eqref{ur}.
Namely, since $|\braket{\dot{\hat{M}}}|=|\braket{[\hat{H},\hat{M}]}|
=|\braket{[h\hat{\sigma}_1^x,\hat{\sigma}_1^z]}|$, we can consider
$\mc{B}'_\mr{ur}:=2\sqrt{\mbb{V}[h\hat{\sigma}_1^x]\mbb{V}[\hat{\sigma}_1^z]}$.
As shown in Fig.~\ref{figspin}(a,b), this indeed gives a bound for $|\braket{\dot{\hat{M}}}|$.
In this specific case, we can show $\mc{B}'_\mr{ur}\leq \mc{B}_H$.
On the other hand, our speed limits based on the local conservation law of probability (such as $\mc{B}_\mr{Lip}$) have some advantages over $\mc{B}'_\mr{ur}$ for the following reasons.
First, the possibility of the reduction of $\hat{H},\hat{M}\ra h\hat{\sigma}_1^x,\hat{\sigma}_1^z$ relies on the specific observable and the Hamiltonian, and it is difficult to obtain appropriate $\mc{B}'_\mr{ur}$  for general settings. 
Second, our method can be used for obtaining  speed limits for, e.g., macroscopic coherence (see the next paragraph), going beyond the bound $\mc{B}'_\mr{ur}$.
Third, even when we focus on the bound on $|\braket{\dot{\hat{M}}}|$, we numerically find that $\mc{B}_\mr{Lip}$ and $\mc{B}_p$ provide better bounds than $\mc{B}'_\mr{ur}$.

To discuss the coherence structure of the dynamics, we next consider the variance of the macroscopic magnetization difference $\mbb{V}[\hat{M}]=\mc{C}/2$ with $c^{ij}=(m_i-m_j)^2$ in Eq.~\eqref{cohdef}.
Note that $\mbb{V}[\hat{M}]=\mr{O}(L^2)$ when the state is a macroscopically superposed (cat) state~\cite{PhysRevLett.95.090401}, such as the 
 Greenberger–Horne–Zeilinger state,
\aln{\label{GHZ}
\ket{\Phi_3}=\frac{1}{\sqrt{2}}\lrs{\ket{\uparrow\cdots\uparrow}+\ket{\downarrow\cdots\downarrow}},
}
which satisfies $\mbb{V}[\hat{M}]=L^2$.
In contrast, non-cat states possess $\mr{o}(L^2)$, such as the equally superposed state $\frac{1}{\sqrt{d}}\sum_i\ket{i}$ with 
$\mbb{V}[\hat{M}]=L$ and the computational-basis states with $\mbb{V}[\hat{M}]=0$.

Figure~\ref{figspin}(d) shows a time evolution of the variance from the GHZ state $\ket{\Phi_3}$.
We find that the speed limit $\mc{B}_\mr{var}$ in \eqref{varbound} works well for $\mbb{V}[\hat{M}]$, especially at $t\simeq p\pi/2\:(p=1,2,\cdots)$.
We also note that our inequality~\eqref{varfint} enables us to obtain a rigorous lower bound of the time $T$ for a cat state to relax to a non-cat state (or vice versa) via the unitary time evolution,
{
\aln{
T\geq \frac{\lrv{\sqrt{\mbb{V}[\hat{M}(t)]}-\sqrt{\mbb{V}[\hat{M}(0)]}}}{\|\nabla M\|_\infty \av{\sqrt{C_H^2-E_\mr{trans}^2}}}=\frac{\mr{O}(L)}{2},
}
}
which is extensive with respect to the system size.

\section{Classical Markovian systems and irreversibility bound\label{SecV}}
This section turns our attention toward classical Markovian systems and shows that the irreversible entropy production bounds macroscopic transition speed.
Entropy production has recently been 
found to play an important role~\cite{PhysRevLett.105.170402,PhysRevLett.107.140404,PhysRevLett.117.190601,PhysRevLett.114.158101,PhysRevLett.116.120601,horowitz2020thermodynamic,dechant2018current,PhysRevE.97.062101,PhysRevLett.121.070601,funo2019speed,PhysRevE.93.052145,PhysRevE.96.020103,proesmans2017discrete,PhysRevLett.119.140604,PhysRevLett.123.110602,PhysRevLett.120.090601,dechant2020fluctuation,PhysRevE.99.062126,PhysRevLett.125.140602} for thermodynamic bounds, e.g., certain speed limits and the thermodynamic uncertainty relation.
We here show that a similar useful speed limit for macroscopic transitions at any coarse-grained level under the (local) detailed balance condition by applying
 our general method in Sec.~\ref{SecII}.
 Our bound can be qualitatively better for macroscopic transitions than that proposed in Ref.~\cite{PhysRevLett.121.070601}.
Furthermore, we newly derive a  modified classical speed limit based on the Hatano-Sasa entropy production rate~\cite{PhysRevLett.86.3463}, which works well for long times even when the  detailed balance condition is not satisfied.

\subsection{General setting for discrete systems}
We introduce a finite-dimensional state space $\mc{S}$ and consider the Markovian equation 
\aln{
\fracd{\rho_x}{t}=\sum_y\mc{W}_{xy}\rho_y,
}
where $x,y\in\mc{S}$, $\rho$ is the classical probability distribution and $\mc{W}_{xy}$ is the transition rate  matrix satisfying $\mc{W}_{xy}\geq 0$ for $x\neq y$ and $\sum_{x}\mc{W}_{xy}=0$.

We arbitrarily decompose $\mc{S}$ as
$
\mc{S}=\bigoplus_n\mc{S}_n
$
and define the coarse-grained probability
\aln{
p_n=\sum_{x\in\mc{S}_n}\rho_x
}
as well as the coarse-grained transition rate  matrix
\aln{
{W}_{nm}=\sum_{x\in\mc{S}_n}\frac{\sum_{y\in\mc{S}_m} \mc{W}_{xy}\rho_y}{\sum_{y\in\mc{S}_m}\rho_y}.
}
Then, the original Master equation becomes the coarse-grained one~\cite{PhysRevE.85.041125},
\aln{
\fracd{p_n}{t}=\sum_m{W}_{nm}p_m=-\sum_{m(\sim  n)} J_{mn}^c, 
}
where 
\aln{
J_{mn}^c=-{W}_{nm}p_m+{W}_{mn}p_n.
}
and the graph $\mc{E}$ is defined as $\mc{E}=\lrm{(n,m)\:|\:(n\neq m)\text{ and } (W_{nm}\neq 0 \text{ or } W_{mn}\neq 0)}$.
We note that ${W}$ depends on $p(t)$ in general and that $\sum_nW_{nm}=0$.

\subsection{Bound for macroscopic observables}
We consider a macroscopic observable that is given by
\aln{
A=\sum_na_np_n.
}
Using the general method introduced in Sec.~\ref{SecII}, we obtain
$\braket{\dot{A}}=-\frac{1}{2}\sum_{n\sim m}(a_n-a_m)J_{mn}^c
$
and inequalities such as
\aln{\label{classJ}
|\braket{\dot{A}}|\leq \mc{B}_\mr{cur}^c:=\frac{\|\nabla A\|_\infty}{2}\sum_{n\sim_A m}|J_{mn}^c|.
}

\subsubsection{Standard entropy production bound}
Let us first consider a bound based on the standard entropy production rate.
For simplicity, we start from the case where the system is attached to a single heat bath at temperature $\beta^{-1}$ and $\mc{W}$ satisfies the detailed balance condition, i.e., $\mc{W}_{xy}/\mc{W}_{yx}=e^{-\beta (E_x-E_y)}$, where $E_x$ is the energy of the state $x$.
Denoting the Shannon entropy of $\rho$ as $S(\rho)$, we can write down the entropy production rate of the system $\dot{\Sigma}:=\dot{S}(\rho_x)+\sum_{x\neq y}\mc{W}_{yx}\rho_x\beta(E_x-E_y)$ as~\cite{seifert2012stochastic}
\aln{\label{EP}
\dot{\Sigma}=\sum_{x\neq  y}\mc{W}_{xy}\rho_y\ln\frac{\mc{W}_{xy}\rho_y}{\mc{W}_{yx}\rho_x}.
}
Similarly, we can define the following entropy production rate defined from the coarse-grained Markovian dynamics~\cite{PhysRevE.85.041125}
\aln{\label{EPcg}
\dot{\Sigma}_\mr{cg}:=\sum_{n\sim_Am}W_{nm}p_m\ln\frac{W_{nm}p_m}{W_{mn}p_n}.
}
This quantity is smaller than the bare entropy production rate, i.e.,
$\dot{\Sigma}_\mr{cg}\leq \dot{\Sigma}$,
which is shown via the log-sum inequality (see Appendix~\ref{Appcs}).

Now, as proven in Appendix~\ref{Appcs}, we find the following bound  for the instantaneous speed of the expectation value of $A$:
\aln{~\label{classcg}
|\braket{\dot{A}}|\leq\sqrt{\frac{\Theta_A\dot{\Sigma}_\mr{cg}}{2}}
\leq \|\nabla A\|_\infty\sqrt{\frac{\mc{A}_\mr{cg}\dot{\Sigma}_\mr{cg}}{2}}
}
and 
\aln{~\label{classucg}
|\braket{\dot{A}}|\leq \|\nabla A\|_\infty\sqrt{\frac{\mc{A}\dot{\Sigma}}{2}}
}
where $\Theta_A:=\sum_{n\sim m}(a_n-a_m)^2W_{nm}p_m$ is the second moment of the transition speed of $A$ and we define
\aln{~\label{classucgn}
\mc{A}_\mr{cg}:=\sum_{n\sim_Am}W_{nm}
}
as the dynamical activity of the coarse-grained equation.
Note that $\mc{A}_\mr{cg}\leq \mc{A}:=\sum_{x\neq y}\mc{W}_{xy}$, where $\mc{A}$ is the bare dynamical activity~\cite{PhysRevLett.121.070601}.
The proof of inequality~\eqref{classcg} is made 
by taking $r_{nm}=W_{nm}p_m+W_{mn}p_n$ (note that  $\Theta_A=[A^\mathsf{T}\nabla^2_rA]$) and using the inequalities  introduced in Refs.~\cite{PhysRevLett.121.070601,PhysRevLett.117.190601}.
This inequality can also be obtained from the short-time version of the thermodynamic uncertainty relation~\cite{PhysRevE.101.062106}. 
It explicitly shows that the entropy production rate can provide a useful bound for macroscopic systems because of the factor $\|\nabla A\|_\infty$, which is concretely
illustrated in Sec.~\ref{SecVI}.

We can use the instantaneous speed limit obtained above as the bound for the relaxation time. Integrating the inequality in \eqref{classucg} from time 0 to $T$, 
we obtain
\aln{\label{classbound}
T\geq \frac{2|\braket{{A}}(t)-\braket{A}(0)|^2}{||\nabla A||_\mr{\infty}^2{{\av{\mc{A}}\Sigma}}},
}
where we define the total entropy production 
$\Sigma=\int_0^Tdt\dot{\Sigma}$, which is equal to $D(\rho(0)||\rho^\mr{ss})-D(\rho(t)||\rho^\mr{ss})$ for time-independent $\mc{W}$~\footnote{This bound may be further tightened by using the recently found inequality for the Wasserstein distance~\cite{dechant2021minimum}.}.

We note that the above inequalities can be much tighter than that proposed in Ref.~\cite{PhysRevLett.121.070601}.
Indeed, the reference considers the speed limit for total variation distance, from which we  obtain $|\braket{\dot{A}}|\leq \|A\|_\infty \sqrt{2\mc{A}\dot{\Sigma}}$ and 
$T\geq \frac{|\braket{{A}}(t)-\braket{A}(0)|^2}{2\|A\|_\mr{\infty}^2{{\av{\mc{A}}\Sigma}}}$.
For observables satisfying $\|\nabla A\|_\infty\ll\|A\|_\infty$, which often appears for macroscopic transitions (see Sec.~\ref{SecVI}), inequalities~\eqref{classucgn} and \eqref{classbound} become considerably better than those in Ref.~\cite{PhysRevLett.121.070601}.

The speed limits are obtained even for the entropy production rate itself.
Indeed, as detailed in Appendix~\ref{Appcs},
we have $\dot{\Sigma}_\mr{cg}\leq \Theta_{\Sigma}/2$, where 
$\Theta_\Sigma:=\sum_{n\sim m}W_{nm}p_m\lrs{\ln\frac{W_{nm}p_m}{W_{mn}p_n}}^2$.
Using this, one obtains 
\aln{\label{classtheta}
|\braket{\dot{A}}|\leq\frac{\sqrt{\Theta_A\Theta_\Sigma}}{2}.
}

While the above discussion has been done for the single heat bath, 
we can consider the case with multiple heat baths labeled by $\nu$ with temperatures $\beta_\nu^{-1}$.
In this case, $\mc{W}_{xy}=\sum_\nu\mc{W}_{xy}^\nu$ and we can assume the local detailed balance condition, $\mc{W}_{xy}^\nu/\mc{W}_{yx}^\nu=e^{-\beta_\nu(E_x-E_y)}$ for any $\nu$ and $x\neq y$.
Then, we can introduce $\dot{\Sigma}^\nu$ and $\dot{\Sigma}_\mr{cg}^\nu$ by replacing $\mc{W}_{xy}$ with $\mc{W}^\nu_{xy}$ in the expressions \eqref{EP} and \eqref{EPcg} (note the implicit dependence of $W$ on $\mc{W}$).
Redefining $\dot{\Sigma}$ and $\dot{\Sigma}_\mr{cg}$ by $\dot{\Sigma}=\sum_\nu\dot{\Sigma}^\nu$ and 
$\dot{\Sigma}_\mr{cg}=\sum_\nu\dot{\Sigma}_\mr{cg}^\nu$,
we can reproduce the above bounds inequalities~\eqref{classcg},~\eqref{classucg},~\eqref{classbound}, and~\eqref{classtheta}, which are based on the entropy production rate.
We also note that~\eqref{classcg},~\eqref{classucg},~\eqref{classbound}, and~\eqref{classtheta} still hold for general Markovian dynamics even without the local detailed balance condition
by defining $\Sigma$ and $\Sigma_\mr{cg}$ through Eqs.~\eqref{EP} and \eqref{EPcg},
although  $\Sigma$ and $\Sigma_\mr{cg}$ may no longer be identified  as the physical entropy production in this case.

\subsubsection{Hatano-Sasa entropy production bound}
While the above bounds based on the standard entropy production rate is useful under the detailed balance condition, it may not provide a tight bound for long times without this condition.
This is because of the possible existence of a stationary current, which leads to the nonzero entropy production rate.

Here, we show that a modified speed of the transition of an observable is bounded using the Hatano-Sasa entropy production~\cite{PhysRevLett.86.3463}, which is
the entropy production where the entropy generated by the stationary dissipation is subtracted.
Assuming that the instantaneous stationary state $\rho^\mr{ss}$ of $\mc{W}(t)$ is unique, the Hatano-Sasa entropy production rate for general Markovian systems is given by 
\aln{\label{HSEP}
\dot{\Sigma}^\mr{HS}=\sum_{x,y}\mc{W}_{xy}(t)\rho_y(t)\ln\frac{\mc{W}_{xy}(t)\rho_y(t)}{\tilde{\mc{W}}_{yx}(t)\rho_x(t)},
}
where
\aln{
\tilde{\mc{W}}_{yx}(t)=\frac{\mc{W}_{xy}(t)\rho_y^\mr{ss}(t)}{\rho_x^\mr{ss}(t)}
}
is the generator of the dual process~\cite{PhysRevLett.121.070601}.
The important property of $\dot{\Sigma}^\mr{HS}$ is that it vanishes for the stationary state unlike $\dot{\Sigma}$.
In addition, $\dot{\Sigma}^\mr{HS}=-\fracd{}{t}D(\rho(t)||\rho^\mr{ss})$ for time-independent $\mc{W}$.
To derive general inequalities, we also introduce the coarse-grained version of $\dot{\Sigma}^\mr{HS}$,
\aln{
\dot{\Sigma}^\mr{HS}_\mr{cg}=\sum_{n,m}W_{nm}(t)p_m(t)\ln\frac{W_{nm}(t)p_m(t)}{\tilde{W}_{mn}(t)p_n(t)},
}
where
$
\tilde{W}_{mn}(t)=\frac{W_{nm}(t)p_m^\mr{ss}(t)}{p_n^\mr{ss}(t)}
$.

Now, to derive a speed limit, we notice that
\aln{
-J_{mn}^c=W_{nm}p_m-\tilde{W}_{mn}p_n-J_{mn}^{c,\mr{ss}}\frac{p_n}{p_n^\mr{ss}},
}
where
\aln{
J_{mn}^{c,\mr{ss}}=-W_{nm}p_m^\mr{ss}+W_{mn}p_n^\mr{ss}
}
is the stationary probability current.
From this, we have
\aln{
\braket{\dot{A}}=&-\frac{1}{2}\sum_{(n,m)\in\mc{E}^c}(a_n-a_m)J_{mn}^{c,\mr{ss}}\frac{p_n}{p_n^\mr{ss}}\nonumber\\
&+\frac{1}{2}\sum_{(n,m)\in\mc{E}^c}(a_n-a_m)(W_{nm}p_m-\tilde{W}_{mn}p_n).
}
Then, introducing
\aln{
V_A
&=-\frac{1}{2}\sum_{(n,m)\in\mc{E}^c}a_nJ_{mn}^{c,\mr{ss}}\lrs{\frac{p_n}{p_n^\mr{ss}}+\frac{p_m}{p_m^\mr{ss}}},
}
we have
\aln{\label{HSbound}
|\braket{\dot{A}}-V_A|\leq\sqrt{{\frac{\tilde{\Theta}_A\dot{\Sigma}_\mr{cg}^\mr{HS}}{2}}}\leq \|\nabla A\|_{\infty}\sqrt{\frac{\mc{A}_\mr{cg}\dot{\Sigma}_\mr{cg}^\mr{HS}}{2}},
}
where
$
\tilde{\Theta}_A=\max[\sum_{n\sim_A m}(a_n-a_m)^2W_{nm}p_m$, \:$\sum_{n\sim_A m}(a_n-a_m)^2\tilde{W}_{nm}p_m]$.
As a special case, we have the non-coarse-grained version of the inequality,
\aln{\label{HSbound2}
|\braket{\dot{A}}-V_A|\leq \|\nabla A\|_{\infty}\sqrt{\frac{\mc{A}\dot{\Sigma}^\mr{HS}}{2}}.
}

We can regard the (modified) speed limits in inequalities~\eqref{HSbound} and ~\eqref{HSbound2}  as bounds for the speed where the contribution of the background current is subtracted.
In fact, in the stationary state, $V_A$ becomes a stationary current for ${A}$.
Since the right-hand side of inequality~\eqref{HSbound2} becomes zero for the stationary state, the inequality can be much tighter than the bound using the standard entropy production rate if the detailed balance condition is absent.

\begin{figure*}
    \centering
  \includegraphics[width=\linewidth]{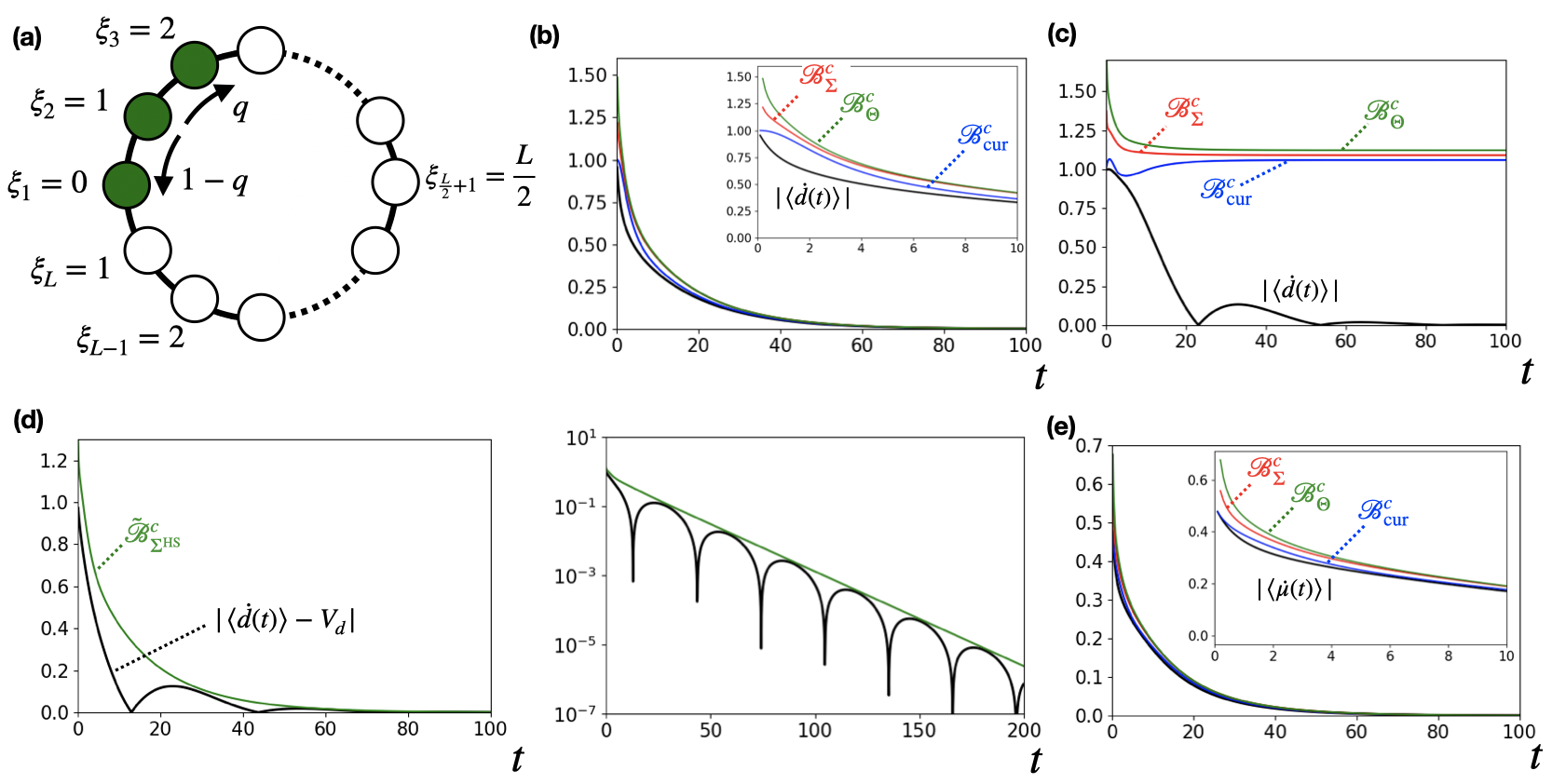}
    \caption{
    (a) Schematic illustration for our model, which describes the symmetric/asymmetric hopping model. We consider $L$ sites and $M$ particles ($M=3$ is shown), which hop between neighboring sites at right and left with  rates $q$ and $1-q$, respectively. We consider the distance $\xi_l$ of particles from the first site in Eq.~\eqref{obsdist}.
    (b) Speed of the sum of $\xi_l$, $|\braket{\dot{d}(t)}|$ (black), and our speed limits for the symmetric simple  exclusion process ($q=0.5$). The main and inset panels show results for the long and short times, respectively.
     The speed is well bounded by the current bound $\mc{B}_\mr{cur}^c$ (blue), which is further bounded by the entropic bound $\mc{B}_\Sigma^c$ (red) and the bound $\mc{B}_\Theta^c$ (green).
     (c) 
     Speed limits of $|\braket{\dot{d}(t)}|$ for the asymmetric simple  exclusion process ($q=0.7$).
     While $\mc{B}_\mr{cur}^c$ works well for short times, none of the bounds become tight for long times due to the stationary current.
     (d) Modified speed limit of $|\braket{\dot{d}(t)}-V_d|$ (black) for the asymmetric simple exclusion process ($q=0.7$).
     we find that the bound $\tilde{\mc{B}}^c_{\Sigma^\mr{HS}}$ (green) is useful to evaluate the speed of the transition both for  (left) short and (right) long  times.
     (e) Speed  of the maximum distance, $|\braket{\dot{\mu}(t)}|$ (black), and the speed limits for the symmetric simple exclusion process ($q=0.5$). The main and inset panels show results for the long and short times, respectively.
     As seen from the almost collapse of the curves, the speed is tightly bounded  by the  bounds $\mc{B}_\mr{cur}^c$ (blue),  $\mc{B}_\Sigma^c$ (red), and $\mc{B}_\Theta^c$ (green), which are obtained from the coarse-grained variables.
          We use $L=18$ and $M=3$ for all the cases.
    }
    \label{fig6}
\end{figure*}

\section{Classical Dynamics: Example\label{SecVI}}
In this section, we discuss concrete examples of the classical speed limits discussed in the previous section.
Specifically, we consider the symmetric and asymmetric simple exclusion processes~\cite{derrida1998exactly} to demonstrate the usefulness of the bound (see Fig.~\ref{fig6}(a)).
The generator describing these dynamics can be written in the operator form as 
\aln{
\mc{W}=\sum_{l=1}^Lq\hat{b}_{l+1}^\dag\hat{b}_l+(1-q)\hat{b}_l^\dag\hat{b}_{l+1}+
\hat{n}_l\hat{n}_{l+1}-\hat{n}_l,
}
where $\hat{b}_l$ is the annihilation operator of a hardcore particle at site $l$ and $\hat{n}_l=\hat{b}_l^\dag\hat{b}_l$ is the number operator of the particle at site $l$.
We here employ the periodic boundary condition with the particle number $M$ and take even $L$.
When $q=0.5$ ($q\neq 0.5$), the dynamics becomes the symmetric (asymmetric) simple exclusion process.
In this model, while the stationary current exists only for $q\neq 0.5$, the stationary state is the uniform state $\rho^\mr{ss}_{x}=1/\mr{dim}[\mc{S}]$ irrespective of $q$. 
Consequently, $\dot{\Sigma}$ in Eq.~\eqref{EP} becomes equivalent to the rate of the Shannon entropy $\dot{S}$ only for $q=0.5$ and $\dot{\Sigma}_\mr{HS}$ in Eq.~\eqref{HSEP} becomes equal to $\dot{S}$ for all $q$.

We first consider an observable describing the sum of the distances from the first site over all particles:
\aln{\label{obsabs}
{d}=\sum_{l=1}^L\xi_l\hat{n}_l
}
with 
\aln{\label{obsdist}
\xi_l=L/2-|L/2-l+1|.
}
If we use the bare (i.e., non-coarse-grained) basis of the many-body Fock-state basis,  parameterized by the number of particles at each site, we have
$
||\nabla d||_\infty=1.
$
In the following, we take an initial state as
a state for which $M$ particles reside in sites  $l=1,\cdots,M$ (Fig.~\ref{fig6}(a)).

Figure~\ref{fig6}(b) shows the speed of $\braket{d}$ for $q=1/2$ (symmetric simple exclusion process).
We find that the speed is well bounded by the current bound $\mc{B}_\mr{cur}^c$ in inequality~\eqref{classJ}, which is further bounded by the entropic bound $\mc{B}_\Sigma^c$ in Eq.~\eqref{classucg}  and the bound $\mc{B}_\Theta^c$ in Eq.~\eqref{classtheta}.
Note that $\mc{B}_\Sigma^c$ and $\mc{B}_\Theta^c$ are almost the same except for early times.

Next, Fig.~\ref{fig6}(c) shows the speed of $\braket{d}$ for $q=0.7$ (asymmetric simple exclusion process).
We find that, while the bound based on the current works well for short times, none of the bounds become tight for long times because of the finite stationary current.
In this case, we find that the bound based on the Hatano-Sasa entropy, $\tilde{\mc{B}}^c_{\Sigma^\mr{HS}}$ in Eq.~\eqref{HSbound2} is useful to evaluate the speed of the transition even for long times (see Fig.~\ref{fig6}(d)).

Let us discuss another observable $\mu$, defined as the particle's distance farthest from the first site.
Namely, for a given state for which the particles are located at $l_1,\cdots,l_M$, $\mu$ takes a value given by 
\aln{
\max_{m=1,\cdots,M}\xi_{l_M}.
}
In contrast with the discussion for $d$, we here consider the coarse-grained subspaces $1,\cdots,L/2$, characterized by the value of $\mu$ (the corresponding graph becomes one-dimension).
Figure~\ref{fig6}(e) shows the  speed of $\braket{\mu}$ for $q=0.5$ and the bounds $\mc{B}_\mr{cur}^c$ in Eq.~\eqref{classJ},  $\mc{B}_\Sigma^c$ in Eq.~\eqref{classcg}  and  $\mc{B}_\Theta^c$ in Eq.~\eqref{classtheta} obtained from the coarse-grained variables.
We see that these bounds give  quite tight speed limits for this observable.

\section{Speed limit for macroscopic quantum systems with Markovian dissipation\label{SecVII}}

In this section, we discuss speed limit for open quantum systems described the Gorini-Kossakowski-Sudarshan-Lindblad master equation~\cite{gorini1976completely,lindblad1976generators}.
In this case, the quantum state obeys 
\aln{
\fracd{\hat{\rho}}{t}=-i[\hat{H},\hat{\rho}]+\sum_\eta\lrs{\hat{L}_\eta\hat{\rho} \hat{L}_\eta^\dag-\frac{1}{2}\lrm{\hat{L}_\eta^\dag \hat{L}_\eta,\hat{\rho}}},
}
where $\eta$ labels the type of the dissipation.
We obtain the following continuity equation for this dynamics:
\aln{
\fracd{p_n}{t}=-\sum_{m, n}J_{mn}^q(t)-\sum_\eta K^\eta_n,
}
where $J_{mn}^q$ is given by Eq.~\eqref{quantcurrent} and 
\aln{
K^\eta_n=&\Tr[\sum_{ml}(\hat{L}_\eta)_{nm}\hat{\rho}_{ml}(\hat{L}_\eta^\dag)_{ln}-\frac{1}{2}(\hat{L}_\eta^\dag)_{nm}(\hat{L}_\eta)_{ml}\hat{\rho}_{ln}\nonumber\\
&-\frac{1}{2}\hat{\rho}_{nm}(\hat{L}_\eta^\dag)_{ml}(\hat{L}_\eta)_{ln}].
}

For an observable written as $\hat{A}=\sum_na_n\hat{P}_n$, we have
\aln{\label{oq1}
|\braket{\dot{\hat{A}}}|\leq \mc{B}_\mr{Uni}+\lrv{\sum_\eta\sum_na_nK_n^\eta},
}
where the first term in the right hand side,
$\mc{B}_\mr{Uni}=\mr{min}\lrm{\mc{B}_\mr{gL},\mc{B}_\mr{Lip}}$, is the  contribution from the Hamiltonian part (unitary dynamics) which we have discussed in Sec.~\ref{SecIII}.
Note that the graph $\mc{E}_\mr{Uni}$ to evaluate 
$\mc{B}_\mr{Uni}$ is defined from the nonzero off-diagonal elements of the Hamiltonian  as in Sec.~\ref{SecIII}, which is different from 
$\mc{E}_\mr{D}$ in the following discussion.

To bound the second term in inequality~\eqref{oq1}, which comes from the dissipation operators $\hat{L}_\eta$,
 we impose one assumption about the jump operator: we require that each jump $\eta$ moves a state in the subspace $\mc{H}_m$ to that in $\mc{H}_{n=f_\eta(m)}$ with an injective function $f_\eta$.
In other words, when $(\hat{L}_\eta)_{nm}$ is nonzero for some $n$  and $m$, 
$(\hat{L}_\eta)_{n'm}\:(n'\neq n)$ and
$(\hat{L}_\eta)_{nm'}\:(m'\neq m)$ should be zero.
Under this assumption, we introduce the (coarse-grained) transition rate from $m$ to $n$,
\aln{
T_{m\ra n}^\mr{cg}=\sum_\eta\Tr[(\hat{L}_\eta^\dag)_{mn}(\hat{L}_\eta)_{nm}\hat{\rho}_{mm}],
}
and the edge set $\mc{E}_\mr{D}=\lrm{(n,m)\:|\:T_{m\ra n}^\mr{cg}\neq 0}$.

Then, we find (see Appendix~\ref{lindapp})
\aln{\label{oq2}
\lrv{\sum_\eta\sum_na_nK_n^\eta}&\leq \max_{(n,m)\in\mc{E}_\mr{D}}|a_n-a_m|\sqrt{\frac{\mc{A}_\mr{cg}^d\dot{\Sigma}_\mr{cg}^d}{2}}, 
}
where we have defined
\aln{
\mc{A}_\mr{cg}^d=\sum_{n\neq m}T_{m\ra n}^\mr{cg},
}
and
\aln{
\dot{\Sigma}_\mr{cg}^d =\sum_{n,m}T_{m\ra n}^\mr{cg}\ln\frac{T_{m\ra n}^\mr{cg}}{T_{n\ra m}^\mr{cg}}\geq 0.
}
We note that the non-coarse-grained version of $\dot{\Sigma}_\mr{cg}^d$,
\aln{
\dot{\Sigma}^d =\sum_{x,y}\sum_\eta |(\hat{L}_\eta)_{xy}|^2p_y\ln\frac{\sum_\eta|(\hat{L}_\eta)_{xy}|^2p_y}{\sum_\eta|(\hat{L}_\eta)_{yx}|^2p_x},
}
reduces to Eq.~\eqref{EP} when
$
\hat{L}_\eta=\hat{L}_{(xy)}=\delta_{\eta,(xy)}\sqrt{\mc{W}_{xy}}\ket{x}\bra{y}.
$
In addition, we can regard $\dot{\Sigma}^d$ as the physical entropy production of open quantum systems by assuming $\{\hat{L}_\eta\}$ as the Lindblad operators driven by baths that satisfy the local detailed balance condition~\cite{funo2019speed}.

To summarize, we have a bound~\eqref{oq1} with inequality~\eqref{oq2}, which represents the bound based on 
 the gradient of $\hat{A}$.
Thus, our speed limit is again valuable for macroscopic transitions for which the coherent dynamics and incoherent dissipation process coexist.
One of the examples of such processes can be the transport process in an  extended quantum system with the stochastic hopping~\cite{eisler2011crossover,temme2012stochastic,PhysRevLett.118.070402,PhysRevLett.127.070402}, which is proposed to be realized in a cold atomic system with engineered dissipation~\cite{diehl2008quantum,PhysRevLett.127.070402}.

\section{Discussions\label{SecVIII}}
Before concluding this paper, we briefly discuss several implications and possible generalizations of our approach that we did not mention in the previous sections.
While we leave detailed investigation as a future problem, we here propose basic ideas, especially for the quantum setting.

\subsection{Absence of the  tradeoff relation between time and energy fluctuation in our formalism}

\begin{figure}
    \centering
    \includegraphics[width=\linewidth]{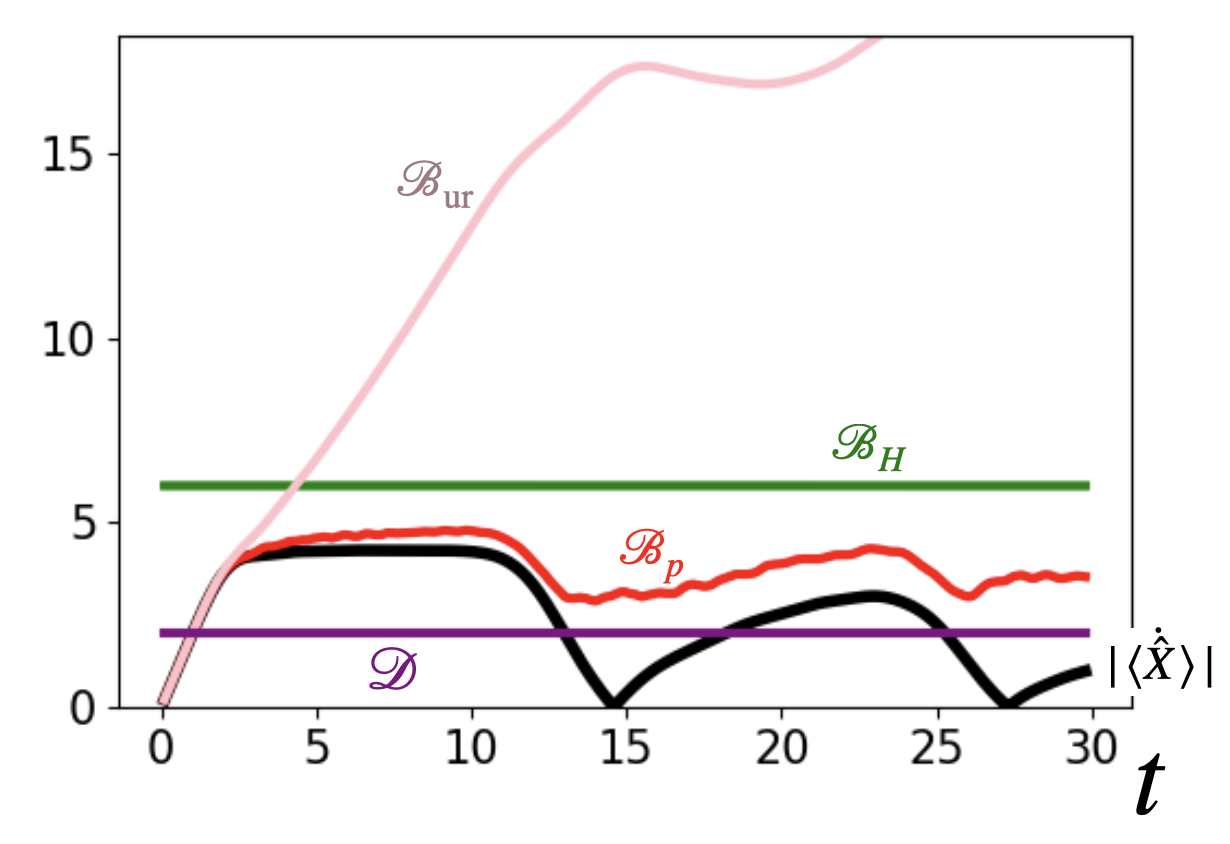}
    \caption{Time dependence of the speed of the sum of the positions of the particles, $|\braket{\dot{\hat{X}}}|$ (black),
from the initial state $\ket{\Psi_1}$ (the same situation discussed in Fig.~\ref{fig24_3}(a)). 
In contrast with the bounds $\mc{B}_p$ in Eq.~\eqref{glRpos} (red) and  $\mc{B}_H$ in Eq.~\eqref{inftycon} (green), the quantity $\mc{D}$ (purple) in Eq.~\eqref{Dc} with $c=2$ cannot bound $|\braket{\dot{\hat{X}}}|$.
This indicates that the tradeoff relation between time and energy fluctuation does not hold in our formalism.
We note that the bound $\mc{B}_\mr{ur}$ in \eqref{ur} (gray), which is based on the standard tradeoff relation between time and energy fluctuation, provides a rather loose upper bound for $|\braket{\dot{\hat{X}}}|$.
}
    \label{figtradetime}
\end{figure}

We have discussed that the speed limit for macroscopic quantum systems 
are given in inequalities \eqref{glRpos} and \eqref{inftycon}.
This inequality indicates the tradeoff relation between time and quantum phase difference, in contrast with the tradeoff relation between time and energy fluctuation in, e.g., inequality \eqref{ur}.
On the other hand, one may wonder if the energy fluctuation also plays the important role in our formalism.
In particular, we ask whether the speed $|\braket{\dot{\hat{A}}}|$ is bounded by following quantity
\aln{\label{Dc}
\mc{D}=c\cdot\|\nabla A\|_\infty \cdot\Delta H
}
with some universal constant $c$.

We argue that such a quantity cannot be the bound for $|\braket{\dot{\hat{A}}}|$, which means that the tradeoff relation between time and energy fluctuation does not hold in our formalism.
To see a simple example, we consider 
 nonequilibrium dynamics of the many hardcore-boson system discussed in Sec.~\ref{SecIV}.
In particular, we consider time dependence of the speed of the sum of the particles positions, $|\braket{\dot{\hat{X}}}|$,
from the initial state $\ket{\Psi_1}$.
As shown in Fig.~\ref{figtradetime},
in contrast with our bounds $\mc{B}_p$ and $\mc{B}_H$ as well as the rather loose bound $\mc{B}_\mr{ur}$,
the quantity $\mc{D}$ with $c=2$ (which is obtained by replacing $\Delta A$ in \eqref{ur} with $\|\nabla A\|_\infty$) cannot be an upper bound for $|\braket{\dot{\hat{X}}}|$.
In general, if we start from $\ket{\Psi_1}$ for $M$-particle dynamics, 
$\Delta H$ does not depend on $M$ whereas $|\braket{\dot{\hat{X}}}|$ can increase with $M$.
Since $\|\nabla X\|_\infty=1$ in this case, there is no universal $M$-dependent constant $c$ such that $|\braket{\dot{\hat{X}}}|\leq \mc{D}$ always holds.

\begin{figure}
    \centering
    \includegraphics[width=\linewidth]{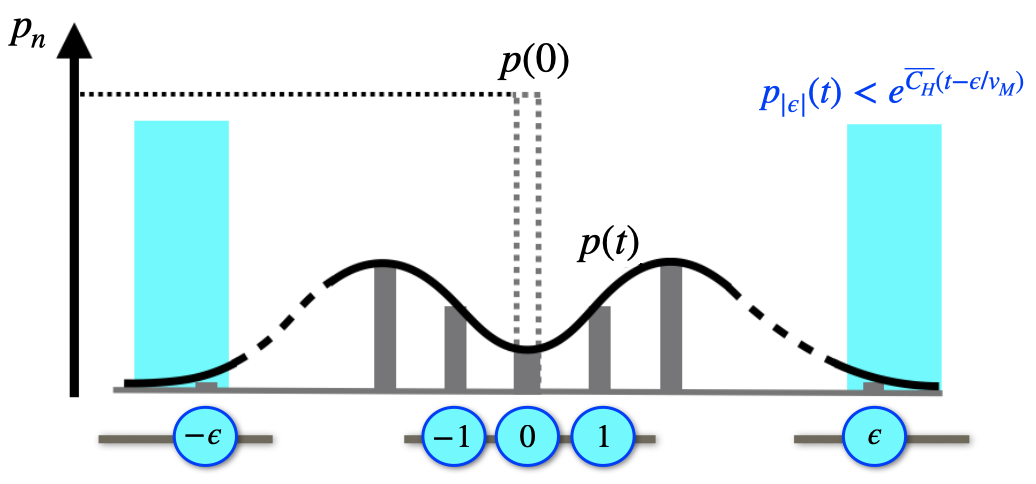}
    \caption{Schematic illustration of the exponential suppression of the tail (shaded in cyan) of the probability distribution $p(t)$ given in Eq.~\eqref{distbound}, where $p_n(0)=\delta_{0n}$ is assumed.}
    \label{figdist}
\end{figure}

\subsection{Concentration and distribution}
Although we have focused on the speed of specific quantities such as the expectation value, we can also discuss the bound on the entire probability distribution $p(t)=\{p_n(t)\}$.
To see this, we first notice Chebyshev's inequality,
$
\mbb{P}_t[|A-\braket{A}|>\epsilon]\leq \frac{\mbb{V}[A]}{\epsilon^2}
$
for $\epsilon>0$,
where $\mbb{P}_t$ is the probability with respect to $p(t)$.
To simplify the discussion, let us consider unitary quantum dynamics.
{
Then, $\mbb{V}$ is bounded as (see Eqs.~\eqref{varbound} and \eqref{varfint})
$\mbb{V}[\hat{A}(t)]\leq \int_0^t\mc{B}_\mr{var}(t')dt'+\mbb{V}[\hat{A}(0)]$ and $\mbb{V}[\hat{A}(t)]\leq \|\nabla A\|_\infty^2(\Delta A(0)+\av{\sqrt{C_H^2-E_\mr{trans}^2}}t)^2$.
Employing the latter inequality, for example, we obtain
\aln{
\mbb{P}_t\lrl{\lrv{\frac{A-\braket{\hat{A}}}{\|\nabla A\|_\infty}}>\epsilon }\leq \frac{\lrs{\Delta A(0)+\av{\sqrt{C_H^2-E_\mr{trans}^2}}t}^2}{\epsilon^2}.
}
To see an example of the physical implication of this inequality, let us consider a situation for which the underlying graph is one dimensional as in Fig.~\ref{fig_graph}(c). We also assume that $\Delta A(0)=0$ for simplicity.
Then, taking $\hat{A}=\hat{n}$ leads to
\aln{
\sum_{|n-\braket{\hat{n}}|>\epsilon}p_n(t)\leq \frac{(\av{\sqrt{C_H^2-E_\mr{trans}^2}}t)^2}{\epsilon^2}.
}
Thus, the probability of the tail for $|n-\braket{n}| \gg \av{\sqrt{C_H^2-E_\mr{trans}^2}}t$ is polynomially suppressed.
}

Furthermore, we find the following nontrivial exponential form of the  inequality
\aln{\label{concent}
&\mbb{P}_t\lrl{\lrv{\frac{A-A'}{\|\nabla A\|_\infty}}>\epsilon }\leq
2\braketL{\cosh\frac{\lambda(\hat{A}-A')}{\|\nabla A\|_\infty}}_0e^{\av{C_H}t-\lambda \epsilon}
}
for $0\leq \lambda\leq \lambda_M:= 2\log\frac{\sqrt{5}+1}{2}\simeq 0.96$ and an arbitrary real value $A'$ (see Appendix~\ref{Appconc}), 
where $\braket{\cdots}_0$ denotes the average with respect to $\hat{\rho}(0)$.
Let us again consider the one-dimensional graph (where we assume $n=\cdots, -1,0,1,\cdots$ instead of $n=1,2,\cdots$) and take $\hat{A}=\hat{n}$.
For simplicity, we assume that $p_n(0)=\delta_{n0}$ and take $A'=0$.
Then, we obtain
\aln{\label{distbound}
\sum_{|n|>\epsilon}p_n\leq 2e^{\av{C_H}t-\lambda \epsilon}.
}
This indicates that  $p_n$ is exponentially suppressed for $t\lesssim t_{n}:=n/{v_M}$ with $v_M=\av{C_H}/\lambda_M$ (see Fig.~\ref{figdist})~\footnote{As an example, if we consider the system with $M$ particles in Sec.~\ref{SecIV} and take $n$ as the sum of the particle positions, $t_{n}$ for $n\sim l_\mr{ave} M$ (i.e., $l_\mr{ave}$ denotes the average position of particles) is $t_n\sim l_\mr{ave}/(2K\lambda_M)$.}.
While this is consistent with the Lieb-Robinson bound stating that
the tail of information propagation (concerning the operator norm of the commutativity of two distant operators) is exponentially suppressed, our setting has advantage in that $t_{n}=n/v_M$ can  be larger in general than that predicted by the Lieb-Robinson bound.
Note that our results hold for time-dependent processes.
Even in such cases, the speed is evaluated only from the time-averaged value of the Hamiltonian structures $C_H$.

\subsection{Other observables}
In the previous sections concerning quantum dynamics, we have considered observables that are written as $\hat{A}=\sum_na_n\hat{P}_n$.
Here we discuss the possibility of extending our results to other observables.

One direction is to diagonalize a general observable $\hat{B}$ as $\hat{B}=\sum_\nu b_\nu\hat{P}_\nu$ and consider a graph whose vertices are labeled by $\{\nu\}$.
In this case, the speed is given by $|\braket{\dot{\hat{B}}}|=\lrv{\sum_{\nu\mu}(b_\nu-b_\mu)J_{\nu\mu}^q}$, where $J_{\nu\mu}^q=\Tr[\hat{H}_{\nu\mu}\hat{\rho}_{\mu\nu}-\hat{\rho}_{\nu\mu}\hat{H}_{\mu\nu}]$. 
Thus, when $|J_{\nu\mu}^q|$ is suppressed for $(\nu,\mu)$ with large $|b_\nu-b_\mu|$, the speed limit can decrease.
We leave the detailed analysis of this condition as a future problem.

Another direction is to obtain the speed limit of certain observables 
from the knowledge of the distribution $p(t)$, which was discussed in the previous subsection.
For example, let us assume the one-dimensional graph  with the label $n=\cdots, -1,0,1,\cdots$ and $p_n(0)=\delta_{n0}$ (see the previous subsection).
We consider a general observable  $\hat{B}$ satisfying 
$\hat{P}_n\hat{B}\hat{P}_{n'}=0$ when $n,n'\leq n_\mr{th}=\mr{O}(N)$.
In this case, using inequality~\eqref{distbound}, we can prove that the speed of $\braket{\hat{B}}$ is small for $t\lesssim n_\mr{th}/v_M$ as
\aln{
|\braket{\dot{\hat{B}}}|\leq c\|\hat{B}\|_\mr{op}\|\hat{H}\|_\mr{op}e^{\av{C_H}(t-n_\mr{th}/v_M)},
}
where $c$ is a constant.
Note that, even when $\|\hat{B}\|_\mr{op}$ and $\|\hat{H}\|_\mr{op}$ are $\mr{O}(N)$, the right-hand side is exponentially suppressed with $N$ for $t\lesssim n_\mr{th}/v_M$.

\subsection{Tightening quantum speed limit for mixed states}
In the previous sections concerning quantum systems, we have mainly focused on pure quantum states.
On the other hand, the loss of quantum coherence can tighten the speed limit for a purity-decreasing process.

As a first example, let us consider the bound on the transition time for finite times in inequality~\eqref{hota}.
Assuming time-independent unitary time dynamics, we obtain the bound for the averaged speed (using the discussion similar to Appendix~\ref{Appderexp})
\aln{
\frac{|A_\mr{fin}-A_\mr{ini}|}{T}\leq {\|\nabla A\|_\infty}\sqrt{R_\mr{av}^2-\av{E_\mr{trans}}^2},
}
where $R_\mr{av}=\sum_{n\sim_Am}\Tr[\av{\hat{\rho}}_{nm}\hat{H}_{mn}]$.
The right-hand side can be suppressed for finite $T$ compared with the case for $T\simeq 0$ when starting from a pure state, since $R_\mr{av}\leq \sum_{n\sim_Am}\|\av{\hat{\rho}}_{nm}\|_1\|\hat{H}_{nm}\|_\mr{op}$
and $\|\av{\hat{\rho}}_{nm}\|_1$ can be smaller than $\sqrt{p_np_m}$ for mixed states.

Another example is the case for the dissipative quantum dynamics discussed in Section~\ref{SecVII}.
In this case, we have, e.g., 
\aln{\label{bhcoh}
\mc{B}_\mr{H}\leq 
\sqrt{2\frac{[A^\mathsf{T}\nabla^2_{r^\mr{coh}}A]}{{R_\mr{coh}}}}\sqrt{R_\mr{coh}^2-{E_\mr{trans}^2}},
}
where $r_{nm}^\mr{coh}=\|\hat{\rho}_{nm}\|_1\|\hat{H}_{nm}\|_\mr{op}$ (see Eq.~\eqref{rcoh}) and $R_\mr{coh}=\sum_{n\sim_Am}r_{nm}^\mr{coh}$.
Since $r_{nm}^\mr{coh}\leq r_{nm}^p$ for mixed states
and a dissipative time evolution leads to the loss of purity, 
the right-hand side of inequality~\eqref{bhcoh} is in general smaller than $\mc{B}_p$ for such a process.

\section{Conclusion and outlook\label{SecIX}}
This paper has presented a fundamental framework for deriving speed limits in processes with macroscopic transitions.
Our strategy is to employ the local conservation law of probability
on a general graph, which describes general systems including many-body ones.
We prove the bound with the local probability current and the gradient of the quantity of interest,
which is concisely expressed by the notions of the graph theory.
Our rigorous bounds are qualitatively tighter for extended and many-body systems than  conventional speed limits based on, e.g., the uncertainty relation.

Our framework applies to various dynamics since it relies only on the local conservation law of probability, leading to many novel and practical bounds.
Applying it to quantum systems, we obtain previously unknown bounds that become smaller when the expectation value of the transition Hamiltonian increases for unitary quantum dynamics.
This is intuitively understood as the novel tradeoff relation between time and quantum phase difference instead of that between time and energy fluctuation in the Mandelstam-Tamm bound.
Our results provide first general state-dependent speed limits useful for macroscopic systems;
it can achieve the equality condition for some situations, in stark contrast with the state-independent Lieb-Robinson bound.
In addition, we demonstrate that our speed limits can apply to quantities that are not written as the expectation value of an observable, such as macroscopic quantum coherence.
We have also discussed that speed limits are described by the entropy production rate for macroscopic classical stochastic processes.
In particular, we derive a valuable speed limit for long times even  without the detailed balance condition.
Furthermore, our method enables us to derive a speed limit for macroscopic dissipative quantum systems. 

There are many open questions for future studies.
First, while our starting point is to use the local conservation law of probability, which holds for generic systems, it is interesting to discuss how additional conservation laws may tighten the speed limit.
For example, a time-dependent Hamiltonian in an isolated system leads to the local conservation law of energy density, from which a different type of speed limits can arise.
Moreover, more nontrivial conservation laws (such as the dipolar moment~\cite{PhysRevX.9.021003,PhysRevX.10.011047} or the ones associated with integrability~\cite{PhysRevX.6.041065,PhysRevLett.119.220604}) have recently been known to slow down the dynamics, and our approach may uncover corresponding speed limits.
We also note that our method can be helpful for a more coarse-grained level by starting from the continuity equation of the (possibly discrete) quantum/classical hydrodynamics~\cite{spohn2012large,tsubota2013quantum}.

Second, it is important to thoroughly investigate the relationship with quantum thermalization in isolated many-body systems.
Timescale for thermalization for quantum chaotic systems is an active field of research~\cite{PhysRevLett.111.140401,goldstein2015extremely,reimann2016typical,de2018equilibration,PhysRevB.99.174313,PhysRevX.7.031027}, but it is not simple to obtain rigorous results on a relevant timescale for macroscopic transitions.
While our approach succeeds in obtaining rigorous bounds for such a macroscopic process,
we may obtain even tighter bounds by taking in the complexity of 
many-body dynamics.
For this purpose, we may need to combine our method and the property of quantum chaotic systems, such as the eigenstate thermalization hypothesis~\cite{PhysRevLett.54.1879,PhysRevA.43.2046,PhysRevE.50.888,rigol2008thermalization} and the local random-matrix theory~\cite{PhysRevX.7.031016,PhysRevX.8.021014,PhysRevX.8.021013,PhysRevX.8.031057,PhysRevX.8.031058,PhysRevLett.120.080603,PhysRevLett.126.120602}.

Third, it will be intriguing to compare our results with the knowledge in the field of statistics and mathematics~\cite{villani2009optimal,boucheron2013concentration,van2014probability}.
While our general approach evades the explicit  introduction of the Wasserstein distance, which may complicate the discussion and concrete calculation, the obtained bounds can be tied to this distance as indicated by the case for one dimension (see inequality~\eqref{Was1} and Appendix~\ref{wasapp}). 
It is a future outlook to relate our results with recent development of the notion of quantum Wasserstein distance~\cite{carlen2014analog,datta2020relating,deffner2017geometric,chow2019discrete,PhysRevLett.126.010601,de2021quantum,PhysRevA.101.042107} and the optimal-transport-based inequalities for stochastic systems~\cite{dechant2019thermodynamic,PhysRevLett.126.010601,nakazato2021geometrical,dechant2021geometric,dechant2021minimum}.
Another mathematical perspective is the connection with the spectral graph theory~\cite{chung1997spectral}.
While our speed limits are described in the terminology of graph theory, the graph's structure can also be uncovered from the spectra of the graph Laplacians.
It is thus interesting to discuss our speed limit in light of the spectral property of the graph.

Finally, it is a significant challenge to investigate the implications of our results to the engineering of quantum systems.
For example, it is interesting to discuss how our speed limits are related to the quantum sensing~\cite{toth2014quantum} and shortcuts to adiabaticity~\cite{RevModPhys.91.045001} for many-body systems.
In addition, the problem of optimal quantum-state transfer~\cite{PhysRevLett.91.207901,kay2010perfect,PhysRevA.74.030303,PhysRevLett.103.240501,PhysRevA.82.022318,PhysRevA.85.052327} is closely related to our setup, since one needs to transfer a quantum state from one site to another distant site.
Our speed limits may give a new useful bound on the transfer speed, which we leave for future research.

\section{Acknowledgements}
We thank Ryosuke Iritani and Masaru Hongo for helpful discussions.
We are also grateful to Kyosuke Adachi and Mamiko Tatsuta for fruitful comments and discussions on the manuscript. We thank Takeru Matsuda  for notifying us of the graph Laplacian. The numerical calculations were carried out with the help of QUSPIN~\cite{SciPostPhys.2.1.003,weinberg2019quspin}.

\appendix

\section{Analogy to the optimal transport problem}\label{wasapp}
We here discuss the relation between the  speed limit for continuous systems in inequality~\eqref{Was1} and the optimal transport problem~\cite{villani2009optimal,boucheron2013concentration}.
In the optimal transport problem, the Wasserstein distance plays a crucial role in characterizing the distance between two probability distributions $P(\mbf{x})$ and $Q(\mbf{x})$.
The Wasserstein distance takes into account the underlying geometric structure for variables $\{\mbf{x}\}$ by introducing some cost function (i.e., distance) $\mathsf{d}(\mbf{x},\mbf{y})$.
In particular, the (order 1) Wasserstein distance is defined by
\aln{
W_1(P,Q):=\inf_{\Pi\in\mc{C}(P,Q)}\int d\mbf{x}d\mbf{y}\mathsf{d}(\mbf{x},\mbf{y})\Pi(\mbf{x},\mbf{y}),
}
where $\Pi$ is a joint probability distribution in a set defined by
\aln{
\mc{C}(P,Q)=\lrm{\Pi\:|\:P(\mbf{x})=\int d\mbf{y}\Pi(\mbf{x},\mbf{y}), \:Q(\mbf{y})=\int d\mbf{x}\Pi(\mbf{x},\mbf{y})}.
}
Intuitively, the Wasserstein distance measures the optimal expectation value of the cost $\mathsf{d}$ for transportation from $P$ to $Q$.
This distance is also written as~\cite{van2014probability}
\aln{
W_1(P,Q)=\sup_{f\in \mc{L}}\lrv{\int d\mbf{x}f(\mbf{x})(P(\mbf{x})-Q(\mbf{x}))},
}
where $\mc{L}$ is a set of Lipschitz functions whose Lipshitz constant is one, i.e.,
\aln{
\mc{L}=\lrm{f\:|\:\forall x\forall y,\:|f(\mbf{x})-f(\mbf{y})|\leq \mathsf{d}(\mbf{x},\mbf{y})}.
}

In one dimension and for $\mathsf{d}(x,y)=|x-y|$, one can show that the Wasserstein distance between two probability distributions $P$ and $Q$ is given by~\cite{van2014probability}
\aln{
W_1(P,Q)=\int dx\lrv{\int_{-\infty}^xdy(P({y})-Q({y}))}.
}
Substituting $P=P(0)$ and $Q=P(T)$, we have
\aln{
W_1(P,Q)=T\int d{x}\lrv{\av{{J}}(x)},
}
where we have assumed $\av{{J}}(-\infty)=0$.
Thus, the inequality~\eqref{Was1} can be rewritten as
\aln{
|A_\mr{fin}-A_\mr{ini}|\leq \max |\partial_xA({x})|\cdot W_1(P,Q).
}

\section{Brief review of the graph Laplacian}\label{lapapp}
In this Appendix, we review the basics of the graph theory and the graph Laplacian.
A graph $G$ is an object that consists of  vertices $\mc{V}=\lrm{v_1,\cdots,v_{|\mc{V}|}}$ and edges
$\mc{E}=\lrm{e_1,\cdots,e_{|\mc{E}|}}$.
We write each of the edges as
$e_a=(n,m)$ if the ends of the edge $e_a$ is $v_n$ and $v_m$.
When the order between $n$ and $m$ is ignored, the graph is called the undirected graph; otherwise, it is called the directed graph.
In the following, we only discuss the undirected graph for $G$.
In addition, when each of the edges is accompanied by some nontrivial value $r_{nm}$, the graph is called the weighted graph; otherwise, it is called the unweighted graph.

\subsection{Unweighted graph}
Let us first formulate the graph Laplacian for unweighted graphs.
In this case, the $|\mc{V}|\times|\mc{V}|$ matrix that characterizes the connections of each vertex by the edges, called the adjacent matrix $r^\mr{uw}$, is given by
\aln{
r^\mr{uw}_{nm}=
\left\{
\begin{array}{ll}
1 & (n\sim m), \\
0 & (\mathrm{otherwise}).
\end{array}
\right.
}
As in the main text, we use the symbol $n\sim m$ when $(n,m)\in\mc{E}$.

Next, we introduce the incident matrix $\nabla$, which is a 
$|\mc{E}|\times|\mc{V}|$ matrix whose elements are given by
\aln{
(\nabla)_{(n,m),l}=
\left\{
\begin{array}{lll}
1 & (n = l), \\
-1 & (m = l), \\
0 & (\mathrm{otherwise}).
\end{array}
\right.
}
For a vector $A=(a_1,\cdots,a_{|\mc{V}|})^\mathsf{T}$, we have
\aln{
(\nabla A)_{(n,m)}=a_n-a_m.
}
This expression indicates that $\nabla A$ is the discrete version of the gradient for the continuous function.
Also, we note that this representation naturally leads to the definition in the main text,
\aln{
\|\nabla A\|_\infty=\max_{n\sim m}|a_n-a_m|.
}

The graph Laplacian for the unweighted graph is defined by
\aln{
\nabla^2:=\nabla^\mathsf{T}\nabla,
}
whose elements are given by
\aln{
(\nabla^2)_{nm}=-r^\mr{uw}_{nm}+\delta_{nm}\sum_{m'(\sim n)}r^\mr{uw}_{nm'}.
}
We note that
\aln{
(\nabla^2 A)_n=\sum_{m(\sim n)}(a_n-a_m)
}
and
\aln{
[A^\mathsf{T}\nabla^2 A]:=(A,\nabla^2 A)=\frac{1}{2}\sum_{n\sim m}(a_n-a_m)^2.
}

\subsection{Weighted graph}
Next, we consider the case for the weighted graph, whose adjacent matrix is written by
\aln{
r_{nm}=
\left\{
\begin{array}{ll}
r_{nm} & (n\sim m), \\
0 & (\mathrm{otherwise}),
\end{array}
\right.
}
where $r_{nm}=r_{mn}$ for undirected graphs.
We define the graph Laplacian for this case as
\aln{
\nabla^2_r:=\nabla^\mathsf{T}\mathsf{D}_r\nabla,
}
where
$\mathsf{D}_r=\mr{diag}(r_{e_1},\cdots,r_{e_{|\mc{E}|}})$ is an $|\mc{E}|\times |\mc{E}|$ matrix.
The elements of the graph Laplacian are given by
\aln{
(\nabla^2_r)_{nm}=-r_{nm}+\delta_{nm}\sum_{m'(\sim n)}r_{nm'}.
}
We note that
\aln{
(\nabla^2_r A)_n=\sum_{m(\sim n)}r_{nm}(a_n-a_m)
}
and
\aln{
[A^\mathsf{T}\nabla^2_r A]:=(A,\nabla_r^2 A)=\frac{1}{2}\sum_{n\sim m}r_{nm}(a_n-a_m)^2,
}
as discussed in the main text.

\section{Derivation of the quantum speed limits for expectation values}\label{Appderexp}
We here show the detailed derivation of the speed limits for expectation values, especially the inequality~\eqref{hier2}.
Since $|\braket{\dot{\hat{A}}}|\leq \mc{B}_\mr{gL}$ and $|\braket{\dot{\hat{A}}}|\leq \mc{B}_\mr{Lip}$ are obtained using  H\"{o}lder's inequality as discussed in Sec.~\ref{SecII},
we focus on 
\aln{\label{Appieq1}
\mc{B}_\mr{gL}\leq \mc{B}_p,
}
\aln{\label{Appieq123}
\mc{B}_p\leq \mc{B}_{\mr{Lip},p},
}
and
\aln{\label{Appieq3}
\mc{B}_{\mr{Lip},p}\leq \mc{B}_H,
}
where we have introduced another quantity
\aln{
\mc{B}_{\mr{Lip},p}=\|\nabla A\|_\infty\sqrt{R_p^2-E_\mr{trans}^2}.
}
For completeness, we also show that there is another hierarchy of the bound, i.e., 
\aln{\label{Appieq2}
\mc{B}_\mr{Lip}\leq \mc{B}_{\mr{Lip},p}.
}

\subsection{Proof of inequalities~\eqref{Appieq1} and \eqref{Appieq123}}
To show inequality~\eqref{Appieq1}, we first take
\aln{\label{rnm}
r_{nm}=|\Tr[\hat{H}_{nm}\hat{\rho}_{mn}]|
}
for $n\sim m$ (we assume $r_{nn}=0$) and show
\aln{\label{transderu}
{\sum_{n\sim_A m}\frac{|J_{nm}^q|^2}{r_{nm}}}
\leq 4R-\frac{4|E_\mr{trans}|^2}{R},
}
where
\aln{\label{Rdef}
R=\Tr[\nabla^2_{r}]=\sum_{n\sim_Am}r_{nm}.
}
To see this, we note that the following parametrization of the local current is possible by setting $\Tr[\hat{H}_{nm}\hat{\rho}_{mn}]=r_{nm}e^{i\theta_{nm}}$:
\aln{
J_{nm}^q=-i\Tr[\hat{H}_{nm}\hat{\rho}_{mn}-\mr{h.c.}]=2r_{nm}\sin\theta_{nm}.
}
We also introduce
\aln{
Y_{nm}:=\Tr[\hat{H}_{nm}\hat{\rho}_{mn}+\mr{h.c.}]=2r_{nm}\cos\theta_{nm},
}
which is proportional to the local energy concerning the transition Hamiltonian.
We note that
\aln{
\sum_{n\sim_Am}Y_{nm}=2\braketL{\sum_{n\sim_Am}\hat{H}_{nm
}}=2E_\mr{trans}.
}

Here, we employ the following inequality
\aln{\label{sincos1}
\sin^2\theta \leq 1+\frac{d^2}{4}-d|\cos\theta|
}
for arbitrary $d\in\mbb{R}$ and $\theta\in\mbb{R}$.
Then, we have
\aln{
\frac{|J_{mn}^q|^2}{r_{nm}}&\leq 4r_{nm}\lrs{1+\frac{d^2}{4}-d|\cos\theta_{nm}|}\nonumber\\
&\leq 4r_{nm}\lrs{1+\frac{d^2}{4}}-2d|Y_{nm}|
}
and thus
\aln{
&\lrv{\sum_{n\sim_Am}\frac{|J_{mn}^q|^2}{r}}\nonumber\\
&\leq 4\lrs{1+\frac{d^2}{4}}\sum_{n\sim_Am}r-2d\lrv{\sum_{n\sim_Am}Y_{mn}}\nonumber\\
&=4\lrs{1+\frac{d^2}{4}}R-4d|E_\mr{trans}|
}
Since $d$ is arbitrary, we can minimize the right hand side with the optimal $d$ ($d=2|E_\mr{trans}|/R$) and get
\aln{
\lrv{\sum_{n\sim_Am}\frac{J_{mn}^q(t)^2}{r_{nm}}}
\leq 4R-\frac{4|E_\mr{trans}|^2}{R},
}
which is the inequality \eqref{transderu}.
To conclude, we have
\aln{
\mc{B}_\mr{gL}\leq \sqrt{\frac{[A^\mathsf{T}\nabla_r^2A]}{2}}\sqrt{4R-\frac{4|E_\mr{trans}|^2}{R}}.
}

Now, we note that, for $r'_{nm}$ satisfying $r_{nm}\leq r'_{nm}$ for all $n$ and $m$,
we have
\aln{
[A^\mathsf{T}\nabla_r^2A]\leq 
[A^\mathsf{T}\nabla_{r'}^2A]
}
and
\aln{
4R-\frac{4|E_\mr{trans}|^2}{R}
\leq 4R'-\frac{4|E_\mr{trans}|^2}{R'},
}
where $R'=\Tr[\nabla_{r'}^2]=\sum_{n\sim_Am}r'_{nm}\:(\geq R)$. 
Thus, if we can show
\aln{\label{pnisuru}
r_{nm}\leq r^p_{nm}=\|\hat{H}_{nm}\|_\mr{op}\sqrt{p_np_m},
}
we can show inequality~\eqref{Appieq1}.

Inequality \eqref{pnisuru} is proven as follows.
First, using the (matrix version of) H\"{o}lder's inequality, we have
\aln{\label{rcoh}
r_{nm}\leq r_{nm}^\mr{coh}:= \|\hat{H}_{nm}\|_\mr{op}\|\hat{\rho}_{mn}\|_1,
}
where $\|\hat{X}\|_1:=\Tr\lrl{\sqrt{\hat{X}^\dag\hat{X}}}$.
To evaluate $\|\hat{\rho}_{mn}\|_1$, we diagonalize the quantum state as $\hat{\rho}=\sum_k\lambda_k\ket{\lambda_k}\bra{\lambda_k}$.
Then
\aln{
\|\hat{\rho}_{mn}\|_1&\leq
\sum_k\lambda_k\|\hat{P}_m\ket{\lambda_k}\bra{\lambda_k}\hat{P}_n\|_1\nonumber\\
&=
\sum_k\lambda_k\sqrt{\braket{\lambda_k|\hat{P}_m|\lambda_k}\braket{\lambda_k|\hat{P}_n|\lambda_k}}\nonumber\\
&\leq\sqrt{
\sum_k\lambda_k\braket{\lambda_k|\hat{P}_m|\lambda_k}\sum_k\lambda_k\braket{\lambda_k|\hat{P}_n|\lambda_k}}\nonumber\\ &=\sqrt{p_mp_n}.
}
Thus, combining all of the above, we obtain inequality~\eqref{pnisuru} and thus \eqref{Appieq1}.

Now, we can prove inequality~\eqref{Appieq123} by noticing
\aln{
\frac{2[A^\mathsf{T}\nabla_{r^p}^2A]}{R_p}
=\frac{\sum_{n\sim_A m}(a_n-a_m)^2r_{nm}^p}{\sum_{n\sim_A m}r_{nm}^p}\leq \|\nabla A\|_\infty^2.
}

\subsection{Proof of inequality~\eqref{Appieq3}}
We can bound $R_p$ with state-independent quantity as
\aln{
R_p&\leq \sum_{n\sim_Am}\|\hat{H}_{nm}\|_\mr{op}\frac{p_n+p_m}{2}\nonumber\\
&=\sum_{n\in\mc{V}}p_n\sum_{m(\sim_An)}\|\hat{H}_{nm}\|_\mr{op}\nonumber\\
&=\max_{n\in\mc{V}}\sum_{m(\sim_An)}\|\hat{H}_{nm}\|_\mr{op}:= C_H.
}
Thus, inequality~\eqref{Appieq3} follows.

\subsection{Proof of \eqref{Appieq2}}
Inequality \eqref{Appieq2} is obtained in a manner similar to \eqref{Appieq1}.
We begin to show 
\aln{\label{JRbound}
\sum_{n\sim_Am}|J^q_{nm}|\leq 2\sqrt{R^2-E_\mr{trans}^2},
}
where $r_{nm}$ and $R$ are defined in Eqs.~\eqref{rnm} and \eqref{Rdef}.
For this purpose, we employ the same parametrization of $J_{nm}^q$ and $Y_{nm}$ using  $\theta_{nm}$.

Now, Instead of inequality~\eqref{sincos1}, we use the following inequality
\aln{
|\sin\theta|\leq \sqrt{1+b^2}-b|\cos\theta|
}
for $b,\theta\in\mbb{R}$.
Then, we have
\aln{
\sum_{n\sim_Am}|J^q_{nm}|&\leq
\sum_{n\sim_Am}2r_{nm}\sqrt{1+b^2}-b|Y_{nm}|\nonumber\\
&\leq 2\sqrt{1+b^2}R-2b|E_\mr{trans}|.
}
The optimal upper bound is obtained for 
$b=\frac{|E_\mr{trans}|/R}{\sqrt{1-(|E_\mr{trans}|/R)^2}}$, from which we have
the inequality~\eqref{JRbound}.
Thus, we have
\aln{
\mc{B}_\mr{Lip}\leq \|\nabla A\|_\infty\sqrt{R^2-E_\mr{trans}^2}.
}
Finally,
as in the previous subsection, we use $R\leq R_p$ and obtain
\aln{
\mc{B}_\mr{Lip}\leq \|\nabla A\|_\infty\sqrt{R_p^2-E_\mr{trans}^2},
}
which is inequality~\eqref{Appieq2}.

\section{Derivation of the speed limit for acceleration}\label{accapp}

While many bounds in the main text are derived using inequalities~\eqref{entF} and~\eqref{entF2}, we can find a  speed limit for quantities that are not written as $F(p)$.
Specifically, we can show that  the acceleration of the expectation value, $\braket{\ddot{\hat{A}}}$,
can be bounded above provided that $\hat{H}$ is independent of time.
This leads to speed limits of  $\braket{\hat{A}}$ 
that is distinct from $\mc{B}_\mr{gL}$ and $\mc{B}_\mr{Lip}$ discussed in the main text.

Indeed, we find, e.g.,
\aln{
|\braket{\ddot{\hat{A}}}|
&\leq \max_{m,n(\sim_Am),l(\sim_Am)}|a_n+a_l-2a_m|\cdot{Q_p}\nonumber\\
&\leq \max_{m,n(\sim_Am),l(\sim_Am)}|a_n+a_l-2a_m|\cdot {C_H^2},
}
where
\aln{
Q_p=\sum_{m,n(\sim_Am),l(\sim_Am)}
\sqrt{p_np_l}\|\hat{H}_{nm}\|_\mr{op}\|\hat{H}_{ml}\|_\mr{op}.
}
From this acceleration bound, 
we find a new type of the bound of velocity.
If we assume that $\braket{\dot{A}(0)}=0$, we find
\aln{
|\braket{\dot{A}(t)}|\leq \mc{B}_\mr{acc}=\max_{m,n(\sim_Am),l(\sim_Am)}|a_n+a_l-2a_m|\int_0^td\tau {Q_p},
}
which is bounded by the state-independent quantity $\max_{m,n(\sim_Am),l(\sim_Am)}|a_n+a_l-2a_m|C_H^2t$.
Note that $\mc{B}_\mr{acc}$ can be better for short times than $\mc{B}_\mr{gL}$ and $\mc{B}_\mr{Lip}$.
Indeed, we can see that $\mc{B}_\mr{acc}\ra 0$ for $t\ra 0$, while $\mc{B}_\mr{gL}$ and $\mc{B}_\mr{Lip}$ may not be necessarily zero even for  $t\ra 0$.

To prove the above inequalities, We first note that
\aln{
\braket{\ddot{\hat{A}}}&=-i\sum_{n\sim_A m}(a_n-a_m)\Tr[\hat{H}_{nm}\dot{\hat{\rho}}_{mn}]\nonumber\\
&=-i\sum_{n ml}(a_n-a_m)\Tr[\hat{H}_{nm}(\hat{H}_{ml}\hat{\rho}_{ln}-\hat{\rho}_{ml}\hat{H}_{ln})]\nonumber\\
&=-i\sum_{n ml}(a_n-a_m)\Tr[\hat{H}_{nm}\hat{H}_{ml}\hat{\rho}_{ln}+\hat{\rho}_{nl}\hat{H}_{lm}\hat{H}_{mn}]\nonumber\\
&=-\frac{i}{2}\sum_{m,n(\sim_Am),l(\sim_Am)}(a_n+a_l-2a_m)\nonumber\\
&\quad\quad\quad\quad\quad\times\Tr[\hat{H}_{nm}\hat{H}_{ml}\hat{\rho}_{ln}+\hat{\rho}_{nl}\hat{H}_{lm}\hat{H}_{mn}].
}
We then have
\aln{
|\braket{\ddot{\hat{A}}}|&\leq 
\max_{m,n(\sim_Am),l(\sim_Am)}|a_n+a_l-2a_m|\nonumber\\
&\quad\quad\times\sum_{m,n(\sim_Am),l(\sim_Am)}|\Tr[\hat{H}_{nm}\hat{H}_{ml}\hat{\rho}_{ln}]|\nonumber\\
&\leq 
\max_{m,n(\sim_Am),l(\sim_Am)}|a_n+a_l-2a_m|\cdot Q_p,
}
where
\aln{
Q_p=\sum_{m,n(\sim_Am),l(\sim_Am)}\sqrt{p_np_l}\|\hat{H}_{nm}\|_\mr{op}\|\hat{H}_{ml}\|_\mr{op}.
}
We can also show the state-independent inequality
\aln{
|\braket{\ddot{\hat{A}}}|\leq 
\max_{m,n(\sim_Am),l(\sim_Am)}|a_n+a_l-2a_m|\cdot C_H^2,
}
since
\aln{
Q_p&\leq \sum_{m,n(\sim_Am),l(\sim_Am)}\frac{p_n+p_l}{2}\|\hat{H}_{nm}\|_\mr{op}\|\hat{H}_{ml}\|_\mr{op}\nonumber\\
&= \sum_{n\in\mc{V}}p_n\sum_{m(\sim_An)}\|\hat{H}_{nm}\|_\mr{op}\sum_{l(\sim_Am)}\|\hat{H}_{ml}\|_\mr{op}\leq C_H^2.
}

\section{Derivation of the speed limit for coherence for general states}\label{cohapp}
Here, we derive the speed limit for macroscopic coherence for arbitrary states.
Using the assumption $c_{nm}=c_{mn}$, we have
\aln{
 \frac{\mathrm{d} \mathcal{C}}{\mathrm{d} t}
 &=-i\sum_{nml}c^{nm}\Tr[(\hat{H}_{nl}\hat{\rho}_{lm}-\hat{\rho}_{nl}\hat{H}_{lm})\hat{\rho}_{mn}]\nonumber\\
&\quad\quad-i\sum_{nml}c^{nm}\Tr[\hat{\rho}_{nm}(\hat{H}_{ml}\hat{\rho}_{ln}-\hat{\rho}_{ml}\hat{H}_{ln})]\nonumber\\
 &=-2i\sum_{nml}c^{nm}\Tr[\hat{H}_{nl}\hat{\rho}_{lm}\hat{\rho}_{mn}-\hat{\rho}_{ml}\hat{H}_{ln}\hat{\rho}_{nm}]\nonumber\\
 &=-i \sum_{nml}\left(c^{n m}-c^{lm}\right) \operatorname{Tr}\left[\hat{H}_{n l} \hat{\rho}_{l m} \hat{\rho}_{m n}-\hat{H}_{l n} \hat{\rho}_{n m} \hat{\rho}_{m l}\right] \nonumber\\ &=\sum_{n\sim_\mc{C} l,m}\left(c^{n m}-c^{m l}\right) \mathcal{J}_{n l m},
}
where
\aln{
\mathcal{J}_{n l m}=-i\operatorname{Tr}\left[\hat{H}_{n l} \hat{\rho}_{l m} \hat{\rho}_{m n}-\hat{H}_{l n} \hat{\rho}_{n m} \hat{\rho}_{m l}\right].
}
Thus, we have, e.g.,  
\aln{
|\dot{\mc{C}}|&\leq \max_{n\sim_\mc{C}l,m}|c^{nm}-c^{ml}|\sum_{n\sim_\mc{C} l,m}|\mathcal{J}_{n l m}|
}
While we focus on this inequality here, inequalities using $r_{nm}$ are obtained similarly.

Now, as in inequality~\eqref{JRbound} to bound $\sum_{nm}|J_{nm}|$, we can show that
\aln{
\sum_{n\sim_\mc{C} l,m}|\mathcal{J}_{n l m}|
\leq 2\sqrt{\mathtt{R}^2-\mathtt{E}^2},
}
where
\aln{
\mathtt{R}=\sum_{n lm}
|\Tr[\hat{H}_{n l} \hat{\rho}_{l m} \hat{\rho}_{m n}]|
}
and
\aln{
\mathtt{E}=\sum_{n lm}\Tr[\hat{H}_{n l} \hat{\rho}_{l m} \hat{\rho}_{m n}]=
\Tr\lrl{\sum_{n l}\hat{H}_{n l}  \hat{\rho}
^2}.
}
We can further show that
\aln{
\mathtt{R}\leq
\sum_{n lm}\|\hat{H}_{nl}\|_\mr{op}\|\hat{\rho}_{l m} \hat{\rho}_{m n}\|_1.
}
The trace-1 norm is bounded using the representation $\hat{\rho}=\sum_k\lambda_k\ket{\lambda_k}\bra{\lambda_k}$:
\aln{
\|\hat{\rho}_{l m} \hat{\rho}_{m n}\|_1
&\leq \sum_{kq}\lambda_k\lambda_q
\|\hat{P}_l\ket{\lambda_k}\bra{\lambda_k}\hat{P}_m\ket{\lambda_q}\bra{\lambda_q}\hat{P}_n\|_1\nonumber\\
&=
\sum_{kq}\lambda_k\lambda_q
\sqrt{\braket{\lambda_k|\hat{P}_l|\lambda_k}\braket{\lambda_q|\hat{P}_n|\lambda_q}}|\braket{\lambda_k|\hat{P}_m|\lambda_q}|\nonumber\nonumber\\
&\leq\sqrt{\sum_{kq}\lambda_k\lambda_q\braket{\lambda_k|\hat{P}_l|\lambda_k}\braket{\lambda_q|\hat{P}_n|\lambda_q}}\nonumber\\
&\quad\quad\times\sqrt{\sum_{kq}\lambda_k\lambda_q|\braket{\lambda_k|\hat{P}_m|\lambda_q}|^2}\nonumber\\
&=\sqrt{p_lp_n\Tr[(\hat{\rho}_{mm})^2]}.
}
Thus, we have
\aln{
|\dot{\mc{C}}|&\leq 2\max_{n\sim_\mc{C}l,m}|c^{nm}-c^{ml}| \cdot\sqrt{\mathtt{\tilde{R}}^2-\mathtt{E}^2},
}
where 
\aln{
\mathtt{\tilde{R}}=R_p\sum_m\sqrt{\Tr[(\hat{\rho}_{mm})^2]}.
}

When $c_{nm}$ is assumed to be a distance between $n$ and $m$, which satisfies the triangle inequality $|c^{nm}-c^{ml}|\leq c^{nl}$, we have
\aln{
|\dot{\mc{C}}|&\leq \max_{n\sim_\mc{C}l}(c^{nl}) \cdot\sum_{n\sim_\mc{C} l,m}|\mathcal{J}_{n l m}|,
}
For pure states, we recover inequality~\eqref{cohbound} since $\mathtt{\tilde{R}}=R_p$ and $\mathtt{E}=E_\mr{trans}$ in this case.

\section{Derivation of the inequality~\eqref{varfint}}\label{Appvarfint}
We have, using the Cauchy-Schwarz inequality,
\aln{
|\dot{\mbb{V}}[\hat{A}]|\leq &2\cdot\frac{\|\nabla A\|_\infty}{2}\sqrt{\sum_{n\sim_Am}(a_n-\braket{A})^2p_n\|\hat{H}_{nm}\|_\mr{op}}\nonumber\\
&\times\sqrt{\sum_{n\sim_Am}\frac{|J_{nm}^q|^2}{p_n\|\hat{H}_{nm}\|_\mr{op}}}.
}
Here, 
\aln{
\sum_{n\sim_Am}(a_n-\braket{A})^2&p_n\|\hat{H}_{nm}\|_\mr{op}\nonumber\\
=&\sum_{n}(a_n-\braket{A})^2p_n\sum_{m(\sim_An)}\|\hat{H}_{nm}\|_\mr{op}\nonumber\\
\leq &C_H{\mbb{V}}[\hat{A}]
}
and
{
\aln{
{\sum_{n\sim_Am}\frac{|J_{nm}^q|^2}{p_n\|\hat{H}_{nm}\|_\mr{op}}}
&\leq \sum_{n\sim_Am}\frac{4r_{nm}^2\sin^2\theta_{nm}}{p_n\|\hat{H}_{nm}\|_\mr{op}}\nonumber\\
&\leq \sum_{n\sim_Am}\frac{4r_{nm}^2}{p_n\|\hat{H}_{nm}\|_\mr{op}}\lrs{1+\frac{d^2}{4}-d|\cos\theta_{nm}|}\nonumber\\
&\leq \sum_{n\sim_Am}\frac{4r_{nm}^2}{p_n\|\hat{H}_{nm}\|_\mr{op}}\lrs{1+\frac{d^2}{4}-d|\cos\theta_{nm}|}\nonumber\\
&\leq \sum_{n\sim_Am}[(4+d^2)p_m\|\hat{H}_{nm}\|_\mr{op}\nonumber\\&\quad\quad\quad-4dp_m\|\hat{H}_{nm}\|_\mr{op}|\cos\theta_{nm}|]\nonumber\\
&\leq (4+d^2)C_H\nonumber\\&\quad\quad\quad-\sum_{n\sim_Am}2d(p_m+p_n)\|\hat{H}_{nm}\|_\mr{op}|\cos\theta_{nm}|\nonumber\\
&\leq (4+d^2)C_H\nonumber-\sum_{n\sim_Am}4dr_{nm}\cos\theta_{nm}|\nonumber\\
&\leq (4+d^2)C_H\nonumber-4d|E_\mr{trans}|\nonumber
}
with $d\in\mbb{R}$,
where we have used inequality~\eqref{sincos1} and $r_{nm}\leq\sqrt{p_np_m}\|\hat{H}_{nm}\|_\mr{op}\leq \frac{p_n+p_m}{2}\|\hat{H}_{nm}\|_\mr{op}$.
Optimization is carried out by taking $d=2|E_\mr{trans}|/C_H$, which leads to 
\aln{
{\sum_{n\sim_Am}\frac{|J_{nm}^q|^2}{p_n\|\hat{H}_{nm}\|_\mr{op}}}
\leq 4C_H-\frac{4|E_\mr{trans}|^2}{C_H}.
}
Thus
\aln{
|\dot{\mbb{V}}[\hat{A}]|
\leq 2{\|\nabla A\|_\infty}\sqrt{{\mbb{V}}[\hat{A}](C_H^2-|E_\mr{trans}|^2)}
}
and
\aln{
\lrv{\fracd{\Delta A}{t}}\leq {\|\nabla A\|_\infty}\sqrt{C_H^2-E_\mr{trans}^2}.
}
Then, integration from $t=0$ to $t=T$ leads to
\aln{
\lrv{\Delta A(T)-\Delta A(0)}&\leq {\|\nabla A\|_\infty}\int_0^Tdt\sqrt{C_H^2-E_\mr{trans}^2.
}\nonumber\\
&={\|\nabla A\|_\infty}T\av{\sqrt{C_H^2-E_\mr{trans}^2}}
}
}

\section{Derivation of the speed limit for Shannon entropy}\label{entapp}
We consider the speed limit of the Shannon entropy
\aln{
S(p)=-\sum_n p_n\ln p_n.
}
We have
\aln{
|\dot{S}|&=\lrv{\sum_n \dot{p}_n\ln p_n}\nonumber\\
&\leq \lrv{\frac{1}{2}\sum_{n\sim m}J_{nm}^q\ln\frac{p_n}{p_m}}\nonumber\\
&\leq\frac{1}{2}\sqrt{\sum_{n\sim m}r_{nm}\lrs{\ln\frac{p_n}{p_m}}^2\sum_{n\sim m}\frac{|J_{nm}^q|^2}{r_{nm}}}\nonumber\\
&\leq 
\sqrt{\sum_{n\sim m}r^p_{nm}\lrs{\ln\frac{p_n}{p_m}}^2}
\sqrt{R_p-\frac{E_\mr{trans}^2}{R_p}}.
}
To proceed, we notice the following inequality
\aln{
\sqrt{p_np_m}&= \frac{p_n+p_m}{2}-\frac{(\sqrt{p_n}-\sqrt{p_m})^2}{2}\nonumber\\
&\leq\frac{p_n+p_m}{2}-\frac{\sqrt{p_np_m}}{8}\lrs{\ln\frac{p_n}{p_m}}^2,
}
where we have used
\aln{
(\sqrt{a}-\sqrt{b})^2\geq\frac{\sqrt{ab}}{4}\lrs{\ln\frac{a}{b}}^2
}
for $a,b>0$.
Then,
\aln{
\sum_{n\sim m}r_{nm}^p\lrs{\ln\frac{p_n}{p_m}}^2\leq 8(C_H-R_p)
}
and
\aln{
|\dot{S}|\leq \sqrt{8\lrs{\frac{C_H}{R_p}-1}}\sqrt{{R_p^2-{E_\mr{trans}^2}}}
}
By replacing $E_\mr{trans}^2/R_p$ with $E_\mr{trans}^2/C_H$ and completing the square, we have
\aln{\label{comps}
|\dot{S}|\leq \sqrt{2}\lrs{C_H-\frac{E_\mr{trans}^2}{C_H}}.
}

We note that a different type of inequality holds by noting
\aln{
R_p\leq C_H-\frac{1}{8}\sum_{n\sim m}r_{nm}^p\lrs{\ln\frac{p_n}{p_m}}^2.
}
We  introduce the quantity similar to the symmetric version of discrete Fisher information
\aln{
\mc{I}(p)=\frac{1}{2}\sum_{n\sim m}r_{nm}^p\lrs{\ln\frac{p_n}{p_m}}^2,
}
and obtain
\aln{
|\dot{S}|\leq \sqrt{2\mc{I}(p)\lrs{C_H-\frac{E_\mr{trans}^2}{C_H}-\frac{\mc{I}(p)}{4}}}.
}
Completing the square again leads to inequality~\eqref{comps}.

\section{Tightness of the bounds for simple systems}\label{tightapp}
As discussed in Sec.~\ref{SecIV}, for a single-particle one-dimensional system with $W_l=0$, the initial state $\ket{\psi_1}$ leads to
\aln{
\braket{\dot{\hat{x}}}=\mc{B}_\mr{gL}=\mc{B}_\mr{Lip}=\mc{B}_{p}
}
for short times $t\lesssim 2$, indicating the tightness of these bounds.
We here discuss why this tightness holds true in this situation.

We first show $\mc{B}_\mr{Lip}=\mc{B}_{p}$.
The left-hand side can be written as
\aln{
\mc{B}_\mr{Lip}={K}\sum_{l}|\braket{\psi(t)|\hat{a}_{l+1}^\dag\hat{a}_l-\hat{a}_{l}^\dag\hat{a}_{l+1}|\psi(t)}|
}
because $\|\nabla x\|_\infty=1$.
To proceed, we  show that 
\aln{\label{symrel}
\braket{\psi(t)|\hat{a}_{l+1}^\dag\hat{a}_l|\psi(t)}=-\braket{\psi(t)|\hat{a}_{l}^\dag\hat{a}_{l+1}|\psi(t)}
}
for $\ket{\psi_1}$ and $\ket{\psi_2}$ with any $l$ and $W_l=0$.
To see this, we introduce the following anti-unitary symmetry
\aln{
\hat{\mc{T}}=\hat{\mc{K}}\bigotimes_{l=1}^{L/2}(2\hat{n}_{2l}-1),
}
where $\hat{\mc{K}}$ denotes the complex conjugation operator.
We then find, assuming the commutation relation for the hardcore boson~\footnote{Our discussion can be similarly made for fermions.}
\aln{
\hat{\mc{T}}\ket{\psi_{1}}\propto\ket{\psi_{1}},\:\hat{\mc{T}}\ket{\psi_{2}}\propto\ket{\psi_{2}},
}
\aln{
\hat{\mc{T}}e^{-i\hat{H}t}\hat{\mc{T}}^{-1}=e^{-i\hat{H}t},
}
and
\aln{
\hat{\mc{T}}\hat{a}_{l+1}^\dag\hat{a}_l\hat{\mc{T}}^{-1}=-\hat{a}_{l+1}^\dag\hat{a}_l.
}
Then, we have
\aln{
\braket{\psi(t)|\hat{a}_{l+1}^\dag\hat{a}_l|\psi(t)}&=
(\ket{\psi},e^{i\hat{H}t}\hat{a}_{l+1}^\dag\hat{a}_le^{-i\hat{H}t}\ket{\psi})\nonumber\\
&=
(\hat{\mc{T}}\ket{\psi},\hat{\mc{T}}e^{i\hat{H}t}\hat{a}_{l+1}^\dag\hat{a}_le^{-i\hat{H}t}\ket{\psi})^*\nonumber\\
&=
(\ket{\psi},-e^{i\hat{H}t}\hat{a}_{l+1}^\dag\hat{a}_le^{-i\hat{H}t}\ket{\psi})^*\nonumber\\
&=-\braket{\psi(t)|\hat{a}_{l}^\dag\hat{a}_{l+1}|\psi(t)}.
}
From this relation, we find
\aln{
\mc{B}_\mr{Lip}&=2{K}\sum_{l}|\braket{\psi(t)|\hat{a}_{l+1}^\dag\hat{a}_l|\psi(t)}|\nonumber\\
&=2K\sum_l\sqrt{p_lp_{l+1}}=R_p.
}
On the other hand, Eq.~\eqref{symrel} also leads to $E_\mr{trans}=0$.
Since we consider pure states and $\sqrt{2[x^\mathsf{T}\nabla_{r^p}^2x]/R_p}=1$, we then find
\aln{
\mc{B}_p=R_p,
}
which is equal to $\mc{B}_\mr{Lip}$,
Similarly, we have $\mc{B}_\mr{gL}=\mc{B}_\mr{Lip}$.

We next discuss the condition for $\braket{\dot{\hat{x}}}=\mc{B}_\mr{Lip}$.
To see this, we notice that the equality condition of the inequality~\eqref{ho} is
obtained when the signs of $(a_n-a_m)J_{mn}$ are equal (or zero) for all $n\sim_A m$.
For the current situation, this reduces to the condition that the signs of $J_{l,l+1}$ are equal (or zero) for all $l$.
For short times from the initial state $\ket{\psi_1}$ we find that this condition is indeed satisfied with $J_{l+1,l}\geq 0$.
Note that this is consistent with the intuition that the positive current 
indicates the transfer of the particle to the right.

\section{Derivation of the speed limits for classical stochastic systems}\label{Appcs}
Here, we prove the speed limits for classical stochastic systems.
We first derive the bound using the entropy production, which is done with our general framework and the inequalities used in Refs.~\cite{PhysRevLett.117.190601,PhysRevLett.121.070601}.
Then we derive the bound using the Hatano-Sasa entropy production for general Markovian systems.

\subsection{Speed limit using the entropy production rate}
We apply the inequality~\eqref{cs} with $r_{nm}=W_{nm}p_m+W_{mn}p_n$. 
We have
\aln{
|\braket{\dot{A}}|\leq \sqrt{\frac{\Theta_A}{2}\sum_{n\sim_A m}\frac{(W_{nm}p_m-W_{mn}p_n)^2}{W_{nm}p_m+W_{mn}p_n}},
}
where
\aln{
\Theta_A&=\frac{1}{2}\sum_{n\sim m}(a_n-a_m)^2(W_{nm}p_m+W_{mn}p_n)\nonumber\\
&=\sum_{n\sim m}(a_n-a_m)^2W_{nm}p_m
}
is the second moment of the transition speed.
Now, using $\frac{2(a-b)^2}{a+b}\leq (a-b)\log\frac{a}{b}$ for $a,b>0$~\cite{PhysRevLett.117.190601,PhysRevLett.121.070601}, we have
\aln{
\sum_{n\sim_Am}\frac{(W_{nm}p_m-W_{mn}p_n)^2}{W_{nm}p_m+W_{mn}p_n}
&\leq 
\sum_{n\sim_Am}W_{nm}p_m\ln\frac{W_{nm}p_m}{W_{mn}p_n}\nonumber\\
&:=\dot{\Sigma}_\mr{cg}.
}
We then obtain 
\aln{
|\braket{\dot{A}}|\leq\sqrt{\frac{\Theta_A\dot{\Sigma}_\mr{cg}}{2}}\leq \|\nabla A\|_\infty\sqrt{\frac{\mc{A}_\mr{cg}\dot{\Sigma}_\mr{cg}}{2}},
}
where $\mc{A}_\mr{cg}$ is the coarse-grained dynamical activity.

The coarse-grained variables above are bounded from above by the non-coarse-grained variables.
In fact, we have
\aln{
\mc{A}_\mr{cg}=\sum_{x\in\mc{S}_n}\sum_{y\in\mc{S}_m}\mc{W}_{xy}\rho_y\leq \mc{A}
}
and
\aln{\label{lscg}
\dot{\Sigma}_\mr{cg}&\leq\sum_{n\neq m}\sum_{x\in\mc{S}_n}\sum_{y\in\mc{S}_m}\mc{W}_{xy}\rho_y
\ln\frac{\sum_{x\in\mc{S}_n}\sum_{y\in\mc{S}_m}\mc{W}_{xy}\rho_y}{\sum_{y\in\mc{S}_m}\sum_{x\in\mc{S}_n}\mc{W}_{yx}\rho_x}\nonumber\\
&\leq\sum_{n\neq m}\sum_{x\in\mc{S}_n}\sum_{y\in\mc{S}_m}\mc{W}_{xy}\rho_y\ln\frac{\mc{W}_{xy}\rho_y}{\mc{W}_{yx}\rho_x}\nonumber\\
&\leq \sum_{x\neq y}\mc{W}_{xy}\rho_y\ln\frac{\mc{W}_{xy}\rho_y}{\mc{W}_{yx}\rho_x}= \dot{\Sigma},
}
where we have used the log-sum inequality~\cite{cover1999elements}.

Instead of the expectation values of observables, we can consider the speed limit for ${\Sigma}_\mr{cg}$ itself.
Indeed, we have
\aln{
\dot{\Sigma}_\mr{cg}
&=\frac{1}{2}\sum_{n\sim_Am}J^c_{nm}\ln\frac{W_{nm}p_m}{W_{mn}p_n}\nonumber\\
&\leq
\sqrt{\frac{1}{2}\sum_{n\sim_Am}W_{nm}p_m\lrs{\ln\frac{W_{nm}p_m}{W_{mn}p_n}}^2\dot{\Sigma}_\mr{cg}}.
}
Thus,
\aln{\label{sigsig}
\dot{\Sigma}_\mr{cg}&\leq\frac{1}{2}\sum_{n\sim_Am}W_{nm}p_m\lrs{\ln\frac{W_{nm}p_m}{W_{mn}p_n}}^2:=\frac{\Theta_\Sigma}{2},
}
where $\Theta_\Sigma$ is the second moment of $\dot{\Sigma}_\mr{cg}$.
We then obtain
\aln{
|\braket{\dot{A}}|\leq\frac{\sqrt{\Theta_A\Theta_\Sigma}}{2}.
}

When the detailed balance condition is satisfied and the non-coarse-grained version is considered, inequality~\eqref{sigsig}  is further bounded as
\aln{
\dot{\Sigma}\leq{\max_{x\sim_Ay}\mc{W}_{xy}}\cdot\tilde{\mc{F}}(\rho||\rho^\mr{ss}),
}
where 
\aln{
\tilde{\mc{F}}(\rho||\rho^\mr{ss})
=\frac{1}{2}\sum_{x\sim y}\rho_x\lrs{\ln\frac{\rho_x}{\rho_y}-\ln\frac{\rho^\mr{ss}_x}{\rho_y^\mr{ss}}}^2
}
is the discrete version of the Fisher divergence~\cite{lyu2012interpretation}.

\subsection{Speed limit using the Hatano-Sasa entropy production rate}
We next derive the speed limit with the Hatano-Sasa entropy production rate.
For this purpose, we first notice that
\aln{
-J_{mn}^c=W_{nm}p_m-\tilde{W}_{mn}p_n-J_{mn}^{c,\mr{ss}}\frac{p_n}{p_n^\mr{ss}},
}
where
\aln{
J_{mn}^{c,\mr{ss}}=-W_{nm}p_m^\mr{ss}+W_{mn}p_n^\mr{ss}
}
is the stationary current.

From this, we have
\aln{
\braket{\dot{A}}=&-\frac{1}{2}\sum_{n\sim_A m}(a_n-a_m)J_{mn}^{c,\mr{ss}}\frac{p_n}{p_n^\mr{ss}}\\
&+\frac{1}{2}\sum_{n\sim_A m}(a_n-a_m)(W_{nm}p_m-\tilde{W}_{mn}p_n).
}
Then, introducing
\aln{
V_A&=-\frac{1}{2}\sum_{n\sim_A m}(a_n-a_m)J_{mn}^{c,\mr{ss}}\frac{p_n}{p_n^\mr{ss}}\\
&=-\frac{1}{2}\sum_{n\sim_A m}a_nJ_{mn}^{c,\mr{ss}}\lrs{\frac{p_n}{p_n^\mr{ss}}+\frac{p_m}{p_m^\mr{ss}}},
}
whose value becomes a stationary current of ${A}$ for the stationary state,
we have
\aln{
|\braket{\dot{A}}-V_A|= \lrv{\frac{1}{2}\sum_{n\sim_A m}(a_n-a_m)(W_{nm}p_m-\tilde{W}_{mn}p_n)
}.
}
To relate the right-hand side with the Hatano-Sasa entropy production,
we use the following deformation of the inequality:
\aln{
&|\braket{\dot{A}}-V_A|^2\nonumber\\&=
\frac{1}{2}\int_0^1dzz\lrv{\sum_{n\sim_A m}(a_n-a_m)(W_{nm}p_m-\tilde{W}_{mn}p_n)
}^2\nonumber\\
&\leq
\frac{1}{2}\int_0^1dz{\sum_{n\sim_A m}(a_n-a_m)^2(zW_{nm}p_m+(1-z)\tilde{W}_{mn}p_n)}\nonumber\\
&\quad\quad\times{\sum_{n\sim_A m}\frac{z(W_{nm}p_m-\tilde{W}_{mn}p_n)^2}{zW_{nm}p_m+(1-z)\tilde{W}_{mn}p_n}}\nonumber\\
&\leq
\frac{1}{2}\max_{0\leq z\leq 1}\lrl{\sum_{n\sim_A m}(a_n-a_m)^2(zW_{nm}p_m+(1-z)\tilde{W}_{mn}p_n)}\nonumber\\
&\quad\quad\times\int_0^1dz{\sum_{n\sim_A m}\frac{z(W_{nm}p_m-\tilde{W}_{mn}p_n)^2}{zW_{nm}p_m+(1-z)\tilde{W}_{mn}p_n}}\nonumber\\
&\leq{\frac{\tilde{\Theta}_A\dot{\Sigma}_\mr{cg}^\mr{HS}}{2}},
}
where 
\aln{
\tilde{\Theta}_A&=\max_{0\leq z\leq 1}\lrl{\sum_{n\sim_A m}(a_n-a_m)^2(zW_{nm}p_m+(1-z)\tilde{W}_{mn}p_n)}\nonumber\\
&=\max\lrl{\sum_{n\sim_A m}(a_n-a_m)^2W_{nm}p_m,\sum_{n\sim_A m}(a_n-a_m)^2\tilde{W}_{nm}p_m}
}
and
we have used $\int_0^1dz\frac{z(a-b)^2}{za+(1-z)b}=a\ln\frac{a}{b}-a+b$~\cite{PhysRevE.102.062132}.
Since
\aln{
\tilde{\Theta}_A\leq \|\nabla A\|_{\infty}^2\mc{A}_\mr{cg},
}
we have
\aln{
|\braket{\dot{A}}-V_A|\leq\sqrt{{\frac{\tilde{\Theta}_A\dot{\Sigma}_\mr{cg}^\mr{HS}}{2}}}\leq \|\nabla A\|_{\infty}\sqrt{\frac{\mc{A}_\mr{cg}\dot{\Sigma}_\mr{cg}^\mr{HS}}{2}}.
}

\section{Derivation of inequality~\eqref{oq2}}\label{lindapp}
We here show the inequality \eqref{oq2}.
We first note that
\aln{\label{keisantotyu}
\sum_na_nK_n^\eta=\sum_{nml}\lrs{a_n-\frac{a_m+a_l}{2}}\Tr[(\hat{L}_\eta)_{nm}\hat{\rho}_{ml}(\hat{L}_\eta^\dag)_{ln}].
}

As mentioned in the main text, we impose one assumption about the jump operator: we require that each jump $\eta$ moves a state in the subspace $\mc{H}_m$ to that in $\mc{H}_{n=f_\eta(m)}$ with an injective function $f_\eta$.
Put differently, when $(\hat{L}_\eta)_{nm}$ is nonzero for some $n$  and $m$, 
$(\hat{L}_\eta)_{n'm}\:(n'\neq n)$ and
$(\hat{L}_\eta)_{nm'}\:(m'\neq m)$ should be zero.
Under this assumption, we can set $m=l$ in Eq.~\eqref{keisantotyu} and have
\aln{
\sum_\eta\sum_na_nK_n^\eta&=\sum_{nm}(a_n-a_m)\sum_\eta\Tr[(\hat{L}_\eta^\dag)_{mn}(\hat{L}_\eta)_{nm}\hat{\rho}_{mm}]\nonumber\\
&=\frac{1}{2}\sum_{nm}(a_n-a_m)\lrs{T_{m\ra n}^\mr{cg}-T_{n\ra m}^\mr{cg}}
}
with
$T_{m\ra n}^\mr{cg}=\sum_\eta\Tr[(\hat{L}_\eta^\dag)_{mn}(\hat{L}_\eta)_{nm}\hat{\rho}_{mm}].$

We use the same technique discussed in Appendix~\ref{Appcs} to obtain inequality~\eqref{oq2}:
\aln{
&\lrv{\sum_\eta\sum_na_nK_n^\eta}\nonumber\\
&\leq
\frac{1}{2}\max_{(n,m)\in\mc{E}_\mr{D}}|a_n-a_m|\nonumber\\
&\times\sqrt{\sum_{nm}(T_{m\ra n}^\mr{cg}+T_{n\ra m}^\mr{cg})\sum_{nm}\frac{(T_{m\ra n}^\mr{cg}-T_{n\ra m}^\mr{cg})^2}{T_{m\ra n}^\mr{cg}+T_{n\ra m}^\mr{cg}}}\nonumber\\
&\leq \max_{(n,m)\in\mc{E}_\mr{D}}|a_n-a_m|\sqrt{\frac{\mc{A}_\mr{cg}^d\dot{\Sigma}_\mr{cg}^d}{2}},
}
where we have defined
\aln{
\mc{A}_\mr{cg}^d=\sum_{n\neq m}T_{m\ra n}^\mr{cg},
}
and
\aln{
\dot{\Sigma}_\mr{cg}^d =\sum_{n,m}T_{m\ra n}^\mr{cg}\ln\frac{T_{m\ra n}^\mr{cg}}{T_{n\ra m}^\mr{cg}}\geq 0.
}

\section{Derivation of inequality~\eqref{concent}}\label{Appconc}
We start from evaluating the speed of $\braket{e^{\pm\lambda \hat{A}}}$ with $\lambda>0$:
\aln{
\lrv{\fracd{\braket{e^{\pm\lambda\hat{A}}}}{t}}&\leq
\lrv{\sum_ne^{\pm\lambda a_n}\dot{p}_n}\nonumber\\
&\leq
\frac{1}{2}\lrv{\sum_{n\sim_A m}(e^{\pm\lambda a_n}-e^{\pm\lambda a_m})J_{nm}^q}\nonumber\\
&\leq
\lrv{\sum_{n\sim_A m}(e^{\pm\lambda a_n}-e^{\pm\lambda a_m})\sqrt{p_np_m}\|\hat{H}_{nm}\|_\mr{op}}.
}
Now, we use the fact that, for $0<\lambda \leq\lambda_M:=\frac{2}{\|\nabla A\|_\infty}\log\frac{\sqrt{5}-1}{2}$, we have
\aln{\label{anam}
e^{\pm\lambda a_n}-e^{\pm\lambda a_m}\leq e^{\pm\frac{\lambda a_n}{2}}e^{\pm\frac{\lambda a_m}{2}}
}
for arbitrary $n\sim m$.
In fact, inequality~\eqref{anam} is equivalent to 
$y^2-y-1\leq 0$ with $y=e^{\pm\frac{\lambda}{2}(a_n-a_m)}$.
This is satisfied when $\pm \lambda (a_n-a_m)\leq 2\log\frac{\sqrt{5}-1}{2}$.
Since the maximum value of $\pm (a_n-a_m)$ is bounded by $\|\nabla A\|_\infty$, $0<\lambda\leq\lambda_M$ ensures inequality \eqref{anam}.

Then, we have
\aln{
\lrv{\fracd{\braket{e^{\pm\lambda\hat{A}}}}{t}}&\leq
\lrv{\sum_{n\sim_A m}(e^{\pm\frac{\lambda a_n}{2}}e^{\pm\frac{\lambda a_m}{2}})\sqrt{p_np_m}\|\hat{H}_{nm}\|_\mr{op}}\nonumber\\
&\leq \sqrt{\sum_{n\sim_A m}e^{\pm\lambda a_n}p_n\|\hat{H}_{nm}\|_\mr{op}}\sqrt{\sum_{n\sim_A m}e^{\pm\lambda a_m}p_m\|\hat{H}_{nm}\|_\mr{op}}\nonumber\\
&\leq C_H\braket{e^{\pm\lambda \hat{A}}}.
}
Integration from time 0 to $t$ leads to
\aln{
\braket{e^{\pm\lambda \hat{A}}}
\leq \braket{e^{\pm\lambda \hat{A}}}_0e^{\av{C_H}t},
}
where $\braket{\cdots}_0$ is the expectation value with respect to $\hat{\rho}(0)$.

Then, the Markov inequality indicates that
\aln{
&\mbb{P}_t\lrl{\lrv{\frac{A-A'}{\|\nabla A\|_\infty}}>\epsilon }\nonumber\\
&=\mbb{P}_t\lrl{{{A-A'}>{\|\nabla A\|_\infty}}\epsilon }
+\mbb{P}_t\lrl{{A'-A>{\|\nabla A\|_\infty}}\epsilon }\nonumber\\
&\leq \frac{\braket{e^{\lambda(\hat{A}-A')}}}{e^{\lambda \|\nabla A\|_\infty\epsilon}}+\frac{\braket{e^{-\lambda(\hat{A}-A')}}}{e^{\lambda \|\nabla A\|_\infty\epsilon}}\nonumber\\
&\leq \braket{e^{\lambda (\hat{A}-A')}}_0 e^{\av{C_H}t-\lambda \|\nabla A\|_\infty\epsilon}+\braket{e^{-\lambda (\hat{A}-A')}}_0 e^{\av{C_H}t-\lambda \|\nabla A\|_\infty\epsilon}.
}
Setting $\lambda$ as $\lambda \|\nabla A\|_\infty$,
we have
\aln{
&\mbb{P}_t\lrl{\lrv{\frac{A-A'}{\|\nabla A\|_\infty}}>\epsilon }\leq
2\braketL{\cosh\frac{\lambda(\hat{A}-A')}{\|\nabla A\|_\infty}}_0e^{\av{C_H}t-\lambda \epsilon}
}
for $0<\lambda\leq 2\log\frac{\sqrt{5}-1}{2}$.

\bibliography{speed.bib}

\end{document}